\documentclass{aa}  
\usepackage{graphicx}
\usepackage{txfonts}
\usepackage{lipsum}
\usepackage{subcaption}         
\usepackage{lscape}            
\usepackage{placeins} 
\usepackage{multirow}
\usepackage{longtable}
\usepackage{booktabs} 
\usepackage{makecell}
\usepackage[normalem]{ulem}
\usepackage{xcolor}

\renewcommand*{\thesubfigure}{(\arabic{subfigure})}

\begin{document}

   \title{The occurrence rate of planets around subgiant stars from the TESS photometric survey}


%

    \author{A. Dutta\inst{1}
        \and P. Chaturvedi\inst{2, 3}\fnmsep\thanks{Corresponding Author: priyanka.chaturvedi@tifr.res.in}
        }

    \institute{Indian Institute of Science Education and Research Kolkata, Mohanpur, India, 741246
    \and Department of Astronomy and Astrophysics, Tata Institute of Fundamental Research, Mumbai, India, 400005
    \and Th\"uringer Landessternwarte Tautenburg, Sternwarte 5, 07778 Tautenburg, Germany}

   \date{Received 28 February 2026 / accepted 01 October 2026}
 
  \abstract
{Subgiant stars represent a critical phase in stellar evolution, offering an opportunity to investigate how planetary systems respond to the early stages of post-main-sequence expansion. Despite their importance, systematic occurrence-rate studies of this population remain sparse compared with those of main-sequence and red-giant hosts. In this paper, we present a uniform analysis of planet occurrence around a sample of 282,904 subgiant stars selected from the \textit{Gaia} Data Release 3 (DR3) catalogue. We performed an independent re-analysis of Transiting Exoplanet Survey Satellite (TESS) light curves for transit-signal detection and refinement, followed by multi-stage vetting including a visual inspection and statistical false-positive assessment. We quantified the detection sensitivity of the survey through injection-and-recovery experiments, with 565,808 synthetic transit signals injected across the stellar sample. Over the period range of 1--10 days and the planet-radius range for which the detection completeness is well characterised, we measure a completeness-corrected planet occurrence rate of $0.644^{+0.048}_{-0.041}\%$. The occurrence rate is strongly dependent on planet radius, with the highest occurrence contributed by planets in the $1$--$4\,R_{\oplus}$ range and a secondary enhancement at $12$--$15\,R_{\oplus}$. These results provide a demographic baseline for studying the population of close-in planets around stars undergoing early post-main-sequence evolution.}


   \maketitle
   \nolinenumbers


\section{Introduction}

The statistical characterisation of exoplanet occurrence rates has become an essential pillar in understanding the architectures of planetary systems. Over the past two decades, extensive photometric and radial-velocity surveys have enabled robust measurements of planet demographics around main-sequence stars, revealing that planets are both common and diverse across a wide range of orbital periods and sizes. In particular, the \textit{Kepler} mission \citep{Borucki2010} and the Transiting Exoplanet Survey Satellite (TESS; \citealt{2015JATIS...1a4003R}) have established that small planets with radii of 1--4~R$_\oplus$ constitute the dominant population around Sun-like stars, with occurrence rates exceeding 30\% within orbital periods of $\lesssim$100 days \citep{2012ApJS..201...15H, doi:10.1073/pnas.1319909110, 10.1093/mnras/staa2391}. In contrast, giant planets are significantly less common, though their occurrence rates are well constrained and exhibit strong dependencies on stellar mass and metallicity \citep{2005ApJ...622.1102F, 2015ApJ...798..112M}. 

While early occurrence-rate studies primarily focused on solar-type stars, a growing number of dedicated surveys have expanded this effort to include low-mass stars. Both radial-velocity surveys and the advent of space-based transit missions have enabled systematic measurements of planet occurrence around M-dwarf hosts. As the most abundant stellar population in the Galaxy, M dwarfs offer favourable conditions for planet detection, including deeper transit depths due to their small radii and larger planet-to-star mass ratios for radial-velocity measurements. Recent analyses of \textit{Kepler} and TESS data have demonstrated that small, short-period planets are particularly prevalent around M dwarfs, while the occurrence of giant planets is markedly lower than around Sun-like stars \citep{2015ApJ...807...45D, Gan2023, bryant2023occurrence}. These results have provided important constraints on the efficiency of giant-planet formation in low-mass protoplanetary disks and highlighted the strong dependence of planet demographics on stellar mass.

Although occurrence rates are now relatively well established for main-sequence stars across a broad range of stellar masses, significantly less attention has been paid to evolved stellar populations, particularly subgiant stars. Subgiants represent the first evolutionary phase beyond the main sequence; this phase is characterised by modest stellar expansion and decreasing surface gravity, yet precedes the dramatic structural changes associated with the red-giant phase. From a planetary perspective, subgiants offer a unique opportunity to probe the early stages of planetary-system evolution under changing stellar conditions, including variations in irradiation, tidal interactions, and potential orbital reconfiguration. These stars are also expected to be a primary focus of upcoming transit missions such as PLATO \citep{rauer2025plato}. 

Previous studies of planets around evolved stars have largely concentrated on red-giant hosts using radial-velocity surveys, often reporting a deficit of short-period planets and interpreting this as evidence of tidal engulfment or dynamical evolution (e.g. \citealt{2007ApJ...670..833J, 2009ApJ...705L..81V}). However, such surveys are typically limited by increased stellar variability and reduced sensitivity to smaller planets. Subgiant stars, by contrast, remain well suited to high-precision transit photometry and therefore provide a cleaner laboratory for investigating how planet populations transition from the main-sequence to the giant-branch regime. Despite their importance, systematic and completeness-corrected occurrence-rate studies specifically targeting subgiant stars remain sparse, particularly in the context of space-based transit surveys.

In this work, we addressed this gap by measuring the occurrence rate of transiting planets around subgiant stars using data from TESS. We constructed a well-defined sample of subgiant hosts, performed a homogeneous re-analysis of transit signals, applied statistical vetting to planet candidates, and quantified detection completeness through large-scale injection--recovery experiments. The resulting occurrence rates, derived as a function of orbital period and planet radius, provide a demographic baseline for planets orbiting subgiant stars and offer new observational constraints on the evolution of planetary systems as their host stars leave the main sequence.


\section{Stellar sample}

For this study, we constructed a sample of subgiant stars from the \textit{Gaia} Data Release 3 (DR3) catalogue and then cross-matched and retrieved their photometric time series from the Transiting Exoplanet Survey Satellite (TESS; \citealt{pub.1012025846}) quick-look protocol (QLP; \citealt{2020RNAAS...4..204H}). Drawing the initial sample from \textit{Gaia} DR3 ensures that robust, recent estimates of stellar parameters (such as mass, radius, and surface gravity; \citealt{2023A&A...674A..26C}) are available for target selection and downstream analysis.

We defined the subgiant sample using $3.5 \leq \log g \leq 4.2$ and $R_\star \leq 3\,R_\odot$. The surface-gravity range was chosen to include stars that have begun to evolve away from the main sequence while excluding the lower gravity giant population. The additional radius constraint limits the sample to mildly evolved stars and
removes substantially expanded giants, for which the detectability of transiting planets is strongly reduced. Together, these constraints allowed us to select a relatively homogeneous population of early-to-intermediate subgiants suitable for a uniform transit-occurrence-rate analysis. The set of conditions on stellar parameters is summarised in Table~\ref{tab:subgiant_params}.

\begin{table}[h!]
\centering
\caption{Stellar parameter criteria for subgiant sample selection.}
\begin{tabular}{lc}
\hline
\hline
Parameter & Selection \\
\hline
\textit{Gaia} G Magnitude (mag) & $\leq 11.5$ \\
\textit{Gaia} RUWE & $\leq 1.20$ \\
Surface Gravity ($\log g$) & $3.5$--$4.2$ \\
Stellar Radius ($R_\ast$) & $\leq 3.0\,R_{\odot}$ \\
Stellar Mass ($M_\ast$) & Available \\
Processing Pipeline & QLP 600s or 200s cadence \\
\hline
\end{tabular}
\label{tab:subgiant_params}
\end{table}

The adopted magnitude cut ($G \leq 11.5$) is motivated primarily by practical considerations related to the feasibility of future high-resolution spectroscopic follow-up, rather than by detection sensitivity. While this choice reduces the overall sample size, it ensures that any promising planet candidates identified in this study remain amenable to follow-up observations to refine orbital parameters and assess their nature. 

After the initial selection of targets from \textit{Gaia}, we queried the light curves of these \textit{Gaia} targets from TESS with processed photometry from MIT's QLP \citep{2020RNAAS...4..204H} pipeline and collected either 600s or 200s-cadence photometry. 

\begin{figure}[ht!]
    \centering
    \includegraphics[width=\columnwidth]{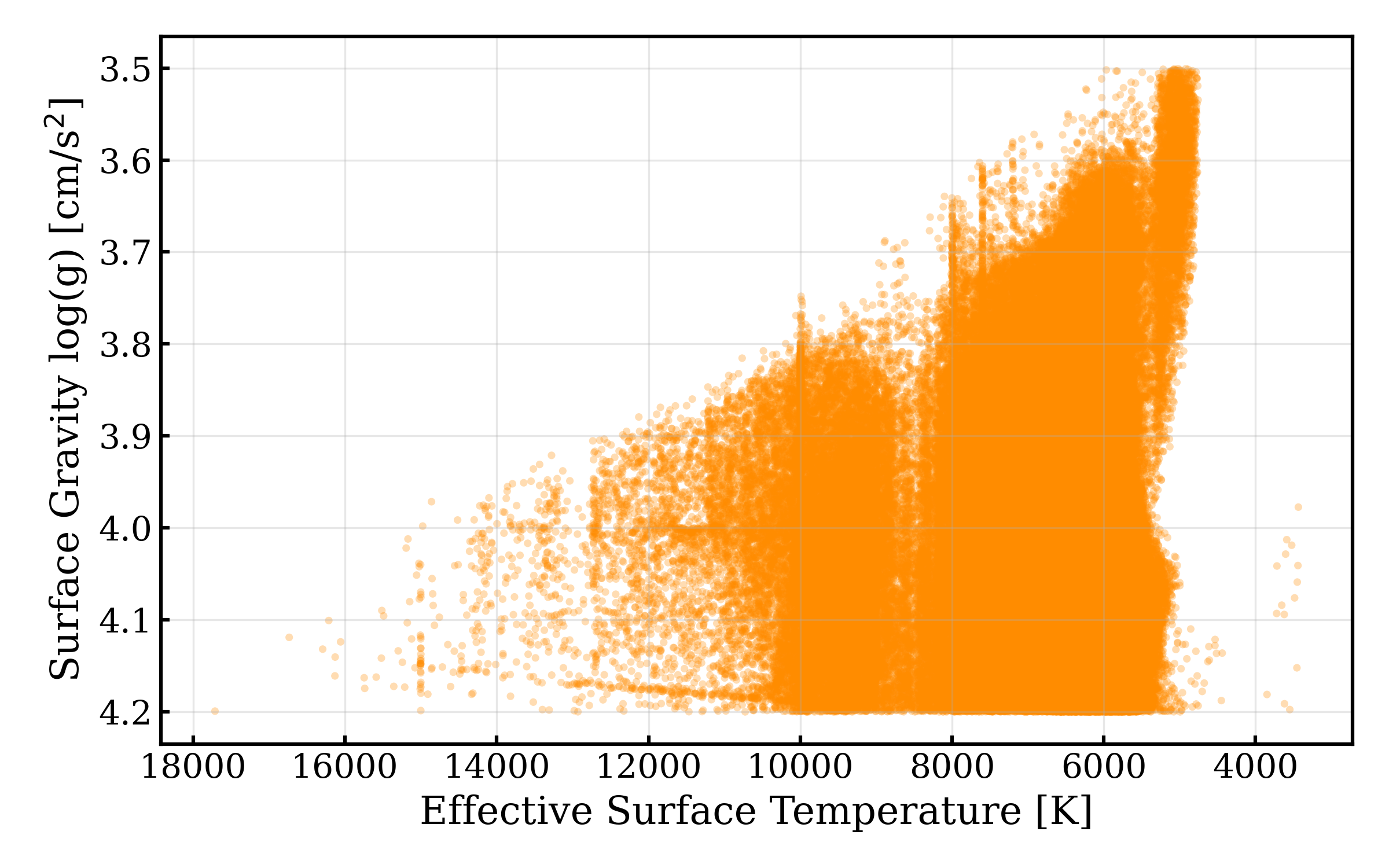}
    \caption{
    Kiel diagram ($T_{\rm eff}$ versus $\log g$) for the final sample of subgiant stars analysed in this work. The selected targets occupy the region between the main sequence and the base of the red-giant branch, which is consistent with mildly evolved stellar populations.}
    \label{fig:kiel}
\end{figure}

\begin{figure}[ht!]
    \centering
    \includegraphics[width=\columnwidth]{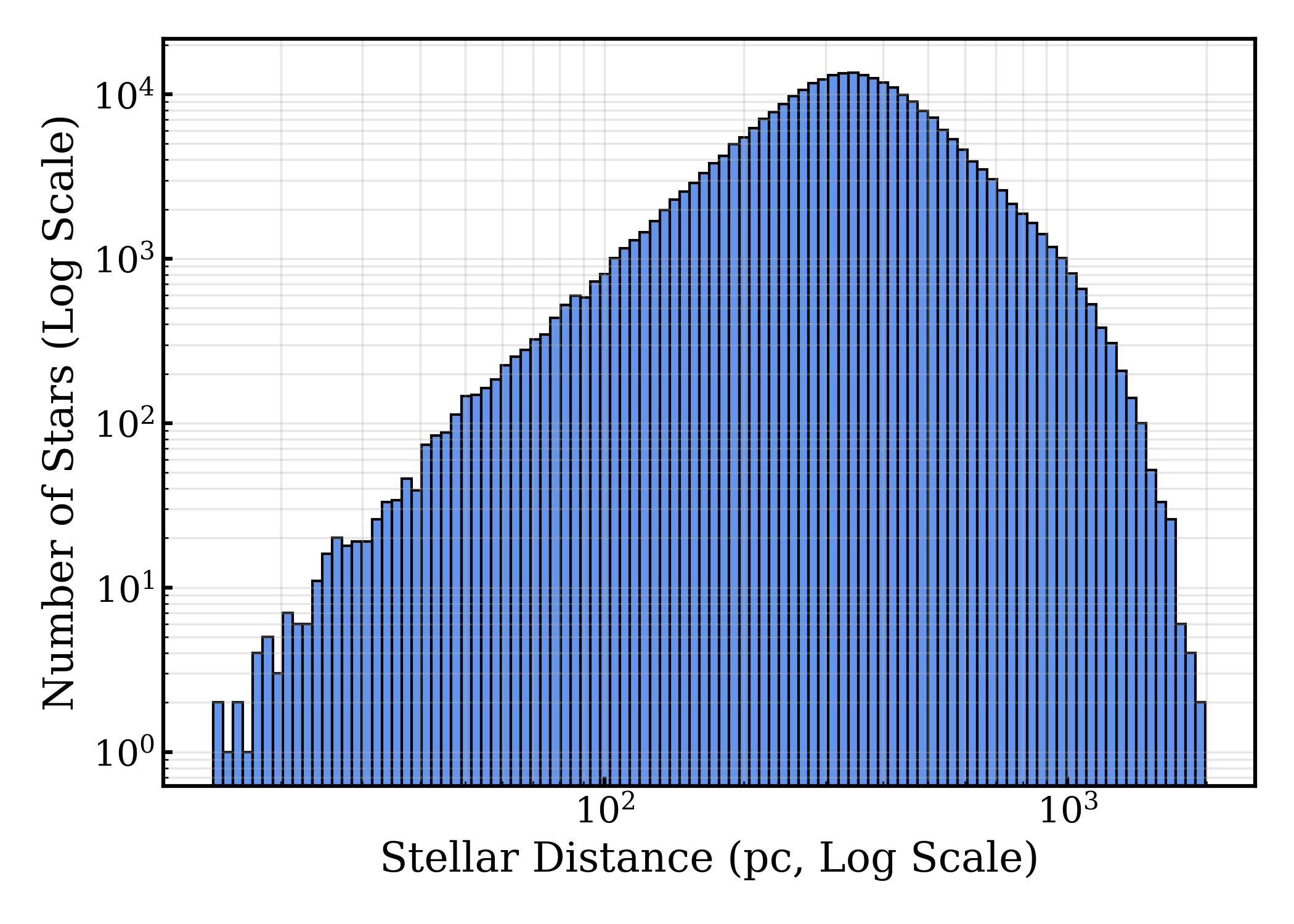}
    \caption{
    Distribution of distances for the subgiant host stars in our sample, derived from the \textit{Gaia} DR3 catalogue. The sample has a median distance of $\sim$331~pc, reflecting the brightness-limited nature of the target selection.
    }
    \label{fig:distance}
\end{figure}

\begin{figure}[ht!]
    \centering
    \includegraphics[width=\columnwidth]{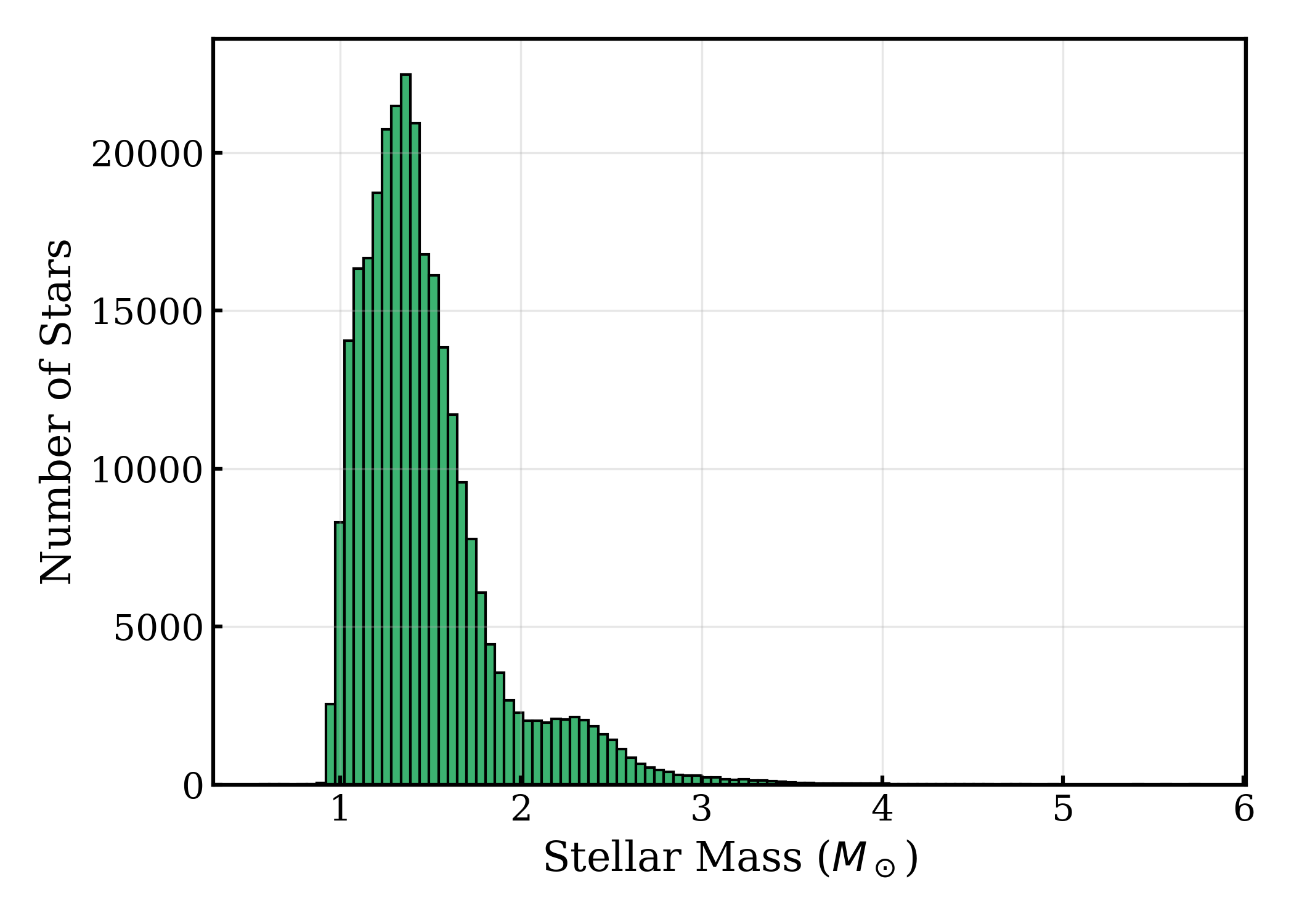}
    \caption{
    Stellar mass distribution of the subgiant stars in the final sample. Mass estimates were taken from the \textit{Gaia} DR3 catalogue, with a median stellar mass of $\sim$1.39~$M_\odot$.
    }
    \label{fig:mass}
\end{figure}

Applying all the above selection criteria yielded a final sample of 282904 subgiant stars. Figure~\ref{fig:kiel} shows the distribution of the sample in the Kiel diagram ($T_{\rm eff}$ versus $\log g$), illustrating the evolutionary state of the host stars. The distribution of stellar distances, derived from \textit{Gaia} DR3 \citep{2023A&A...674A...1G}, is shown in Fig.~\ref{fig:distance}, with a median distance of 331 pc. Figure~\ref{fig:mass} presents the stellar mass distribution of the sample, which has a median mass of $1.39\,M_\odot$ based on \textit{Gaia} DR3 parameters.

\section{Exoplanet search}

Our sample of 282904 subgiant stars was searched for transiting-planet signals using a uniform, automated pipeline built around the \texttt{nuance} framework \citep{2024AJ....167..284G}; the latter performs a joint linear search for box-like transit signals combined with a Gaussian-process (GP) noise model, followed by a periodic search that combines individual transit detections into a phase-folded periodogram. For each target, light curves were retrieved from the MIT QLP \citep{2020RNAAS...4..204H} at 600-second cadence. Where 600-second data were unavailable, 200-second cadence light curves were retrieved instead and binned to an effective 600-second cadence using inverse-variance weighting; thus, all targets entering the search were placed on a uniform 600-second cadence. Up to two sectors were retrieved per target. For each sector, preference was given to the pipeline's most processed flux columns (\texttt{kspsap\_flux}, \texttt{det\_flux}, or \texttt{sys\_rm\_flux}, in that order of preference \citep{kunimoto2021quick}), with simple aperture photometry (SAP) flux being used only when none of these was available. Two sectors were chosen as a practical upper limit to balance detection sensitivity with computational efficiency; each $\sim$27.4-day TESS sector provides at least two to three transit events for the longest periods in our search range, so two sectors guarantee at least three transit observations for all planetary periods of between one and ten days, which is sufficient for reliable detection, while avoiding the substantial computational overhead of processing additional sectors across $\sim$283,000 targets without a proportionate gain in sensitivity for short-period planets. Light curves were normalised by their median flux, concatenated across sectors, and passed through an inter-cadence jump filter and a $12\sigma$ outlier clip (with a symmetric window masked around each flagged point) prior to de-trending.

\subsection{GP de-trending}

Stellar photometric variability was removed using a quasi-periodic Gaussian-process regression, implemented via \texttt{tinygp} \citep{foreman_mackey_2026_19035246} and evaluated through \texttt{nuance}'s GP machinery. We adopted a stochastically driven harmonic oscillator (SHO) kernel, parametrised by an amplitude of $\sigma$, a period, and a quality factor of $Q$, with the GP mean fixed to unity (normalised flux). These three hyperparameters were fitted independently for each target rather than fixed at common values across the sample. Prior to fitting, transit-like dips were identified and masked using a rolling-minimum filter relative to the high-frequency photometric scatter; the GP hyperparameters were thus optimised on the out-of-transit baseline rather than biased by any transit signal that may be present.

The GP was initialised using a data-driven estimate of the photometric variability, computed as the robust (16th--84th percentile) scatter of a median-smoothed flux series and capped at a maximum fractional amplitude. The initial period was seeded from a Lomb--Scargle periodogram (0.5–30 day range), which also provided the stellar rotation period used later for alias rejection. This rotation period was only used for alias rejection when its Lomb--Scargle power exceeded the threshold of 0.05 \citep{2026ApJS..284...75B}; below this threshold, the star was treated as having no detectable rotation signal, and rotation-alias rejection was skipped for that target. Hyperparameters were optimised by minimising the negative log-likelihood of the dip-masked, out-of-transit flux, followed by two iterations of GP re-fitting and rolling sigma-clipping to reject flares and residual outliers. The final flare-masked light curve and best-fit GP model were carried forward into the transit search.

\subsection{Planet search with \texttt{nuance}}

Transit detection was performed using \texttt{nuance}'s linear-and-periodic search architecture. A duration grid of 48 values spanning approximately 0.5-12 hours and a period grid spanning $P_\mathrm{min}=1$ to $P_\mathrm{max}=10$ days were constructed for each target and oversampled relative to the light curve's time baseline in all cases. The linear search evaluated, at every combination of transit epoch and duration, the GP-marginalised likelihood improvement of a signal model over the noise-only model, producing a transit statistic map that was then combined over the period grid by the periodic search to yield a signal-detection-efficiency (SDE) periodogram and an associated signal-to-noise ratio (S/N) periodogram as a function of trial period.

Planet candidates were identified iteratively, up to a maximum of four per target, and a transit-like signal was considered a possible planet signal if both the S/N and SDE were greater than five. To reduce detection outside the range of one to ten days, a period-doubling check was used to test whether the true period was more likely to be twice the detected value; as such we only promoted the candidate to $2P$ when doing so improved both the S/N and SDE by a minimum margin of one and the detected period exceeded half of the maximum searched period (so that the true double-period alias could plausibly fall outside the search window). Each candidate period was also compared to the star's rotation period and its low-order multiples (to flag rotation-induced aliases); harmonics of known signals were required to clear a stricter SDE threshold ($\geq 6$) than genuine new detections since low-level side lobes of a strong existing signal can otherwise mimic an independent periodic dip.

Upon acceptance, its transit depth was re-fitted using \texttt{nuance}'s GP-marginalised depth solver at the best-fit ephemeris, and the corresponding planetary radius was derived from the depth and the host star's radius (taken from the \textit{Gaia} DR3) under the standard $R_p = \sqrt{\delta}\,R_\star$ relation. Its in-transit cadences were then masked from the light curve and the search iterated to look for additional candidates in the residual light curve; this process was repeated until the maximum candidate count was reached, both the S/N and SDE fell below the threshold at the current best peak, or too few cadences remained for further analysis.


\section{Exoplanet candidate vetting}
\label{vetting}

\subsection{Lazy-exoplanet-operations vetter}

From our exoplanet search using \texttt{nuance}, we recovered a total of $42031$ signals out of our $\sim283K$ targets. All transit-like signals returned by the \texttt{nuance} search were first passed through an automated triage using the LEO-Vetter package \citep{2025AJ....170..280K}. This triage was applied in two distinct stages.

In the first stage, each candidate's light curve was flattened using a Savitzky--Golay filter \citep{1964AnaCh..36.1627S} with the in-transit points masked. The \texttt{LEO-Vetter} diagnostic thresholds were calibrated on Savitzky--Golay-flattened light curves \citep{2025AJ....170..280K}, so applying the same preprocessing ensures the classifier operates within its validated regime. Subsequently, a suite of flux-based diagnostics was computed. These tests fall into two categories. False-alarm (FA) tests target signals caused by noise or by instrumental and stellar systematics, whereas false-positive (FP) tests target astrophysical signals that mimic planetary transits, such as eclipsing binaries (EBs). A signal is classified as an FA if it fails any FA test, and as an FP only if it passes all FA tests but fails at least one FP test. The main diagnostics, the scenario each is designed to identify, and the thresholds adopted from \citet{2025AJ....170..280K} are listed below.

\begin{itemize}
    \item Single- and multiple-event statistics (SES and MES): The SES is the S/N of each individual transit event, and the MES is the S/N of the phase-folded signal combining all $N_{\rm tr}$ events. A large discrepancy between the highest SES and the MES indicates that the signal is dominated by a single outlier event rather than by repeated transits. Signals with $N_{\rm tr}\le10$ are flagged as FA if $\mathrm{S/N}_{\rm max}>0.88\times\mathrm{S/N}$.

    \item A model-shift uniqueness test: The best-fit transit model is shifted across all orbital phases to measure the significance of the event at each phase. A genuine planet should be the most significant event in the folded light curve; comparable events elsewhere indicate red noise or systematics (FA), while a significant secondary eclipse indicates an EB (FP).

    \item An odd- and even-numbered transit-depth comparison: The mean depths of odd- and even-numbered transits are compared. A significant difference suggests the true period is twice the detected one, as expected for an EB (FP). Signals are flagged if the odd and even depths differ by more than $3\sigma$.

    \item The SWEET test for ellipsoidal variability or an unresolved secondary eclipse: Sinusoids at the detected period and at half and twice that period are fitted to the out-of-transit flux. A significant sinusoid at short periods indicates stellar variability rather than a transit (FA). Signals are flagged if the significance exceeds 15 and $P\le10$\,d.

    \item A Gaussian-versus-trapezoid shape and AIC comparison: Each signal is fitted with trapezoid, Gaussian, and linear (no-transit) models, which are compared using the Akaike information criterion (AIC). This tests whether the signal has a transit shape rather than a noise-like or V-shaped profile. Signals are flagged as FA if $\Delta\mathrm{AIC}$ (line$-$transit) is $<60$ for $N_{\rm tr}\le10$ or $<30$ for $N_{\rm tr}>10$, or if $\mathrm{SHP}>0.6$.

    \item Individual-transit consistency scores: These test whether all events contribute similarly to the detection. Signals are flagged as FA if the per-event S/Ns are inconsistent with a repeating signal, if a few deep outliers drive the mean depth, or if the events show short-timescale systematics. The corresponding thresholds follow \citet{2025AJ....170..280K}.

    \item Physical plausibility checks: Signals are flagged as FA if the transit duration is unphysical, the transit is asymmetric, or $\ge50\%$ of transits fall near data gaps. Signals are flagged as FP if $R_p>22\,R_\oplus$ or if the transit is V-shaped ($V=R_p/R_\star+b\ge1.5$), indicating a grazing EB.
\end{itemize}

These metrics were combined into an automated FA and FP classification following the standard threshold criteria of \citet{2025AJ....170..280K}. Signals passing all tests were retained as planet candidates. We adopted the standard LEO-Vetter thresholds, except for the S/N threshold, which was set to five rather than the default value of 6.2; the SDE threshold of five was inherited from the \texttt{nuance} search.

Candidates not flagged as FA or FP by the flux-based stage were carried forward to a second, pixel-level stage. Difference images were constructed using TESS target pixel files (TPFs) to assess the spatial origin of the transit signal, and centroid offsets exceeding $3\sigma$ from the target coordinates were used to identify potential contamination from nearby sources. Finally, the FA/FP classification was re-evaluated to include this new pixel-based information.

Signals flagged as FA or FP by either stage of this automated vetting were discarded. In total, $40682$ signals were flagged and removed from further consideration. This left $1349$ candidates that successfully survived LEO-Vetter's automated classification as planet candidates (PCs).

\subsection{Manual vetting and visual inspection}

We then individually visually inspected the 1349 candidates surviving LEO-Vetter by examining the phase-folded transit, the flattened light curve, and the associated diagnostic plots for each signal. This step removed candidates for which the automated pipeline had nonetheless returned an implausible or edge-of-range signal, for example, periods consistent with an alias or harmonic of the true signal (e.g. a factor-of-two sub-harmonic) falling just outside the nominal one-to-ten-day search window, which can occur for sufficiently deep transits even though the \texttt{nuance} search itself was restricted to this range, or candidates associated with targets whose light curves showed clear signs of contamination, poor photometric quality, or non-transit-like shapes inconsistent with a genuine transiting-planet interpretation. After this visual review, 919 candidates remained.

\subsection{Bayesian transit modelling}

The transit signal of each of the 919 surviving candidates was modelled using the analytic transit model implemented in \texttt{batman} \citep{batman}, as wrapped within the \texttt{juliet} modelling framework \citep{juliet}. To account for correlated noise in the light curves, we jointly evaluated a Gaussian Process (GP) model alongside the transit model. We set normally distributed priors on the orbital period and transit epoch, centred on the \texttt{nuance}-derived values with a standard deviation of $0.1$~days for both quantities. This width was set manually and is substantially wider than the typical precision of the \texttt{nuance} ephemeris, providing the sampler with ample volume to refine the ephemeris while preventing convergence to period aliases. For the planet-to-star radius ratio ($R_p/R_\star$) and impact parameter ($b$), we adopted the $(r_1, r_2)$ parametrisation of \citet{juliet} with uniform priors of $\mathcal{U}(0, 1)$. Quadratic limb-darkening coefficients $(q_1, q_2)$ were sampled under uniform priors $\mathcal{U}(0, 1)$ following \citet{2013MNRAS.435.2152K}. The stellar bulk density, $\rho_\star$ was assigned a log-uniform prior between $100$ and $10000\text{ kg m}^{-3}$, and an instrumental white-noise jitter term, $\sigma_w$ was assigned a log-uniform prior between $0.1$ and $1000\text{ ppm}$. Given the short orbital periods ($P \le 10$~days) and lack of constraints on eccentricity from single-band transit photometry alone, orbits were assumed to be circular ($e = 0$, with $\omega = 90^\circ$). To model correlated stellar and instrumental noise, we incorporated a GP using an approximate Mat\'ern-3/2 kernel implemented via \texttt{celerite} within \texttt{juliet} \citep{juliet}, with a wide log-uniform prior on the GP amplitude of $\sigma_{\rm GP} \sim \mathcal{U}_{\log}(10^{-6}, 10^{6})\text{ ppm}$ and a wide log-uniform prior on the GP timescale of $\rho_{\rm GP} \sim \mathcal{U}_{\log}(10^{-3}, 10^{3})\text{ days}$, while the mean out-of-transit flux was assigned a normal prior of $m_{\rm flux} \sim \mathcal{N}(0, 0.1)$.

For targets with multiple candidate signals, we fitted all planet signals simultaneously within a single joint model. However, each planet's transit parameters were obtained from the posterior while accounting for the other planets.

Posterior probability distributions for the transit parameters were sampled using nested sampling as implemented through \texttt{MultiNest} \citep{multinest}, marginalising over the GP noise parameters while obtaining robust estimates of the orbital period, transit epoch, and planet-to-star radius ratio ($R_p/R_\star$), along with their associated uncertainties. We note that the \texttt{juliet} modelling was used strictly for posterior parameter refinement. All candidates converged successfully and were carried forward to the next stage. These refined ephemerides and transit parameters were used in place of the discovery-search values for all subsequent analyses of the planet-candidate population.

\subsection{Automated statistical vetting}

Each of the 919 vetted, \texttt{juliet}-refined candidates was subjected to a statistical false-positive assessment using \texttt{TRICERATOPS} \citep{Triceratops_soft}. 

For each target, TESS aperture masks were retrieved for sectors in which the planet search was performed from the corresponding target pixel files. This was used together with the \texttt{juliet}-refined transit depth to compute the depths TRICERATOPS would expect for each false-positive scenario. The phase-folded, binned light curve around the refined ephemeris was then used to evaluate the false-positive probability (FPP) and the nearby false-positive probability (NFPP) via repeated posterior sampling of TRICERATOPS' scenario likelihoods, with results averaged over 20 independent evaluations to obtain robust mean values and uncertainties. 

For systems with more than one candidate, the transits of all other planets were masked before evaluating each candidate individually. Candidates were classified as selected planet candidates if they satisfied FPP~$<0.5$ and NFPP~$<0.01$. Applying this criterion to the 919 vetted signals yielded 217 planet candidates around 215 host stars, which form the final demographic sample used in the occurrence-rate analysis presented below.

In our demographic analysis, we assumed that the false-negative rate of this multi-tier candidate-vetting process is negligible following arguments similar to those of \citet{Gan2023}. First, the automated diagnostics in \texttt{LEO-Vetter} \citep{2025AJ....170..280K} and the statistical scenario modelling in \texttt{TRICERATOPS} \citep{Triceratops_soft} were designed specifically to reject astrophysical blends, centroid contamination, and systematic noise artefacts while retaining genuine periodic transit signals. Both classifiers have been extensively validated on real and synthetic planet samples. All candidates were additionally examined visually to minimise the inadvertent loss of true planets. Second, even if the vetting stages possessed a low, non-zero false-negative rate, their impact on our final occurrence-rate statistics would be negligible. The dominant source of uncertainty in our measurements is the bootstrap error on the candidate sample, which is substantially larger than any plausible vetting incompleteness. 


\section{Injection and recovery}
\label{injection}

To derive the occurrence rates of transiting planet candidates around subgiant stars, we quantified the detection efficiency of our pipeline using injection--recovery tests. We measured this detection efficiency by injecting synthetic transit signals into the light curves and monitoring to see whether our search pipeline recovered them.

For each target, we first removed every planet candidate already identified in Sects.~3 and 4 from its light curve. Each known transit was masked over a window of 1.25 times its expected transit duration, centred on its ephemeris, so that the injection--recovery test measured sensitivity to new signals rather than re-detecting planets we had already found. 
 
In the next step, we distributed synthetic transit injections over the full period--radius space ($1$--$10$~days in period, $1$--$22$~$R_\oplus$ in radius) uniformly across the entire target sample. To ensure an even and unbiased coverage of the parameter space across all targets, the domain was partitioned into a grid of cells that were sampled systematically across the sample of stars, so that every region of the period--radius space received an equal number of injections. Within whichever grid cell a given injection falls, the actual injected period, planet radius, and transit epoch were drawn uniformly at random from continuous distributions over that cell rather than being fixed at a discrete bin centre. Each target from our sample of subgiants received two such injections, each generated with an analytic transit model and tracked as continuous parameters ($P, R_p, T_0$, and transit depth $\delta$) to evaluate pipeline recovery and calculate the individual geometric transit probabilities.

Each injected light curve was then passed back through the full search pipeline; this is the same linear-and-periodic search architecture described in Sect.~3, run independently for every injection. A signal was only counted as recovered if it satisfied the following four conditions simultaneously: 

\begin{enumerate}
    \item The recovered S/N exceeds our detection threshold of five.
    \item The SDE exceeds our detection threshold of five.
    \item The recovered period agrees with the injected period to within 0.05 days. 
    \item The recovered transit depth, re-fitted at the recovered ephemeris, agrees with the injected depth to within $10\%$.
\end{enumerate}

\noindent These criteria are deliberately stricter than our original planet search cut alone in order to guard against spurious detections that pass our S/N threshold by chance but correspond to the wrong signal.

In total, we injected 565808 synthetic transits across our stellar sample, 371747 of which were recovered under the above criteria, corresponding to an overall recovery fraction of $65.7\%$ with the recovery efficiency varying substantially across the period--radius parameter space. These tests assess the completeness of the search pipeline. We note that we did not perform the vetting step here as discussed in Sect.~\ref{vetting}; we assumed the false-negative to be negligible.

For each (period, radius) grid cell, we computed the completeness as the fraction of injected signals recovered at that cell, pooling injections across the entire stellar sample. This yields a single, sample-wide two-dimensional completeness map across period and planet radius, which we combined with the geometric transit probability to correct our observed candidate counts and derive the completeness-corrected occurrence rates presented in Sect.~6.


\section{Results}

By applying the transit detection, refinement, and vetting procedures described in Sect.~\ref{vetting}, we identified a sample of vetted transiting planet candidates orbiting subgiant stars. These candidates show transit-like signals that pass both automated and manual vetting checks. The final sample consists of 217 planet candidates orbiting 215 subgiant host stars (comprising 97 previously confirmed planets in the literature, 41 previously flagged candidate signals, and 79 new candidate signals). Of these, two systems host multiple planet candidates.

\subsection{Detection efficiency and completeness}

The detection efficiency of our transit search pipeline was quantified using the injection and recovery analysis described in Sect.~\ref{injection}. By measuring the fraction of injected signals recovered as a function of orbital period and planet radius, we characterised the regions of parameter space where our occurrence rate estimates are robust.

\begin{figure}[htb!]
    \centering
    \includegraphics[width=\linewidth]{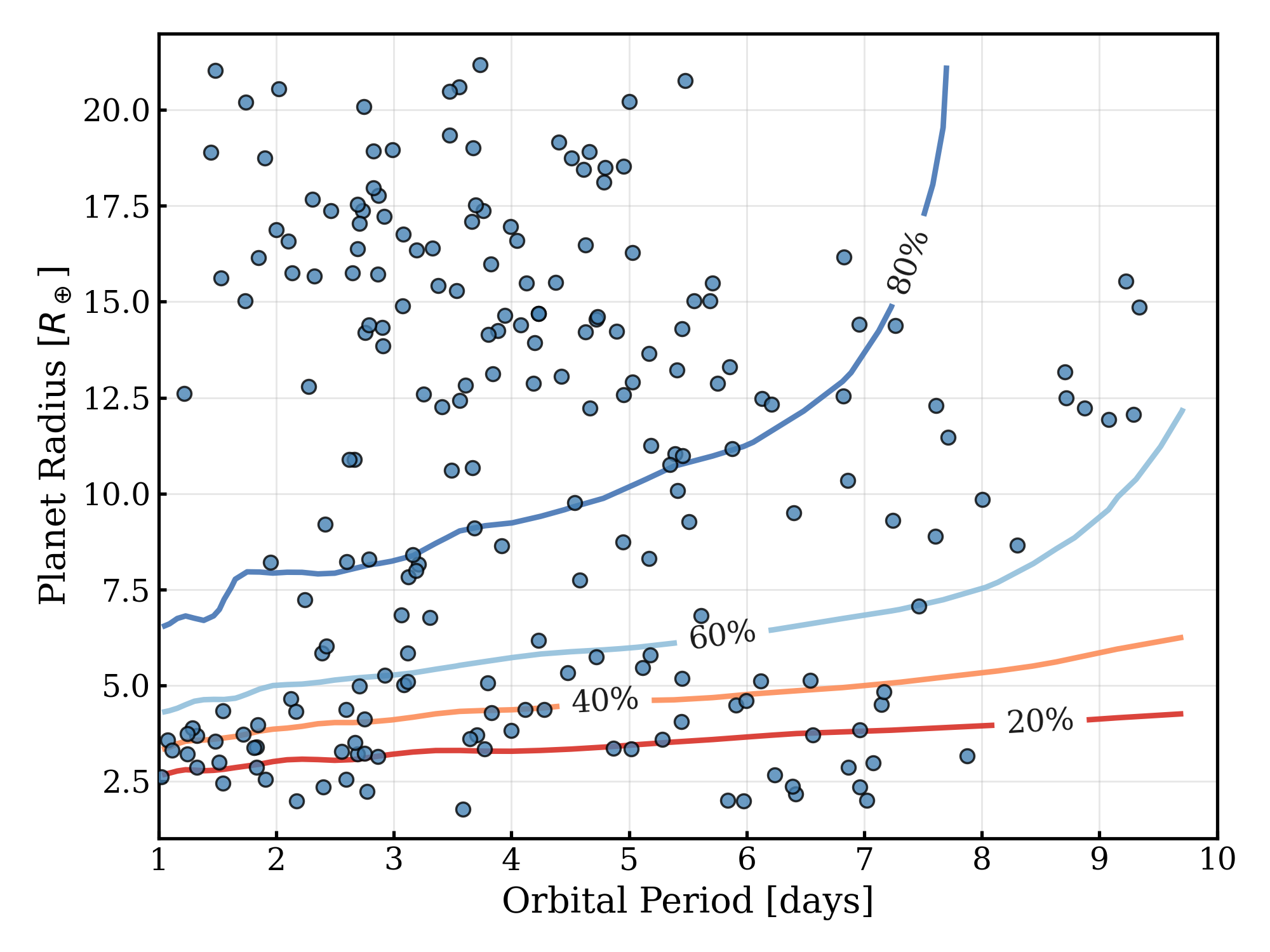}
    \caption{Two-dimensional detection completeness as a function of orbital period and planet radius, derived from injection and recovery experiments on the TESS QLP effective ten-minute cadence, kspsap or DET flux, light curves overlayed with our complete sample. The contours indicate the percentage of injected transit signals that the detection pipeline recovered.}
    \label{fig:completeness_2d}
\end{figure}

Figure~\ref{fig:completeness_2d} shows the two-dimensional detection completeness in period--radius space for the full subgiant sample. The completeness is highest for short-period, larger radius planets and decreases towards longer orbital periods and smaller planet sizes. This behaviour reflects the combined effects of the limited temporal baseline of individual TESS sectors and the reduced transit S/N for planets transiting evolved stars with larger radii.

Based on the recovery statistics shown in Fig.~\ref{fig:completeness_2d}, we restricted our primary occurrence-rate analysis to orbital periods of one to ten days and planet radii where the detection efficiency is well characterised (defined as parameter-space cells where the detection completeness exceeds a threshold of 20\%, which prevented excessive noise amplification from completeness correction).

\subsection{Exoplanet occurrence rates around subgiant stars}

Using the results of the injection--recovery analysis, we estimated the occurrence rate of vetted transiting planet candidates around subgiant stars. The occurrence rate, $f_{\rm occ}$, is defined as

\begin{center}
\begin{equation}
    f_{\rm occ}(P, R_p) = \frac{n_{\rm pl}(P, R_p)}{N_{\rm pr}(P, R_p)},
\end{equation}
\end{center}

where $n_{\rm pl}$ is the effective number of planet candidates and $N_{\rm pr}$ is the number of stars amenable to detecting such planets.

The quantity $N_{\rm pr}$ is calculated as

\begin{center}
\begin{equation}
    N_{\rm pr}(P, R_p) =
    N_\ast \frac{1}{N_{\rm sim}}
    \sum_{i=1}^{N_{\rm sim}}
    \delta_{{\rm det}, i}\,\wp_{{\rm tr}, i},
\end{equation}
\end{center}

where $N_\ast$ is the number of stars in the stellar sample, $N_{\rm sim}$ is the total number of simulated planets over the full parameter space across all stars, $\delta_{{\rm det}, i}$ is a binary detection indicator equal to unity if the $i$-th simulated planet falls within the specific $(P, R_p)$ bin and is successfully recovered by the detection pipeline (and zero otherwise), and $\wp_{{\rm tr}, i}$ is the geometric transit probability given by

\begin{center}
\begin{equation}
    \wp_{{\rm tr}, i} = \frac{R_\ast + R_{p, i}}{a_i}.
\end{equation}
\end{center}

Since the injection--recovery tests were performed for transiting configurations, the resulting completeness represents the conditional detection efficiency for transiting planets; the geometric transit probability was subsequently included in $N_{\rm pr}$ to obtain the occurrence rate per star. By restricting the recovery indicator $\delta_{{\rm det}, i}$ to the $(P, \ R_p)$ bin of interest while normalising by the global injection total $N_{\rm sim}$, Eq.~(2) weights each bin by its parameter-space volume fraction to yield the true effective sample size for that individual bin. The effective number of planet candidates in each $(P, R_p)$ bin was computed using a false-positive-probability-weighted sum,

\begin{center}
\begin{equation}
    n_{\rm pl}(P, R_p) =
    \sum_{j=1}^{N_{\rm cand}} \left( 1 - {\rm FPP}_j \right),
\end{equation}
\end{center}

\noindent where ${\rm FPP}_j$ is the false-positive probability assigned to the $j$th candidate through the vetting procedure described in Sect.~\ref{vetting}.

Uncertainties on the occurrence rates were estimated using a non-parametric, star-level bootstrap re-sampling approach \citep{bayliss2011frequency}. This method captures the counting statistics of detected planets. It also captures the sampling variance across the host-star ensemble.

We generated 1000 bootstrap realisations. Each realisation was formed by drawing $N_*$ host stars from our sample at random with replacements. For each realisation, we re-aggregated the detected planet weight over the re-sampled stars. We also re-aggregated the effective searchable stellar sample over the same re-sampled stars.

The effective sample size was computed from the geometric transit probabilities of the recovered synthetic injections. Only injections associated with the re-sampled stars were included. This sum was normalised by the total number of injections evaluated for those stars. The occurrence rate for each realisation is the ratio of the re-sampled planet weight to the re-sampled effective sample size.

We report the occurrence rate from the full, non-re-sampled sample as the central value. The lower and upper $1\sigma$ uncertainties are given by the 16th and 84th percentiles of the 1000 bootstrap realisations. Specifically, they are defined as the difference between this central value and those percentiles.

\begin{figure}[htb!]
    \centering
    \includegraphics[width=\linewidth]{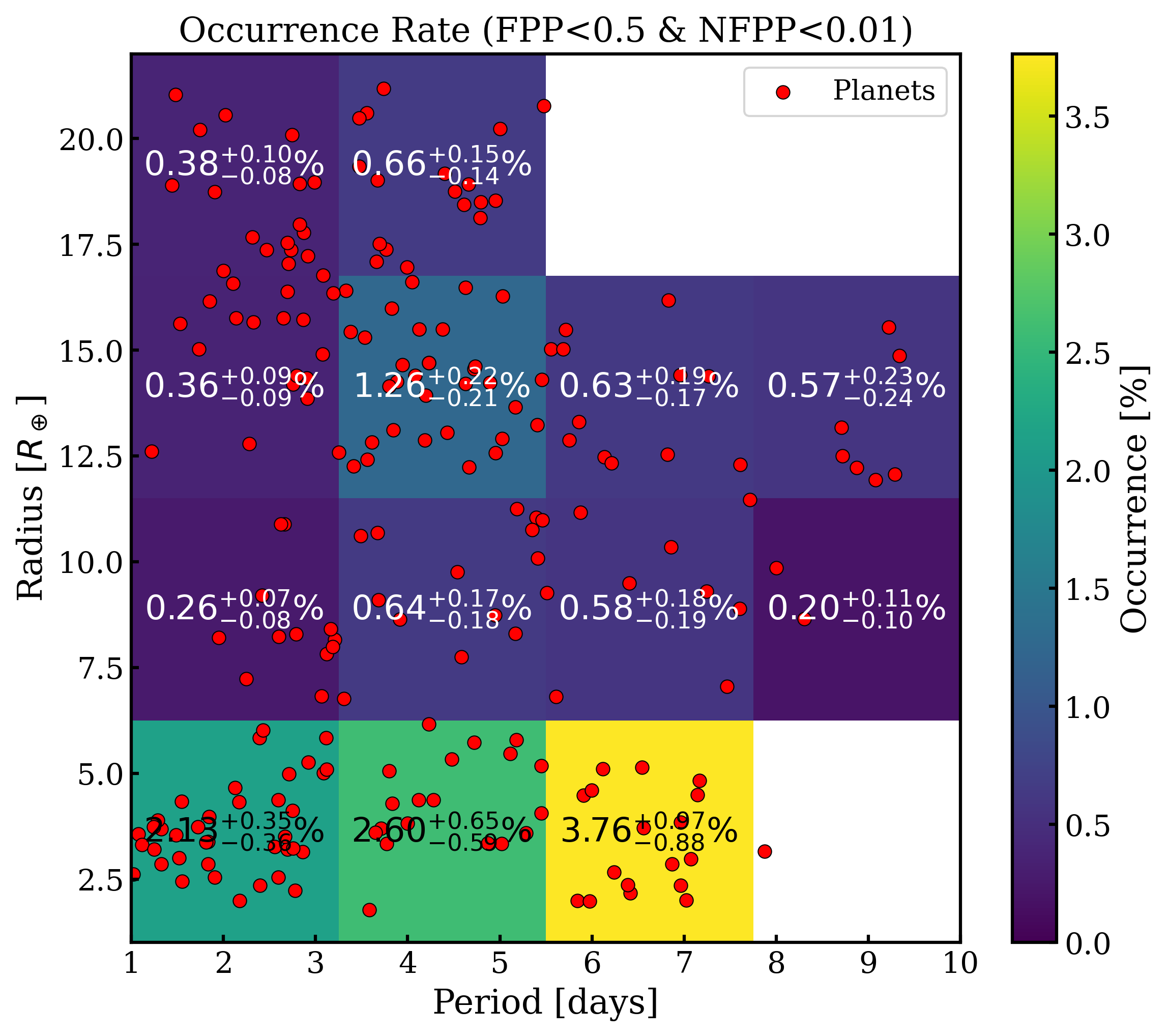}
    \caption{Occurrence rate of transiting planet candidates with orbital periods shorter than ten days around subgiant stars. A completeness cut-off of 0.20, FPP cut-off of 0.5, and NFPP cut-off of 0.01 is used.}
\label{fig:occ_global}
\end{figure}

Figure~\ref{fig:occ_global} shows the occurrence rate of planets with orbital periods shorter than ten days around subgiant stars. This provides a global view of the short-period planet population accessible to TESS in our sample.

The two-dimensional occurrence rate in orbital period--planet radius space is shown in Fig.~\ref{fig:occ_global}. This figure shows the primary occurrence-rate result of this work, with each cell corrected for detection completeness and weighted by the false-positive probabilities of the individual candidates. To avoid unreliable corrections from noise amplification, we imposed a minimum completeness threshold of 20\%, excluding cells below this value from the analysis (see Fig.~\ref{fig:completeness_2d}).

\begin{figure}[htb!]
    \centering
    \includegraphics[width=\linewidth]{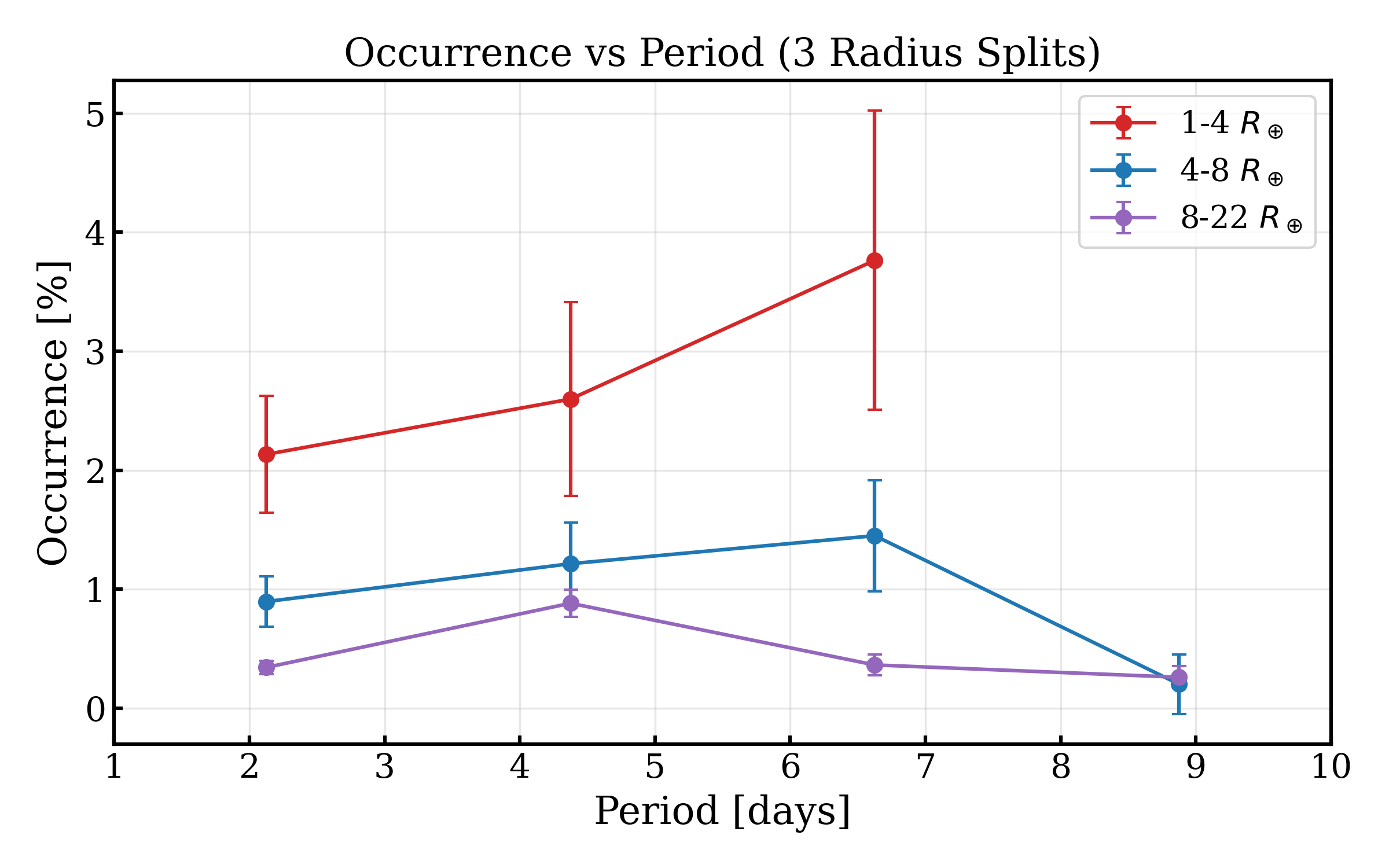}
    \caption{Marginalised occurrence rate as a function of orbital period for different planet radius ranges.}
\label{fig:occ_period}
\end{figure}

\begin{figure}[htb!]
    \centering
    \includegraphics[width=\linewidth]{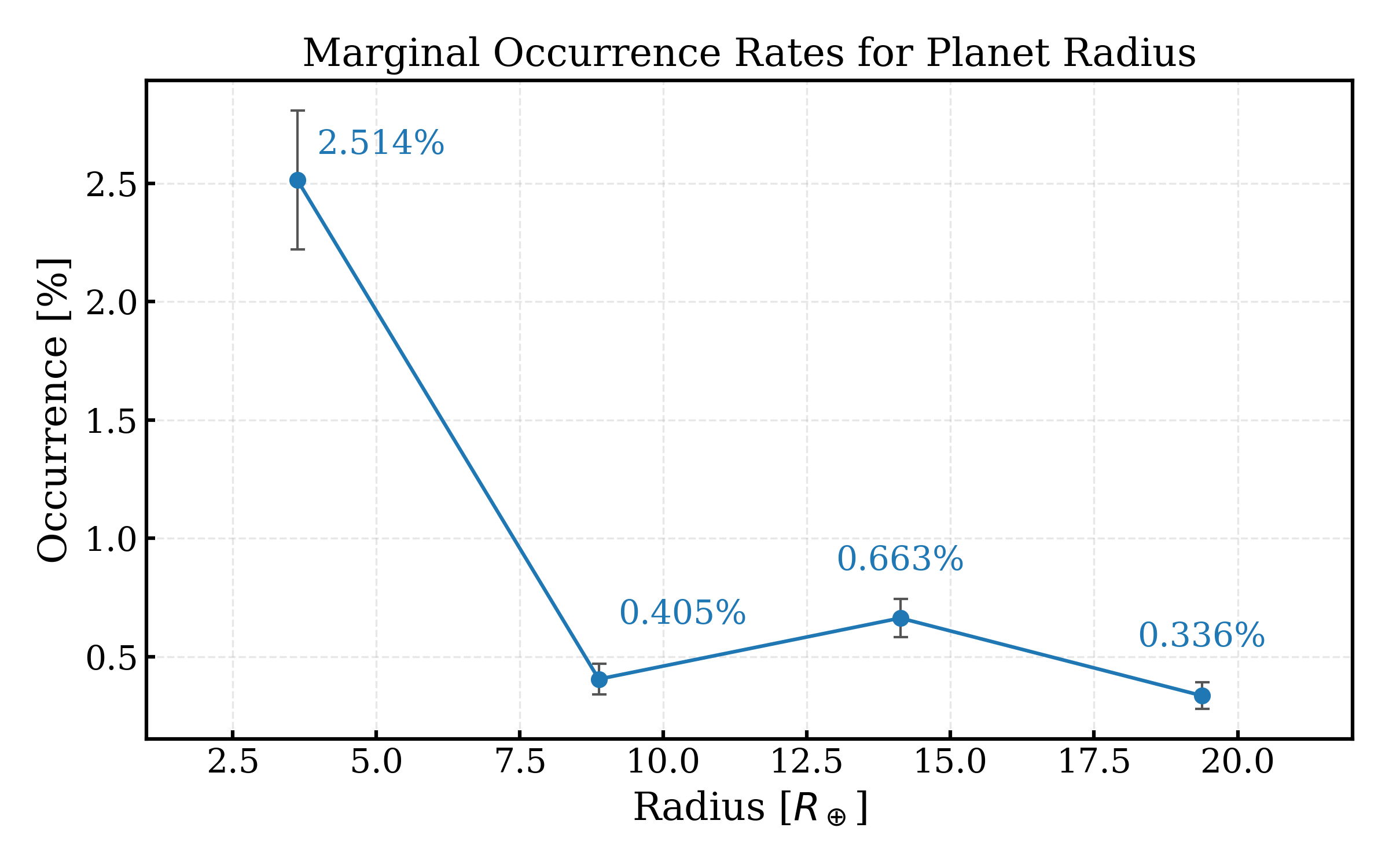}
   \caption{Marginalised occurrence rate as a function of planet radius, integrated over the orbital period range considered in this work.}
\label{fig:occ_radius}
\end{figure}

To aid interpretation, Fig.~\ref{fig:occ_period} shows the occurrence rate as a function of orbital period for different planet radius ranges, while Fig.~\ref{fig:occ_radius} presents the occurrence rate integrated over the orbital period. These projections highlight the dominant contributions to the occurrence rate across the explored parameter space.

Given the high number of injected signals, the uncertainty in $f_{\rm occ}$ is dominated by counting statistics on the planet-candidate sample.

\section{Discussion}

\subsection{Summary of measured occurrence rates}

In this work, we measured the occurrence rate of short-period transiting planet candidates around subgiant stars using TESS data, correcting for detection efficiency, geometric transit probability, and false-positive contamination. The analysis is restricted to orbital periods of one to ten days and to regions of planet radius where the detection completeness is well characterised (Fig.~4).

Integrating over the explored parameter space, we find an overall occurrence rate of $f_{\rm occ}=0.644^{+0.048}_{-0.041}\%$ planets per star for orbital periods shorter than ten days. This value represents the completeness-corrected frequency of short-period planets around subgiant stars in the TESS sample considered here. Figure~7 shows the marginalised occurrence rate as a function of orbital period for different planet radius ranges. The $1$-$4\, R_\oplus$ population exhibits the highest occurrence over much of the explored period range, while the $4$-$8\, R_\oplus$ and $8$-$22\, R_\oplus$ populations show broadly similar period distributions.

The larger radius populations show a modest enhancement at orbital periods of approximately four to seven days. This period range is similar to, though not exactly the same as, the location of the hot-Jupiter pile-up observed around main-sequence stars \citep{dawson2018origins, gaudi2005period}. However, the enhancement is present in both the $4$-$8\, R_\oplus$ and $8$-$22\, R_\oplus$ populations and is not sufficiently distinct in the present sample to establish a separate hot-Jupiter pile-up. A larger sample of evolved stars will be useful for determining whether this feature represents a continuation of the main-sequence period distribution.

In contrast, the small-planet population shows a somewhat different period dependence, with the occurrence increasing from the shortest periods towards intermediate periods. The reduced occurrence at the shortest orbital periods may be related to the evolutionary history of close-in planets around expanding subgiants through tidal interactions and orbital evolution, for example. However, the larger planets do not show an equally strong reduction towards the shortest periods, suggesting that this behaviour cannot be attributed to a simple radius-independent removal process.

The marginalised occurrence rate as a function of planet radius, restricted to $P<10$ days (Fig.~7), shows a prominent occurrence maximum in the $1$--$4\,R_\oplus$ regime, followed by a secondary enhancement at approximately $12$--$15\,R_\oplus$. The prevalence of small planets is broadly consistent with the planet population around main-sequence Sun-like stars, for which \textit{Kepler} occurrence studies have shown that planets with radii of $1$--$4\,R_\oplus$ are substantially more common than Neptune- and Jupiter-sized planets \citep{doi:10.1073/pnas.1319909110, 2014ApJS..210...20M}. Thus, the prominence of small planets in our short-period subgiant sample does not by itself indicate a population unique to evolved stars, although its survival in very short orbital periods provides an interesting constraint on the evolution of close-in planets.

Our principal findings are listed below.

\begin{enumerate}
    \item Our primary results focus on planets with orbital periods of one to ten days, for which completeness corrections can be reliably quantified. In this regime, we find a non-zero population of planets with radii spanning $1$-$22\,R_\oplus$ and an integrated occurrence rate of $0.644^{+0.048}_{-0.041}\%$.
    \item The occurrence-rate distribution shows a broad enhancement in intermediate orbital periods of approximately four to seven days in the larger radius populations spanning a period range similar to the hot-Jupiter pile-up observed around main-sequence stars.
    \item Small planets in the $1$-$4\,R_\oplus$ range contribute the largest occurrence in the short-period population, while a secondary enhancement is present in the giant-planet regime around $12$-$15\,R_\oplus$.
    \item The different period dependencies of small and large planets may provide evidence of evolutionary effects acting on the close-in planet population as subgiant stars expand, although a larger sample is required to distinguish such effects from the intrinsic period distribution.
\end{enumerate}

\subsection{Comparison with previous occurrence-rate measurements}

\begin{figure}[ht!]
    \centering
    \includegraphics[width=\linewidth]{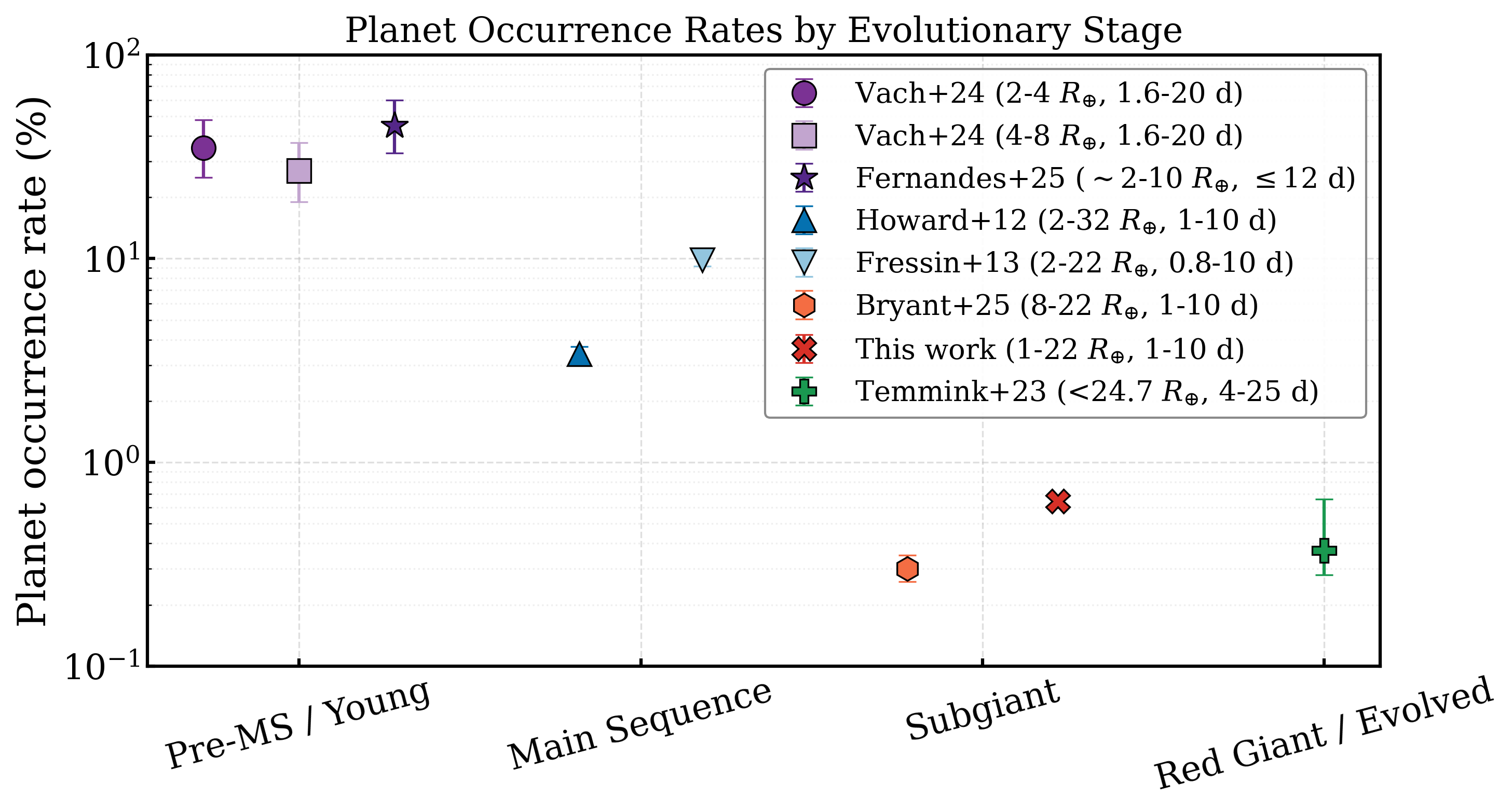}
    \caption{Comparison of our occurrence rate of short-orbital-period planets around subgiant stars with other similar studies of short-orbital-period planets from \textit{Kepler} and TESS for different evolutionary stages of stars}
    \label{fig:comp}
\end{figure}

Occurrence-rate studies of main-sequence stars have established that short-period planets are relatively common, with typical occurrence rates of $\approx3\%$-$10\%$ for orbital periods shorter than ten days \citep{howard2012planet, fressin2013false}. In contrast, the completeness-corrected occurrence rate measured in this work for subgiants is $f_{\rm occ}=0.644^{+0.048}_{-0.041}\%$ over the same period range. This lower occurrence rate may indicate differences in the close-in planet population around subgiant stars, although differences in stellar sample selection and occurrence-rate methodology should also be considered.

A more direct comparison can be made with the recent occurrence-rate measurement of \citet{10.1093/mnras/staf1771}, whose authors reported an occurrence rate of $0.30^{+0.05}_{-0.04}\%$ for giant planets with radii of $8$-$22\,R_\oplus$ and orbital periods shorter than ten days. When restricting our sample to the same radius and period range, we obtained an occurrence rate of $0.49\pm0.04\%$. Our value is somewhat higher --- with a difference of approximately $3\sigma$ --- although differences in stellar sample selection, candidate selection, and completeness modelling may contribute to this difference. Nevertheless, both measurements indicate a relatively low occurrence of short-period giant planets around evolved stars. Unlike \citet{10.1093/mnras/staf1771}, our analysis additionally extends to smaller planets, allowing the radius dependence of the short-period population to be investigated.

Figure~8 compares occurrence rates across different stellar evolutionary stages using measurements from the literature. Young and pre-main-sequence stars exhibit substantially higher occurrence rates, while main-sequence stars show intermediate values \citep{howard2012planet, fressin2013false}. Subgiant stars exhibit lower occurrence rates, including the values measured in this work and by \cite{10.1093/mnras/staf1771}, while red-giant stars show similarly low values \citep{temmink2023occurrence}. This progression suggests a possible decline in the frequency of short-period planets as host stars evolve beyond the main sequence, although differences between the samples and occurrence-rate methodologies should be considered.

These results are consistent with theoretical expectations that tidal interactions, stellar expansion, and orbital evolution can modify close-in planetary systems during post-main-sequence evolution. The radius-dependent period distribution observed in this work may therefore provide additional constraints on the survival of close-in planets around evolving stars, although a larger sample extending to longer orbital periods will be required to distinguish evolutionary effects from the intrinsic planet distribution.

\subsection{Implications for the survival of close-in planets}

The occurrence of close-in planets around subgiant stars constrains the population of planetary systems that survive as their host stars evolve off the main sequence. As the stellar radius increases, tidal interactions between the star and a close-in planet become increasingly important and can lead to orbital decay and eventual engulfment \citep{2009ApJ...705L..81V, 2023ApJ...954..176Y}. The observed population therefore provides an opportunity to test whether the close-in planetary population is modified during the early stages of post-main-sequence evolution.

The reduced occurrence of planets at the shortest orbital periods is qualitatively consistent with such evolutionary effects, although the present analysis cannot distinguish tidal evolution from the intrinsic orbital-period distribution of planets. In addition, stellar expansion and tidal evolution depend on both stellar and planetary properties, so any depletion is not expected to be independent of planet radius. A comparison of occurrence rates across different stellar evolutionary stages and over a wider range of orbital periods will therefore be important for determining whether the observed close-in population has been significantly altered by post-main-sequence evolution.

\subsection{Limitations and future prospects}

Several limitations should be considered when interpreting these results. The restriction to orbital periods shorter than ten days limits the present analysis to the innermost planetary population and prevents us from determining how the occurrence distribution extends to longer orbital periods, where the effects of stellar evolution may differ. In addition, the detection completeness varies across the planet radius--period parameter space, particularly towards smaller planets, and therefore the inferred occurrence rates in these regions remain more sensitive to the completeness correction. The planet sample consists primarily of vetted planet candidates rather than fully confirmed planets. Although false-positive probabilities are incorporated into the occurrence-rate calculation, residual contamination cannot be entirely excluded. The stellar sample is also brightness-limited and selected from the TESS target population, and therefore the measured occurrence rates may not be directly representative of the complete subgiant population.

Future studies using longer TESS baselines and multi-sector observations will facilitate the extension of occurrence-rate measurements to longer orbital periods. Complementary radial-velocity observations will also help confirm individual candidates and characterise the planetary population around evolved stars. A larger sample spanning a wider range of stellar evolutionary stages and orbital periods will ultimately be needed to distinguish intrinsic planet-population differences from evolutionary effects.


\section{Summary and conclusions}

We present a statistical study of the occurrence rate of short-period planets around subgiant stars using TESS photometry. We constructed a sample of 282,904 subgiant stars from the \textit{Gaia} DR3 catalogue and performed a uniform transit search and multi-stage candidate vetting. Injection-and-recovery experiments with 565,808 synthetic transit signals were used to quantify the detection completeness across orbital period and planet radius.

Over the period range of one to ten days and from the regions of parameter space with well-characterised detection efficiency, we measured a completeness-corrected occurrence rate of
\[
f_{\rm occ}=0.644^{+0.048}_{-0.041}\%.
\]
The occurrence rate is strongly radius-dependent, with planets in the $1$--$4\,R_{\oplus}$ range contributing the largest occurrence and a secondary enhancement at approximately $12$--$15\,R_{\oplus}$. The larger radius populations also show a modest enhancement at periods of approximately 4--7 days, but the present sample does not provide sufficient evidence to identify this feature as a distinct hot-Jupiter pile-up.

The observed differences in the period dependence of small and large planets may provide clues to the evolution of close-in planetary systems as their host stars leave the main sequence. However, distinguishing evolutionary effects from the intrinsic planet population requires larger samples spanning a wider range of stellar evolutionary stages and orbital periods. Longer baseline photometry from TESS and future missions such as PLATO, together with improved stellar characterisation and follow-up observations, will enable more stringent tests of the evolution of close-in planets around evolved stars.

\par\addvspace{6pt}
\noindent\textit{Data availability.}
The TESS photometric light curves analysed in this work are publicly accessible via the Mikulski Archive for Space Telescopes (MAST; \url{https://mast.stsci.edu/}) and the quick-look pipeline (QLP). Stellar parameters were retrieved from the AIP \textit{Gaia} DR3 archive (\url{https://gaia.aip.de/cms/data/gdr3/}). \par

\begin{acknowledgements}
AD acknowledges the Tata Institute of Fundamental Research, Department of Astronomy, for its financial support and hospitality during his secondment for an external MS thesis from 2024 to 2025. AD also gratefully acknowledges the computational resources and support provided by the Centre for Computational High-Performance Computing (CCHPC), Tata Institute of Fundamental Research, Mumbai. PC acknowledges the startup research grant numbered 19P0261 from the Department of Atomic Energy.
This research has made use of the Exoplanet Follow-up Observation Program (ExoFOP; DOI: 10.26134/ExoFOP5) website, which is operated by the California Institute of Technology, under contract with the National Aeronautics and Space Administration under the Exoplanet Exploration Program.
This research has made use of the NASA Exoplanet Archive, which is operated by the California Institute of Technology, under contract with the National Aeronautics and Space Administration under the Exoplanet Exploration Program.
\end{acknowledgements}

\bibliographystyle{aa} 
\bibliography{sample701} 

\begin{appendix}




\onecolumn
\section{Phase folded transits of 217 selected targets in Figure~\ref{fig:occ_global}}
\FloatBarrier
\captionsetup[subfloat]{labelformat=simple}

\begin{figure}[h]
    \centering
    \setlength{\tabcolsep}{1.0pt} 
    \renewcommand{\arraystretch}{1.5}
\begin{tabular}{ccccc} 
        \subfloat[TIC 909930\label{fov:1}]{
            \includegraphics[width=0.18\linewidth]{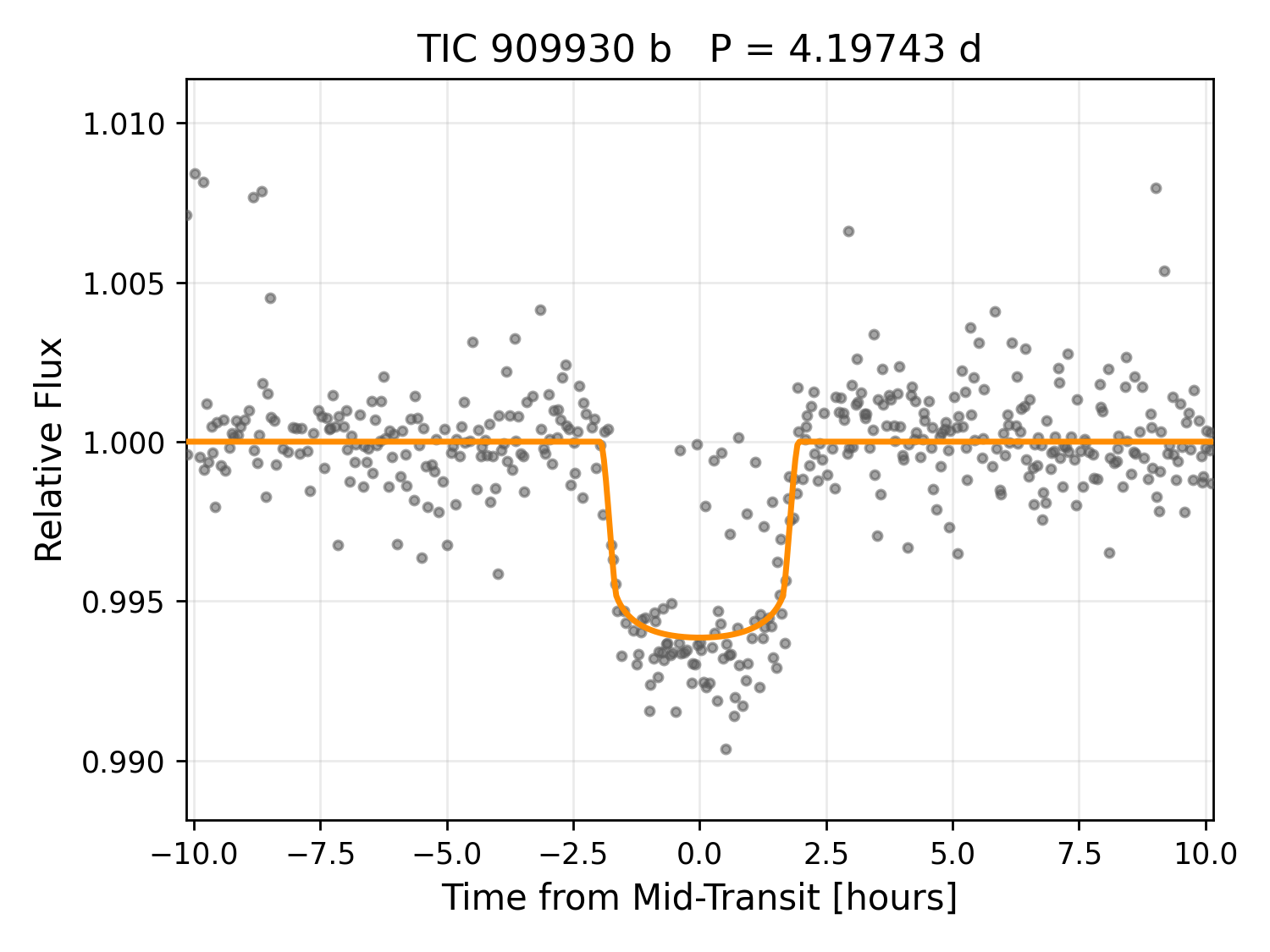}
        } &
        \subfloat[TIC 2688384\label{fov:2}]{
            \includegraphics[width=0.18\linewidth]{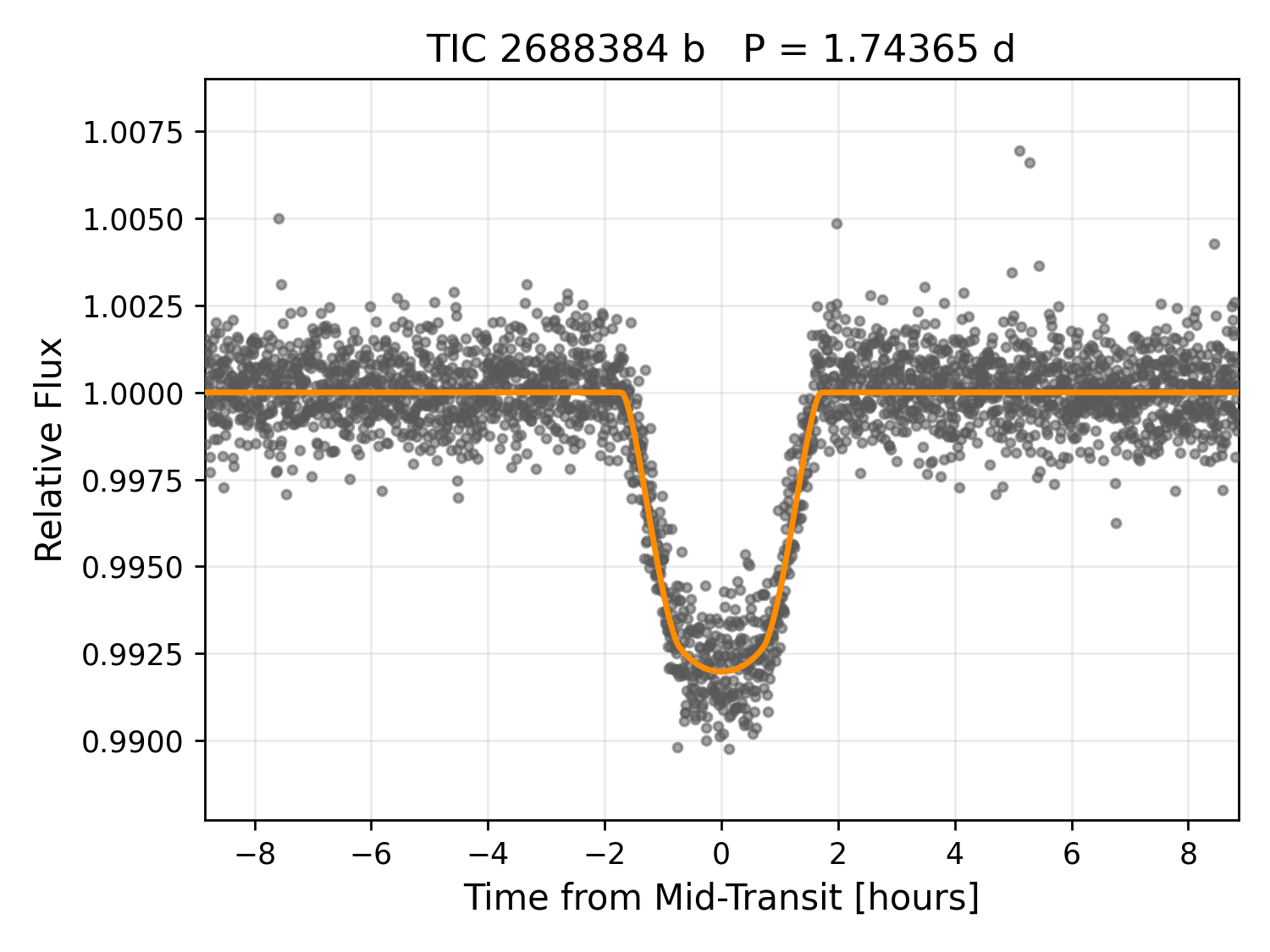}
        } &
        \subfloat[TIC 3317218\label{fov:3}]{
            \includegraphics[width=0.18\linewidth]{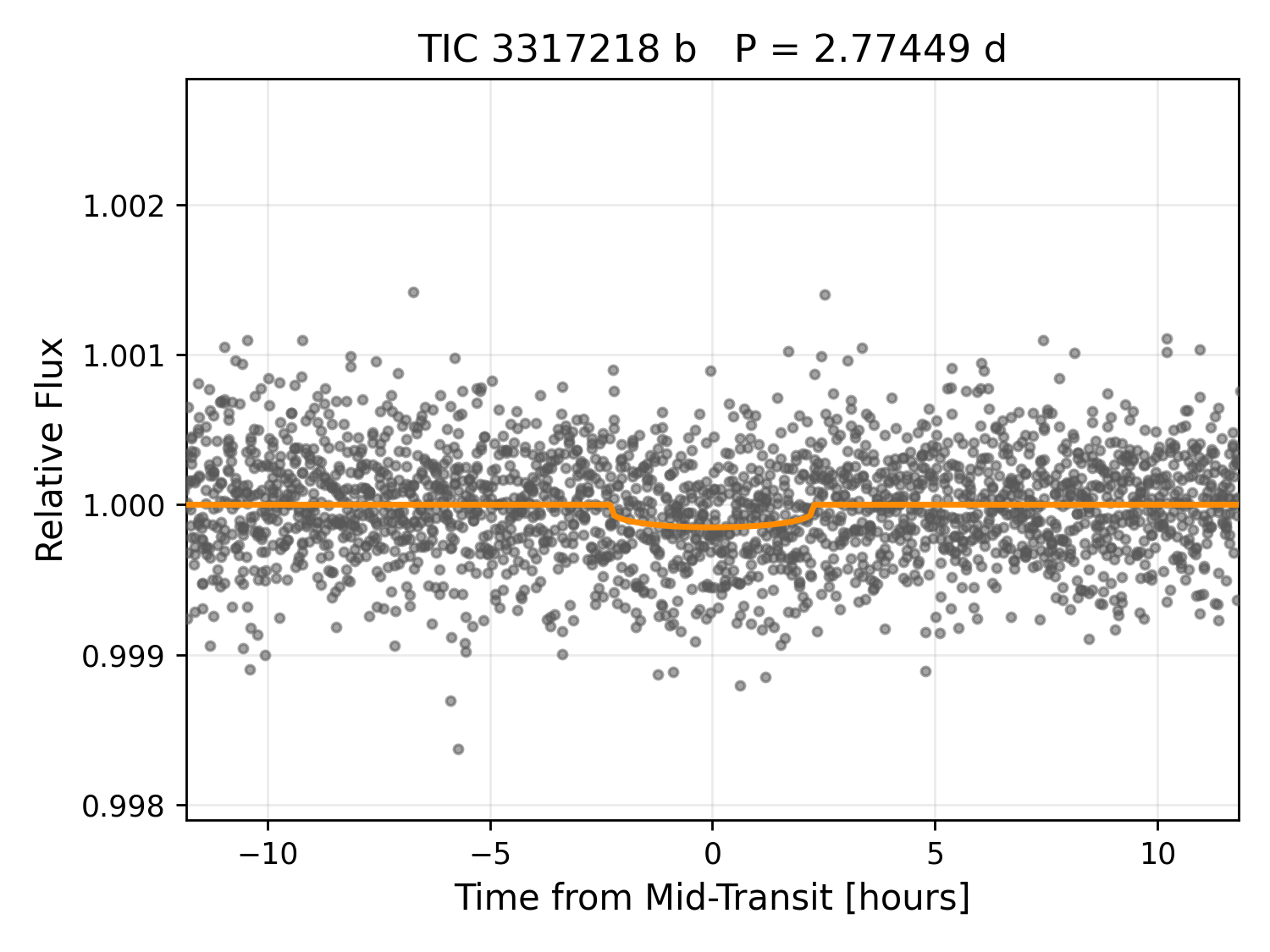}
        } &
        \subfloat[TIC 6893917\label{fov:4}]{
            \includegraphics[width=0.18\linewidth]{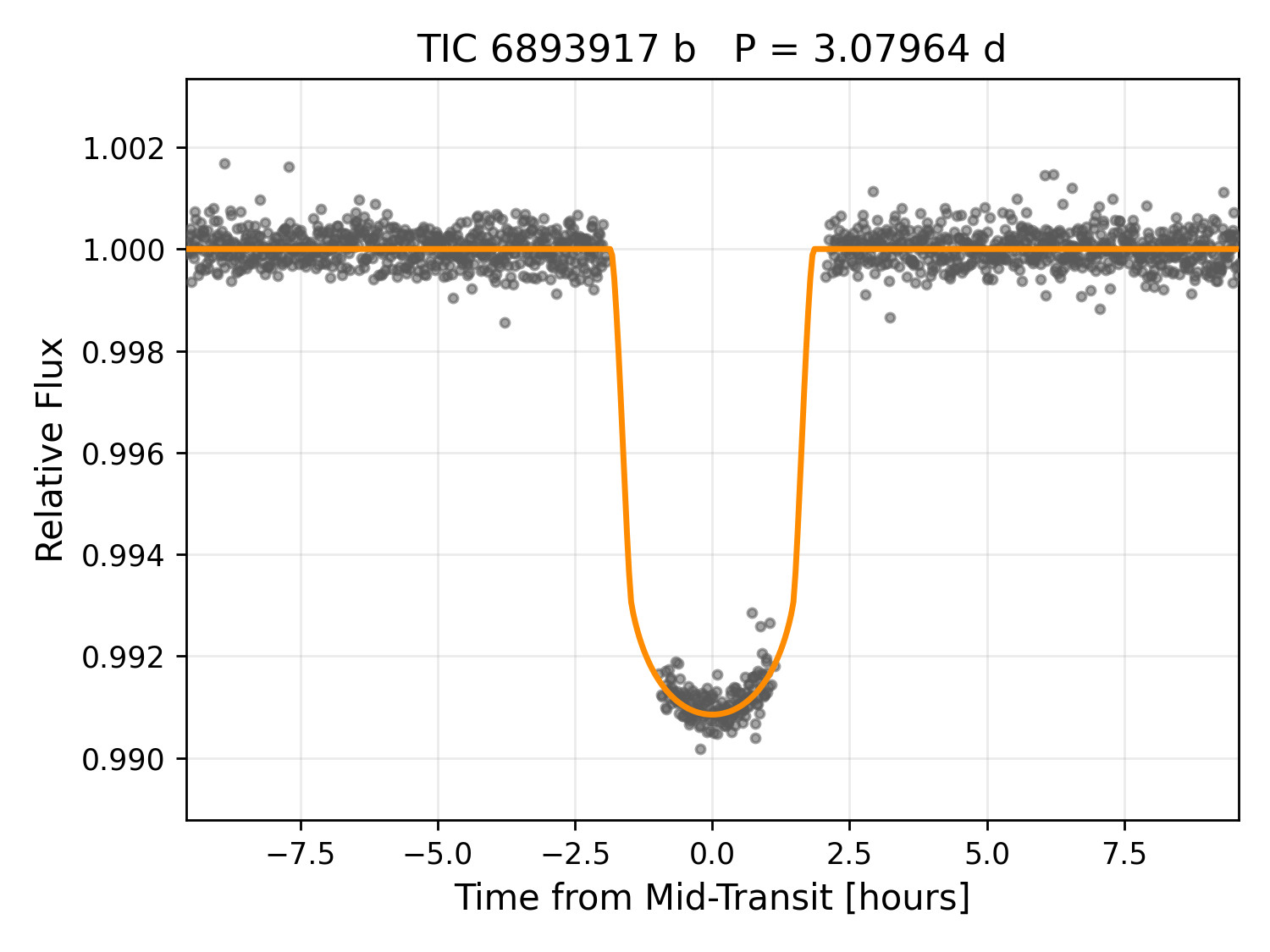}
        } &
        \subfloat[TIC 8400842\label{fov:5}]{
            \includegraphics[width=0.18\linewidth]{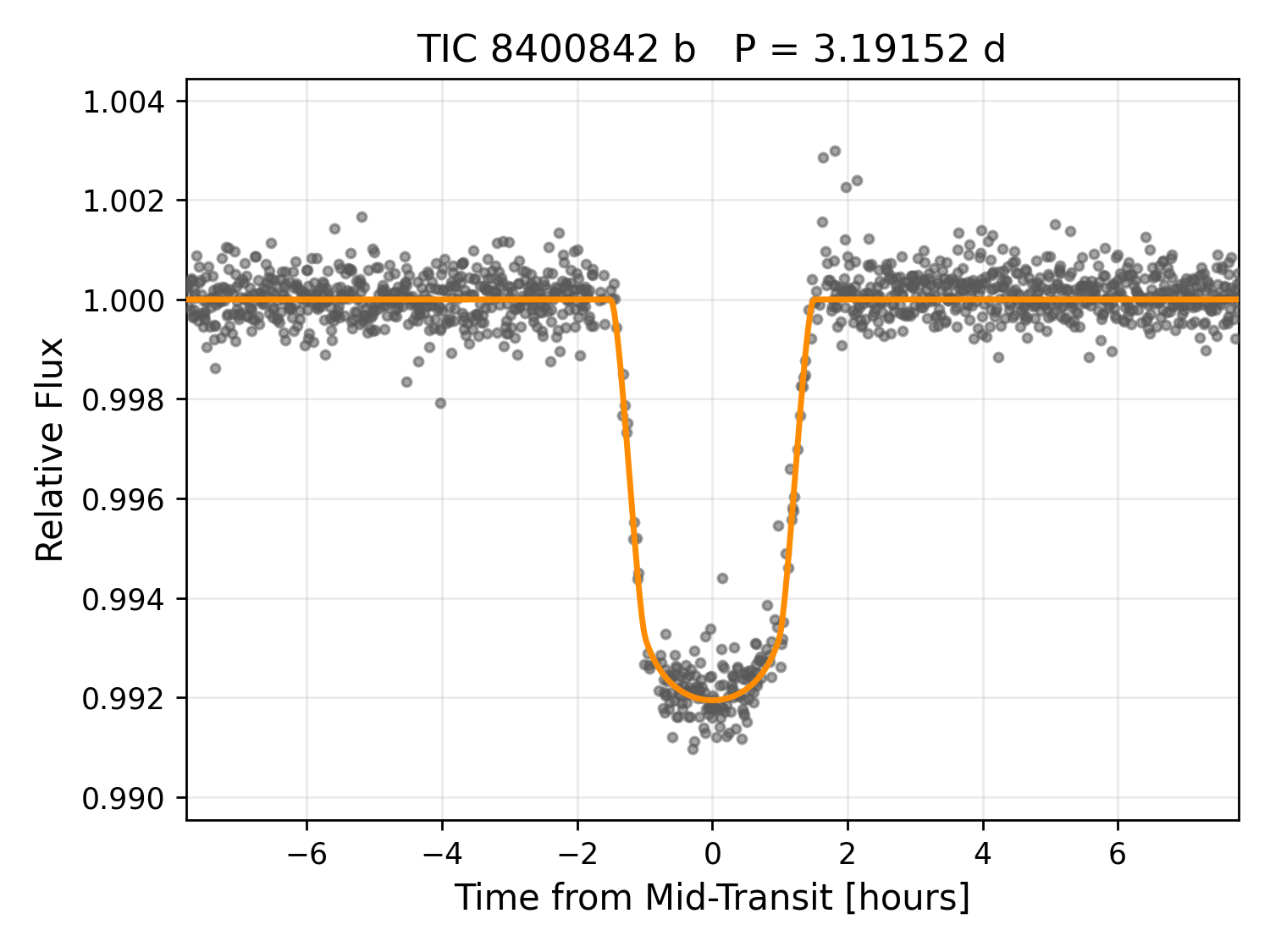}
        }
        \\

        \subfloat[TIC 9828416\label{fov:6}]{
            \includegraphics[width=0.18\linewidth]{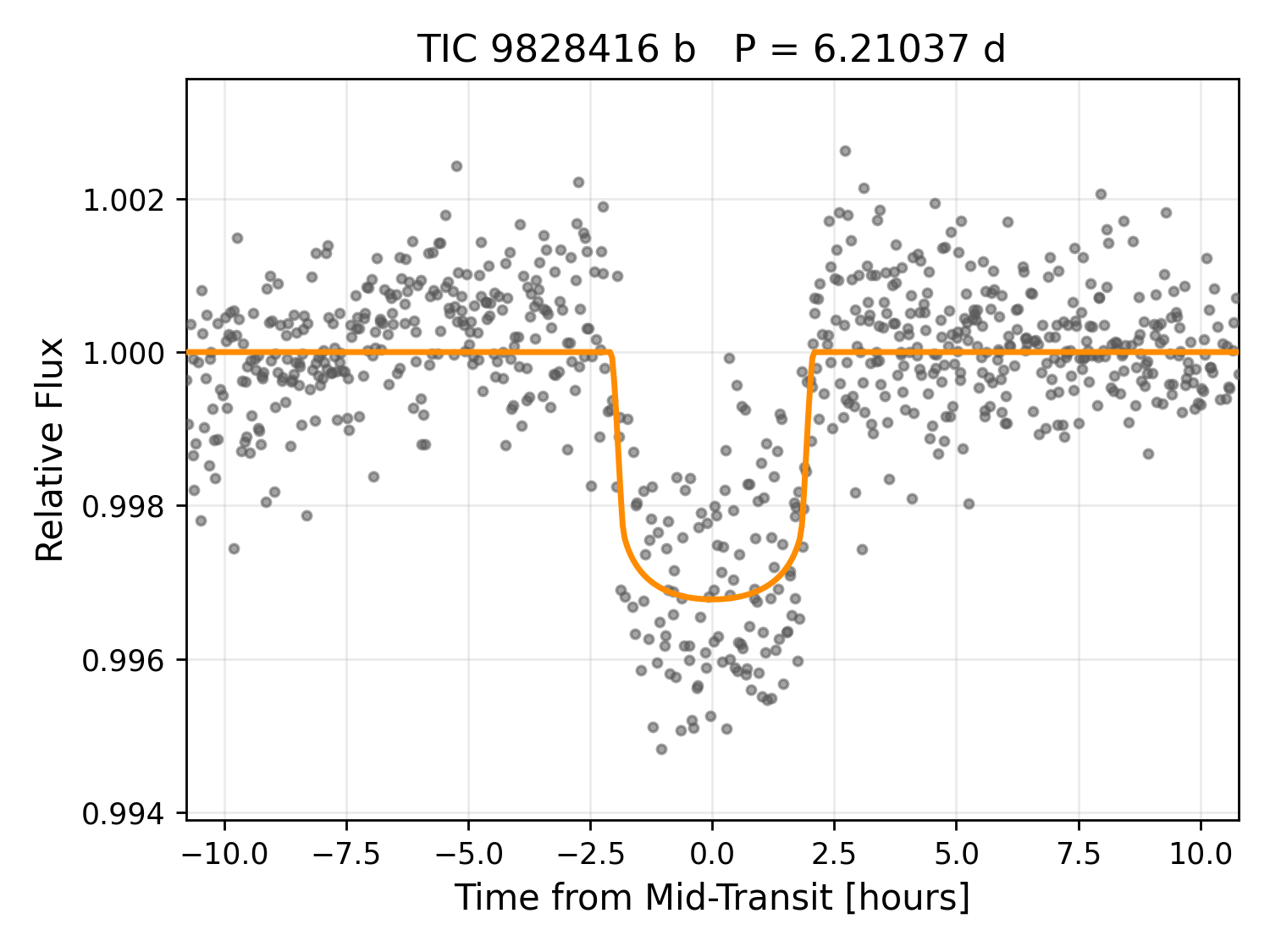}
        } &
        \subfloat[TIC 10394697\label{fov:7}]{
            \includegraphics[width=0.18\linewidth]{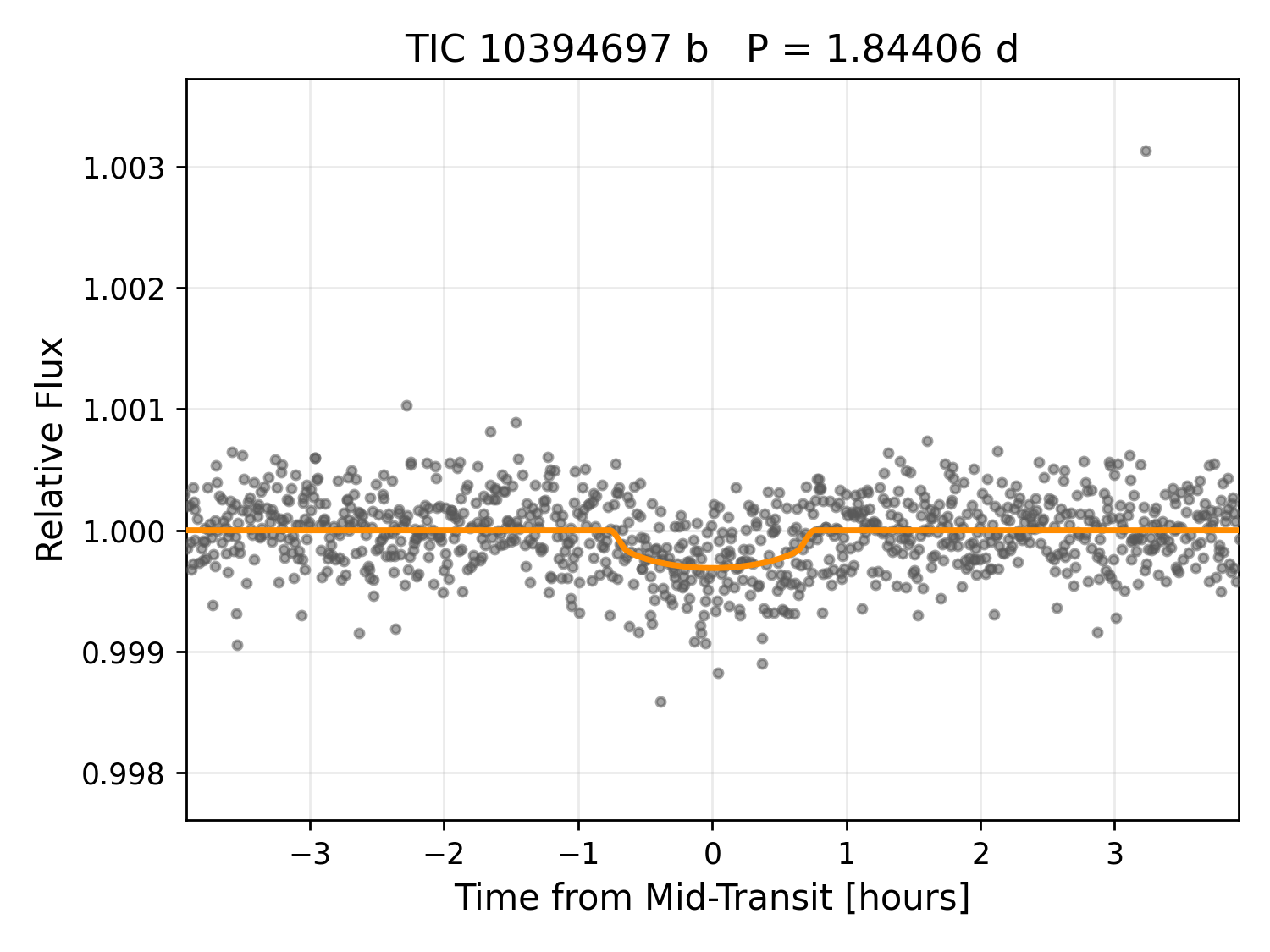}
        } &
        \subfloat[TIC 16740101\label{fov:8}]{
            \includegraphics[width=0.18\linewidth]{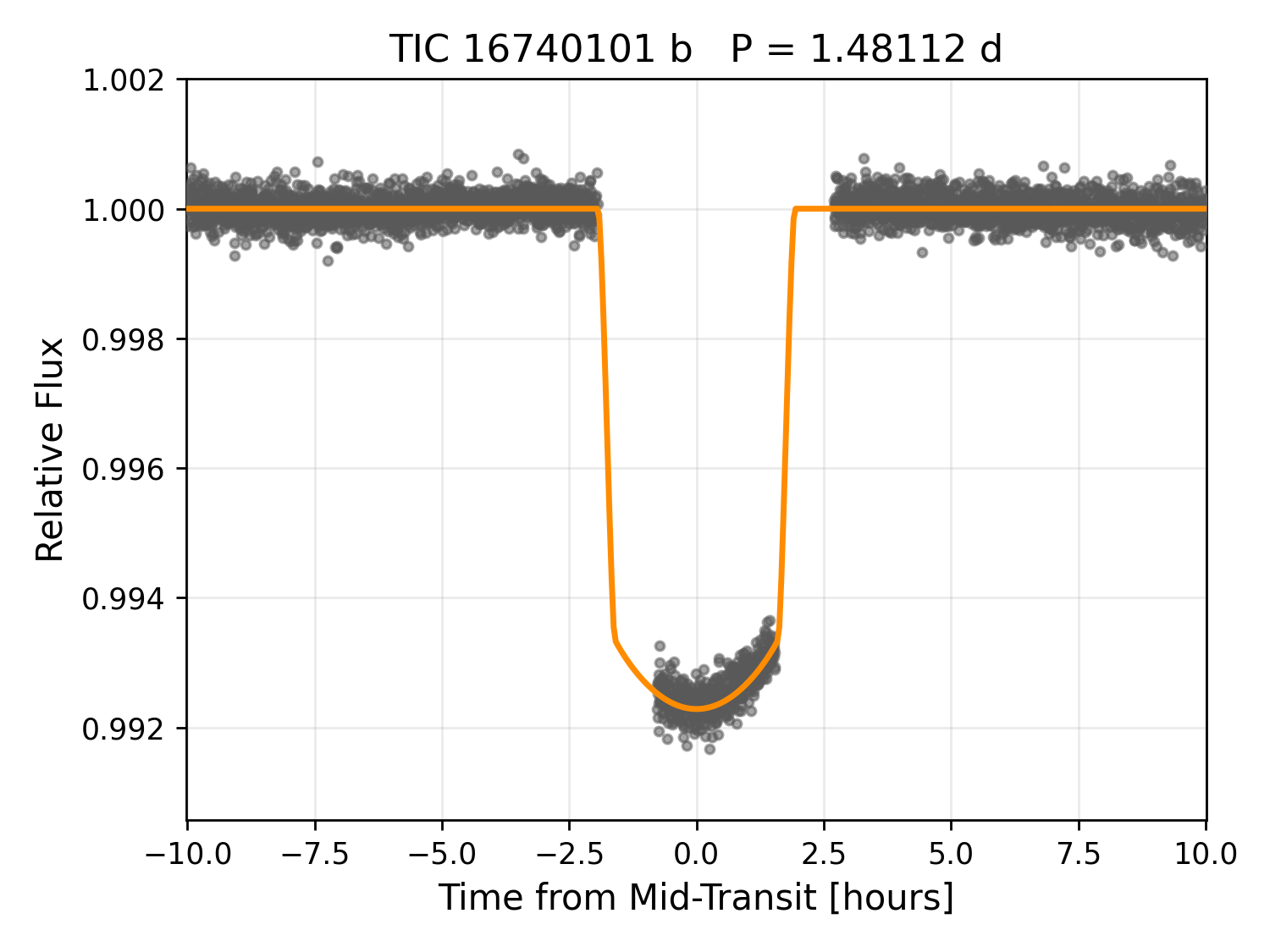}
        } &
        \subfloat[TIC 20096620\label{fov:9}]{
            \includegraphics[width=0.18\linewidth]{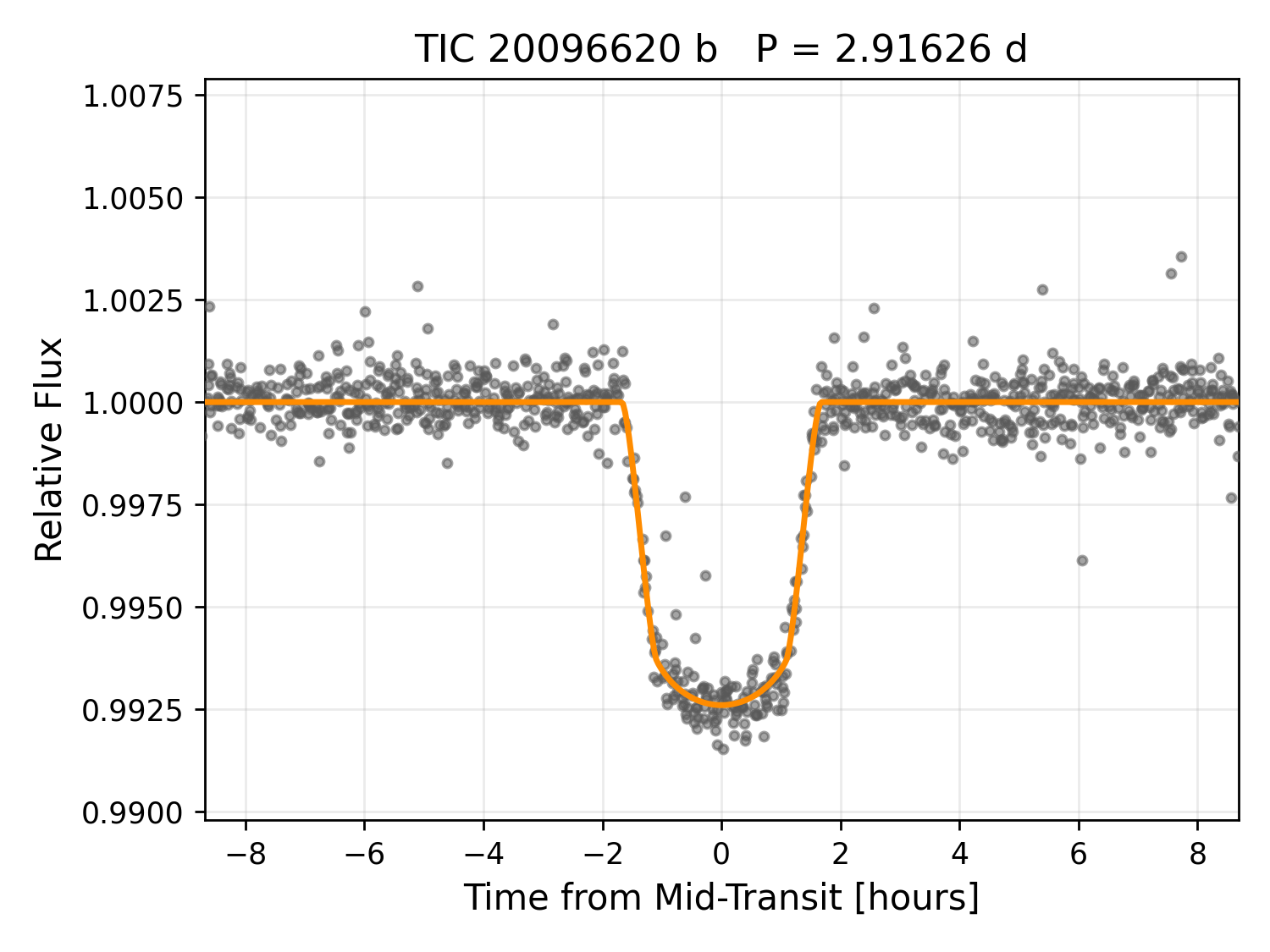}
        } &
        \subfloat[TIC 25375553\label{fov:10}]{
            \includegraphics[width=0.18\linewidth]{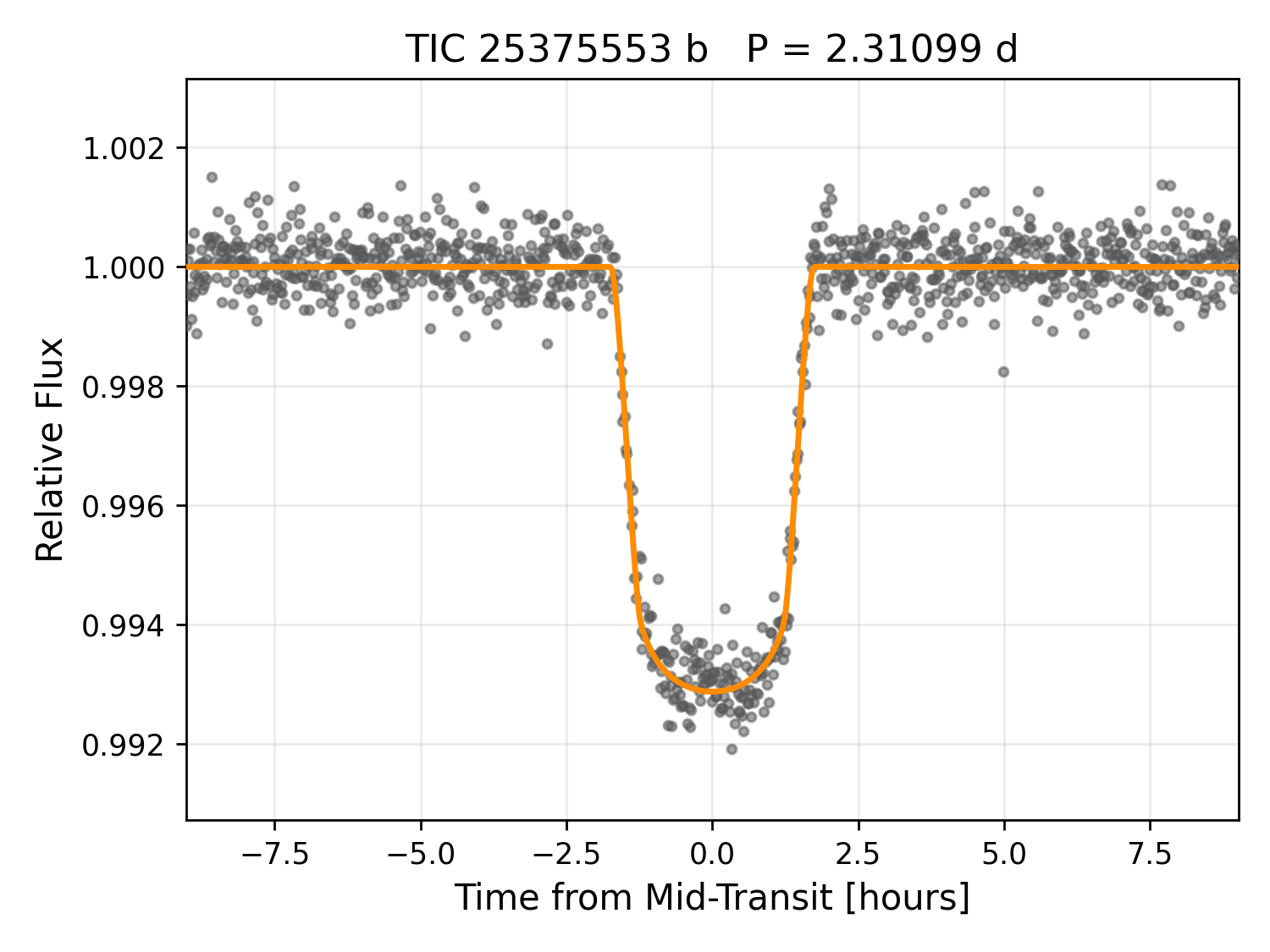}
        }
        \\
\end{tabular}
    \vspace{2pt}
    \caption{Selected Transits}
    \label{fig:appendix_transit}
\end{figure}

\begin{figure}[h]
    \ContinuedFloat
    \centering
    \setlength{\tabcolsep}{1.0pt} 
    \renewcommand{\arraystretch}{1.5}
    \begin{tabular}{ccccc} 
        \subfloat[TIC 77833425\label{fov:36}]{
            \includegraphics[width=0.18\linewidth]{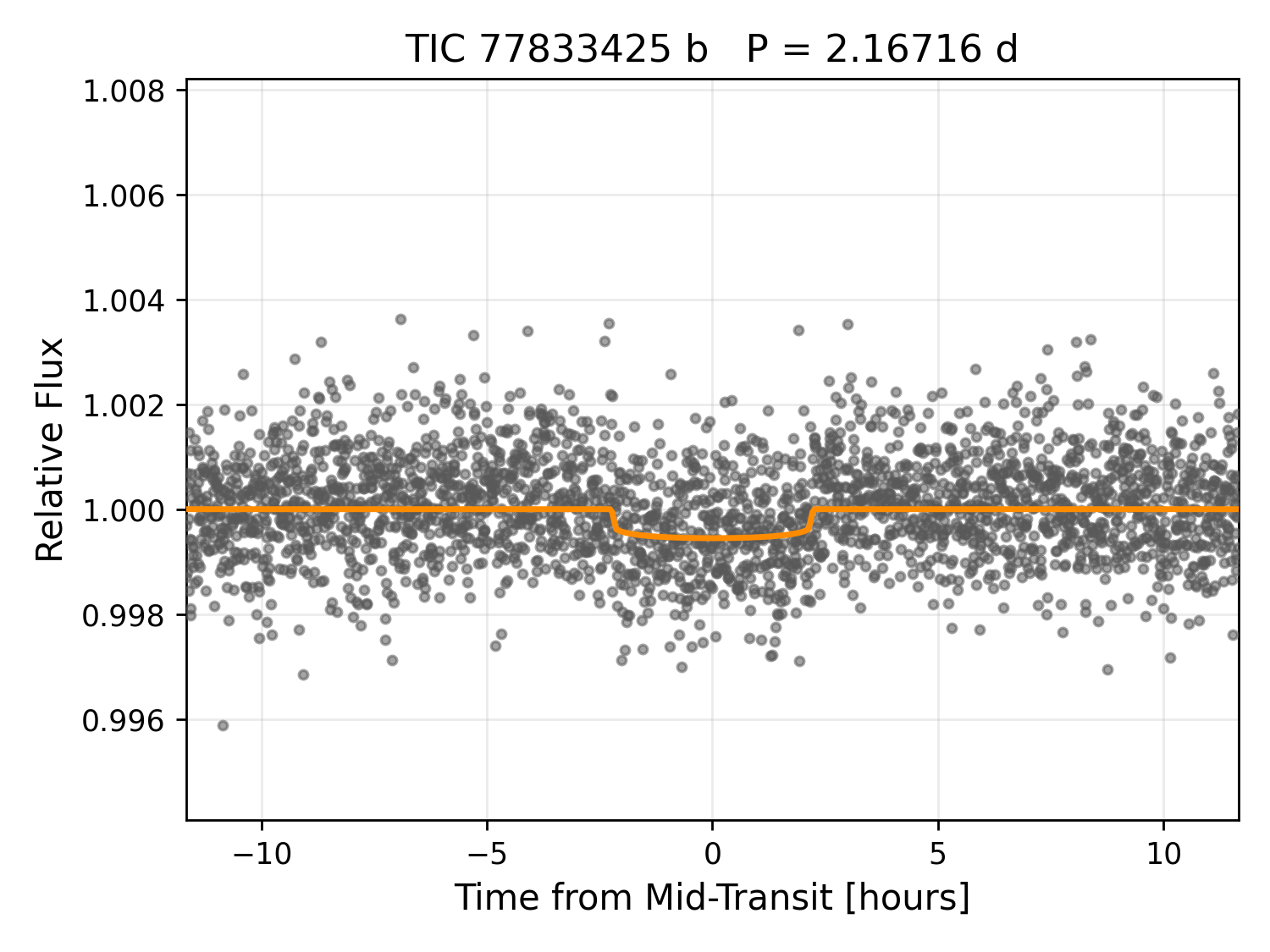}
        } &
        \subfloat[TIC 79292541\label{fov:37}]{
            \includegraphics[width=0.18\linewidth]{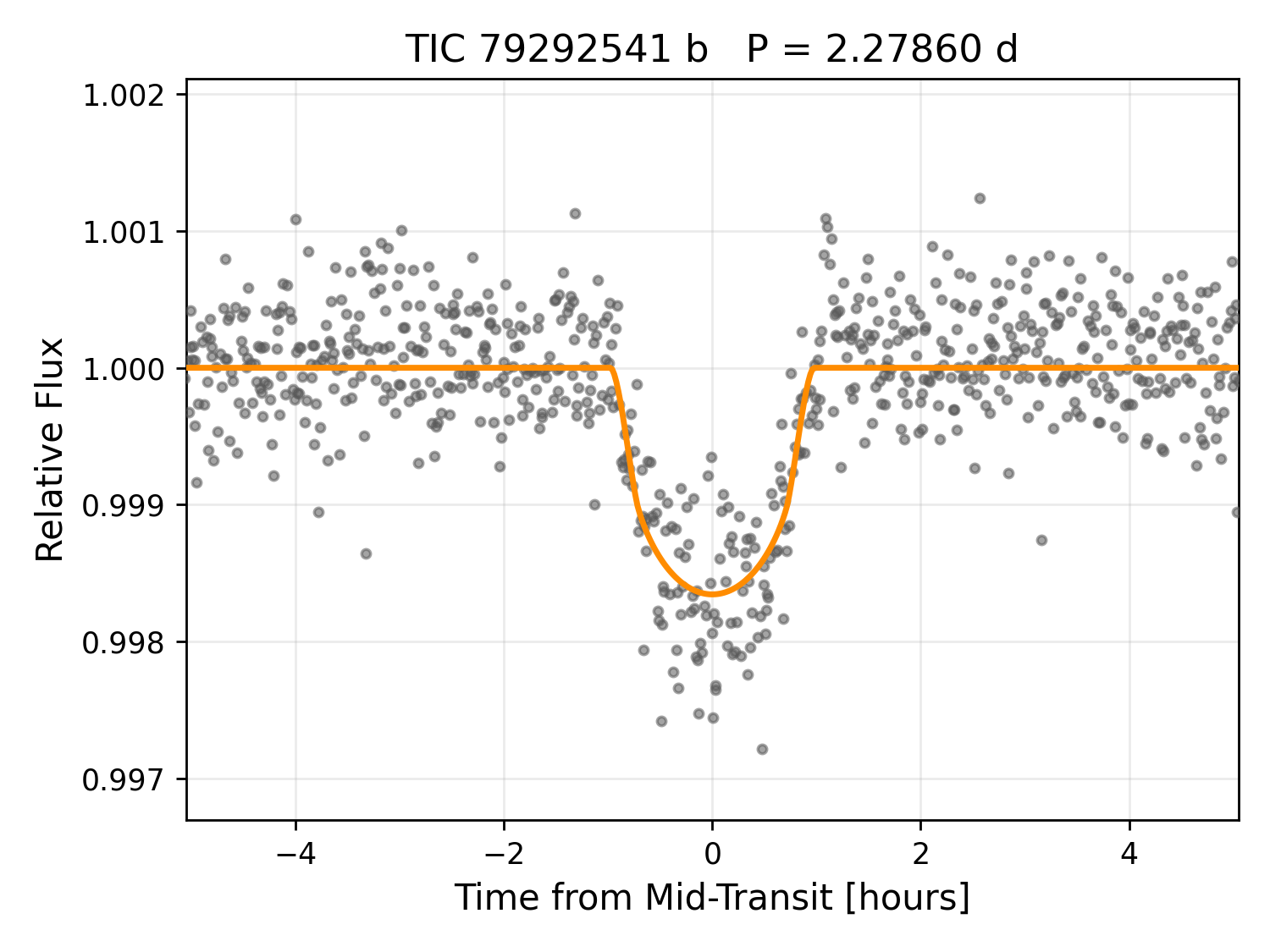}
        } &
        \subfloat[TIC 79605791\label{fov:38}]{
            \includegraphics[width=0.18\linewidth]{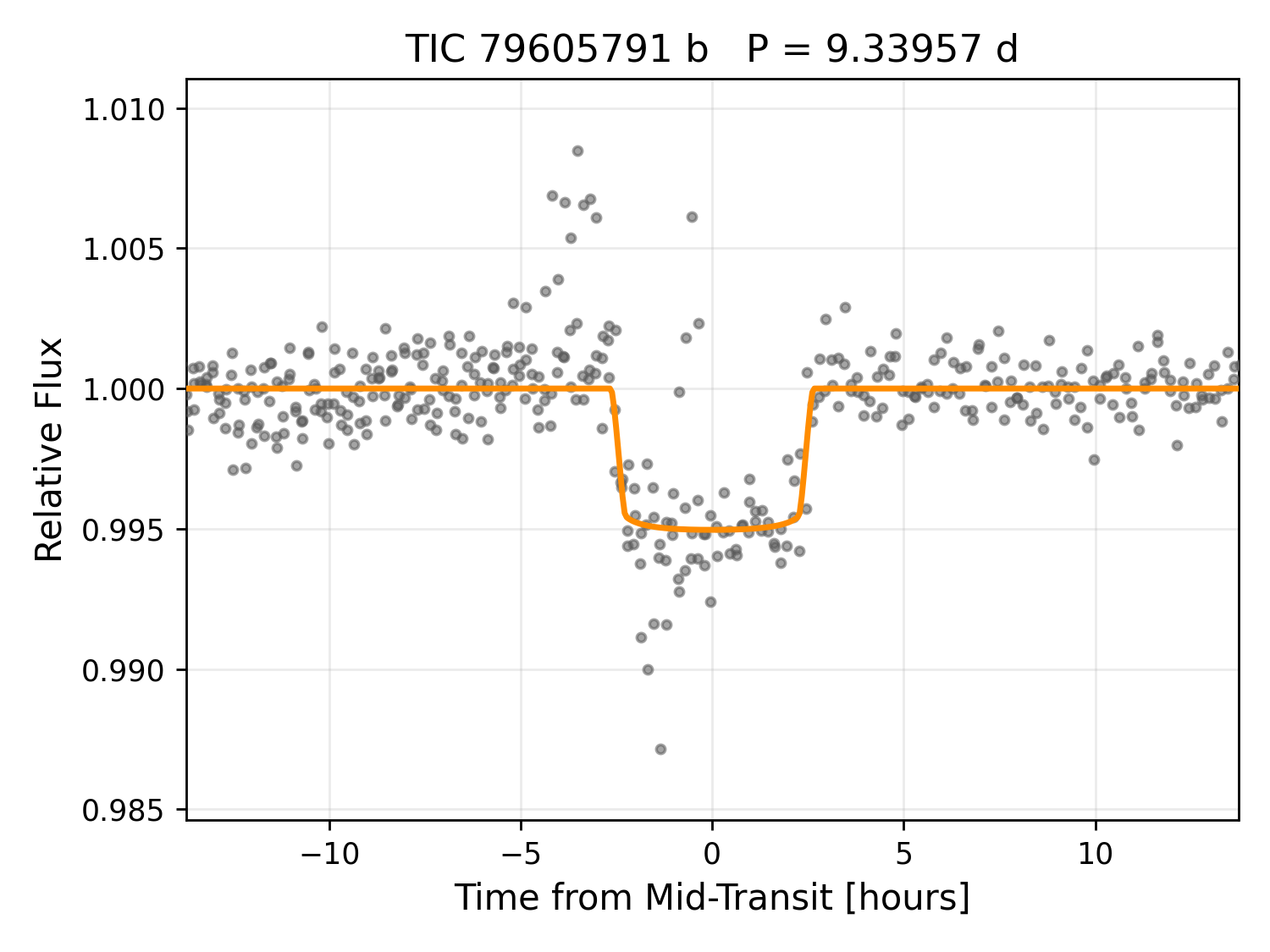}
        } &
        \subfloat[TIC 80166433\label{fov:39}]{
            \includegraphics[width=0.18\linewidth]{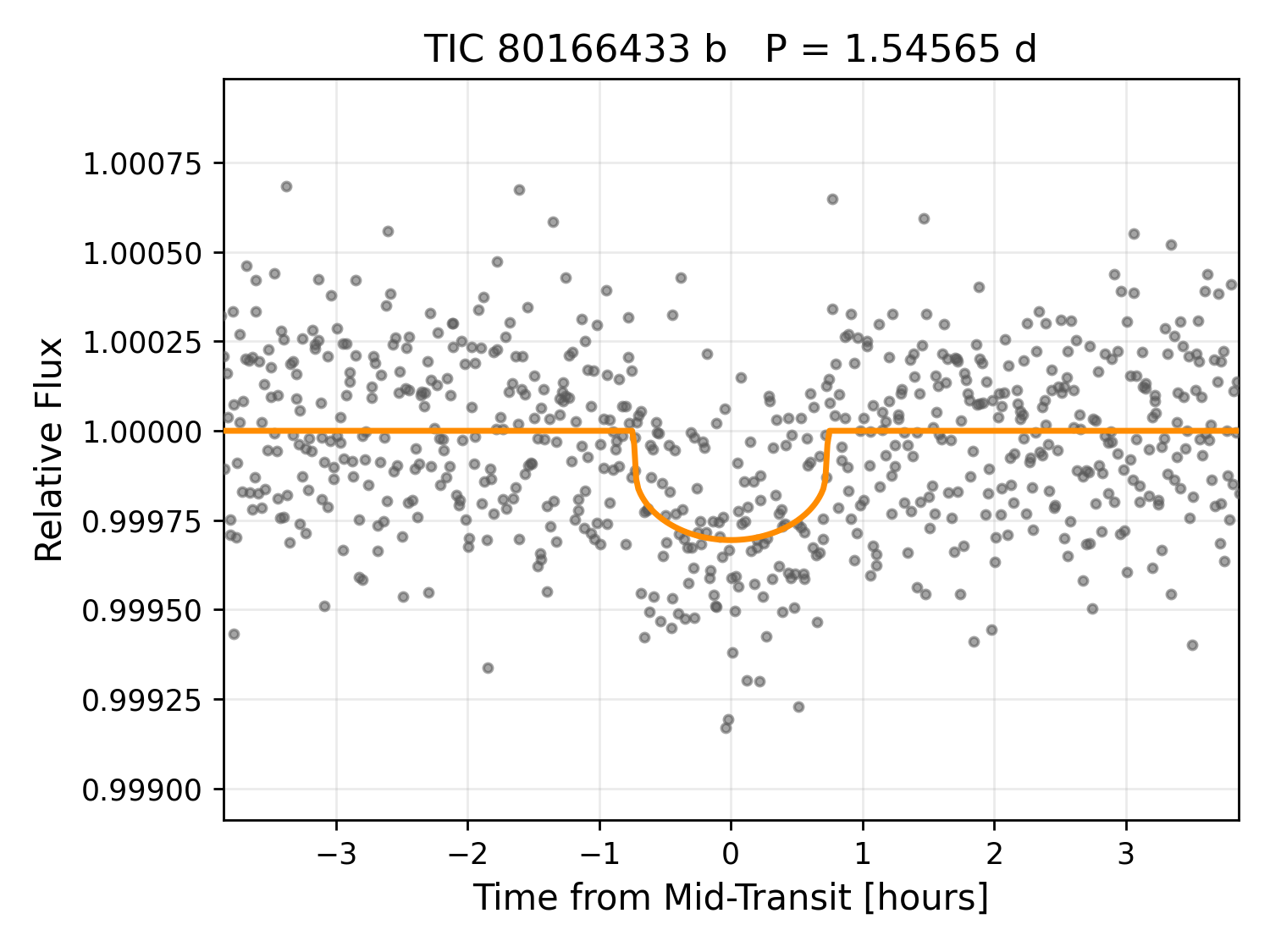}
        } &
        \subfloat[TIC 80832304\label{fov:40}]{
            \includegraphics[width=0.18\linewidth]{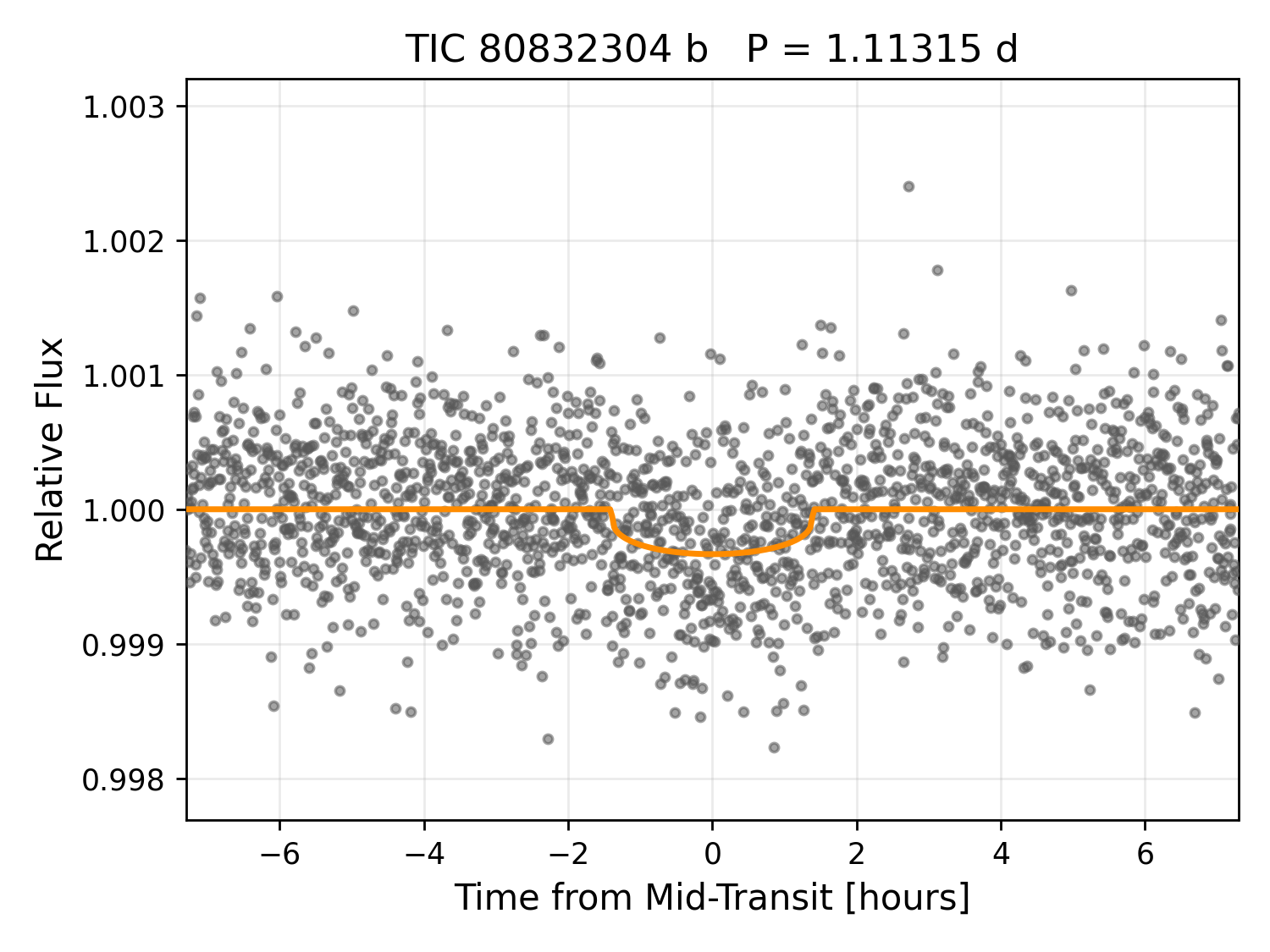}
        }
        \\

        \subfloat[TIC 81077799\label{fov:41}]{
            \includegraphics[width=0.18\linewidth]{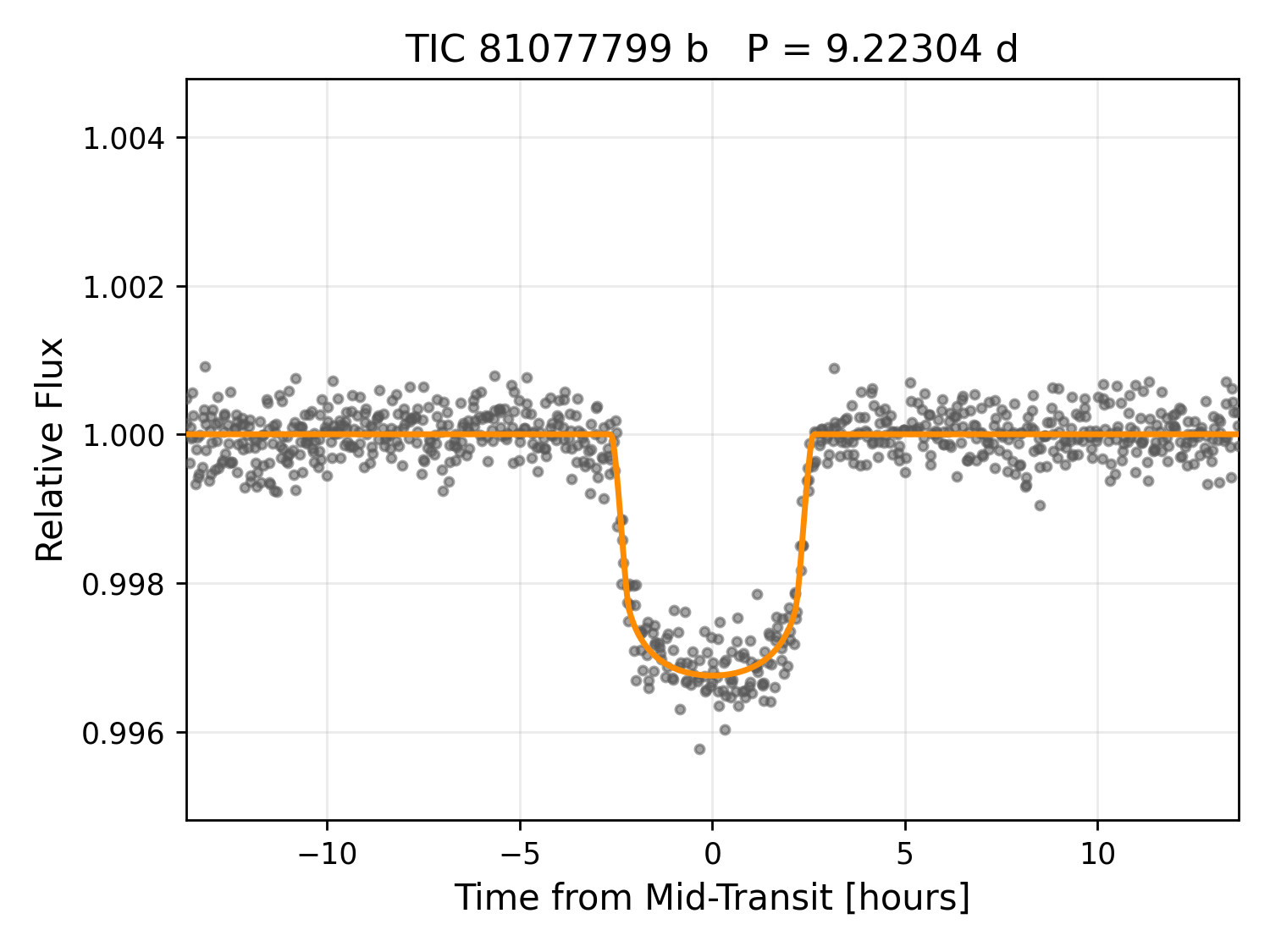}
        } &
        \subfloat[TIC 83092282\label{fov:42}]{
            \includegraphics[width=0.18\linewidth]{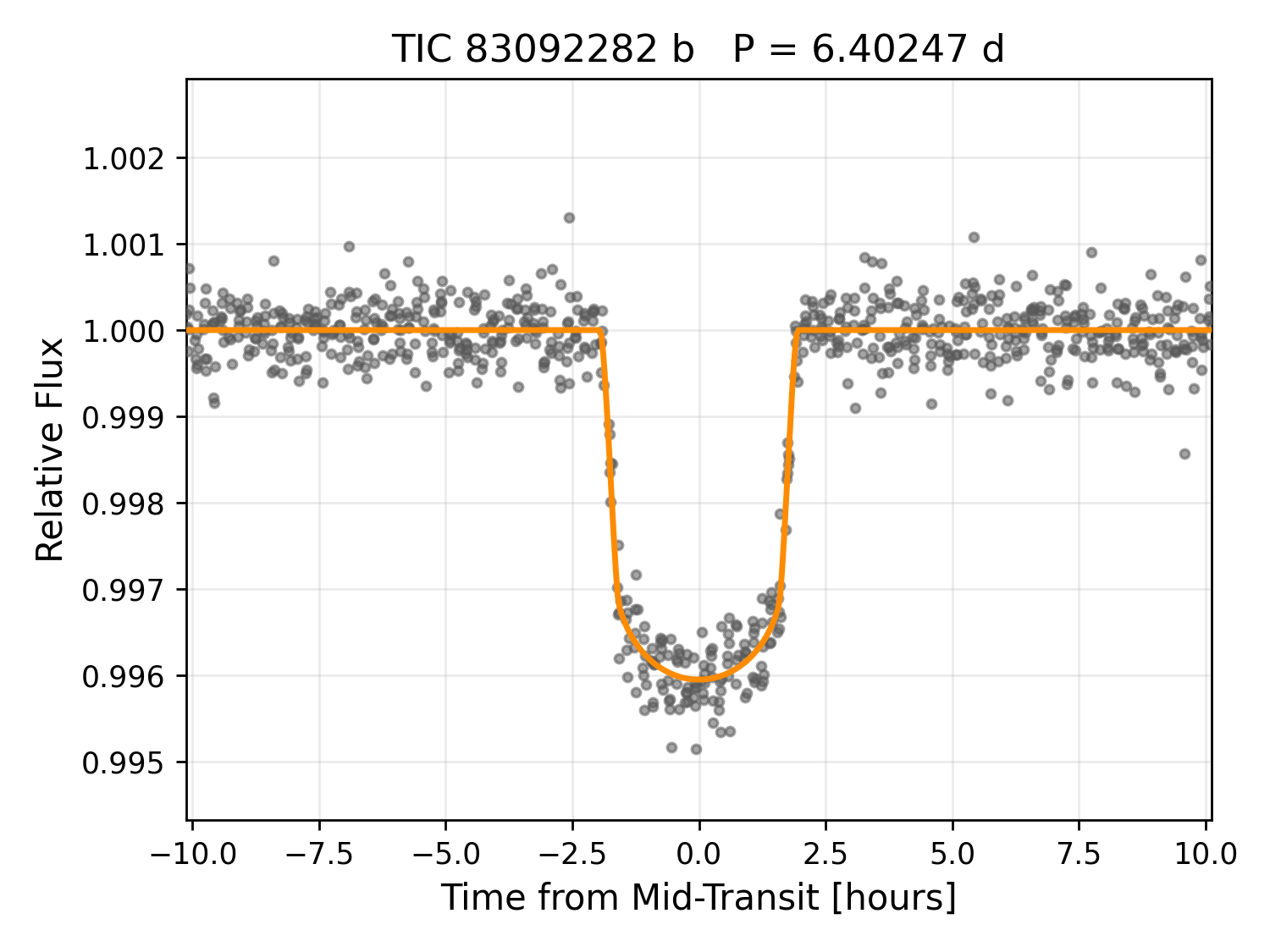}
        } &
        \subfloat[TIC 84339983\label{fov:43}]{
            \includegraphics[width=0.18\linewidth]{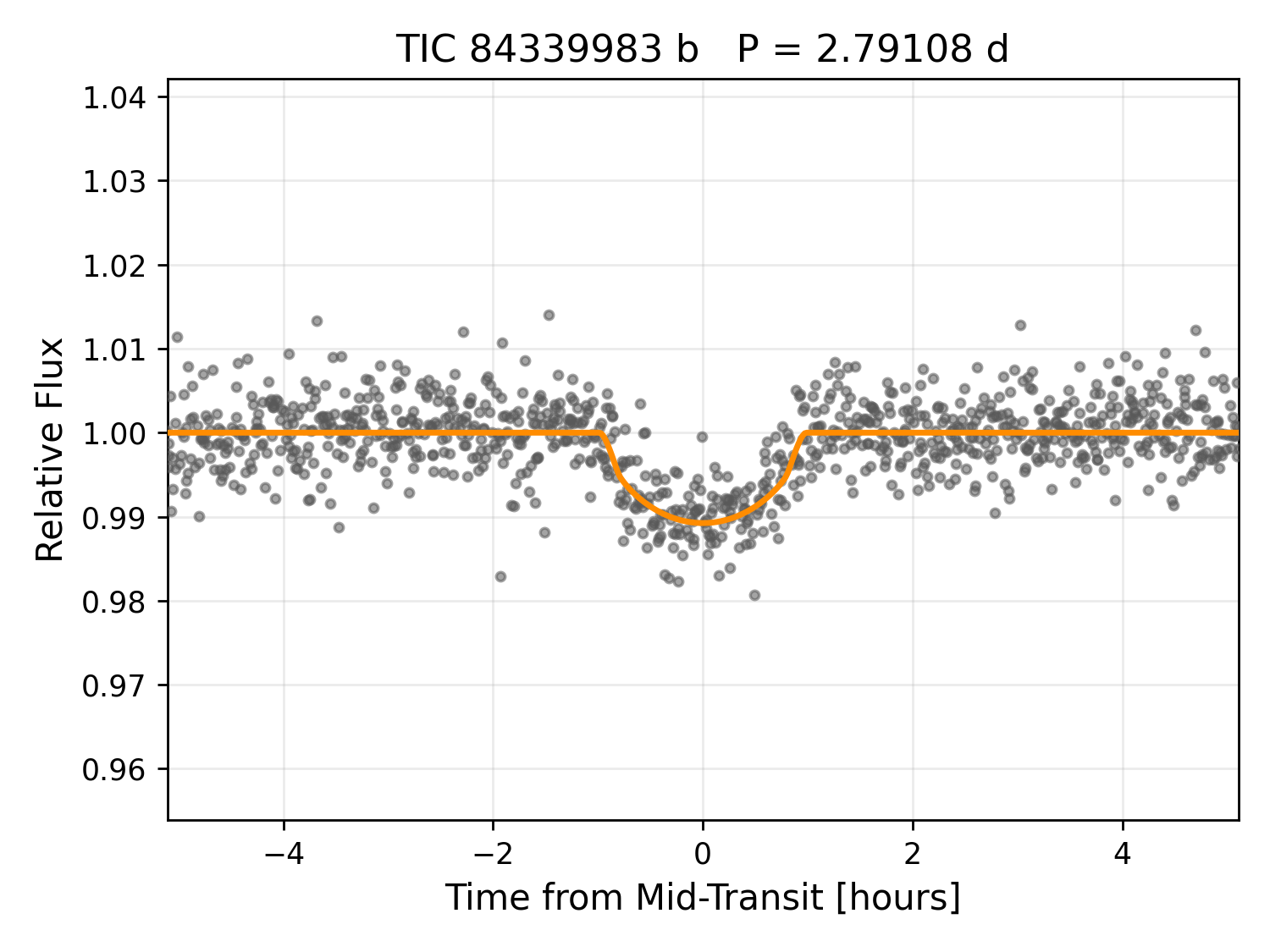}
        } &
        \subfloat[TIC 85266608\label{fov:44}]{
            \includegraphics[width=0.18\linewidth]{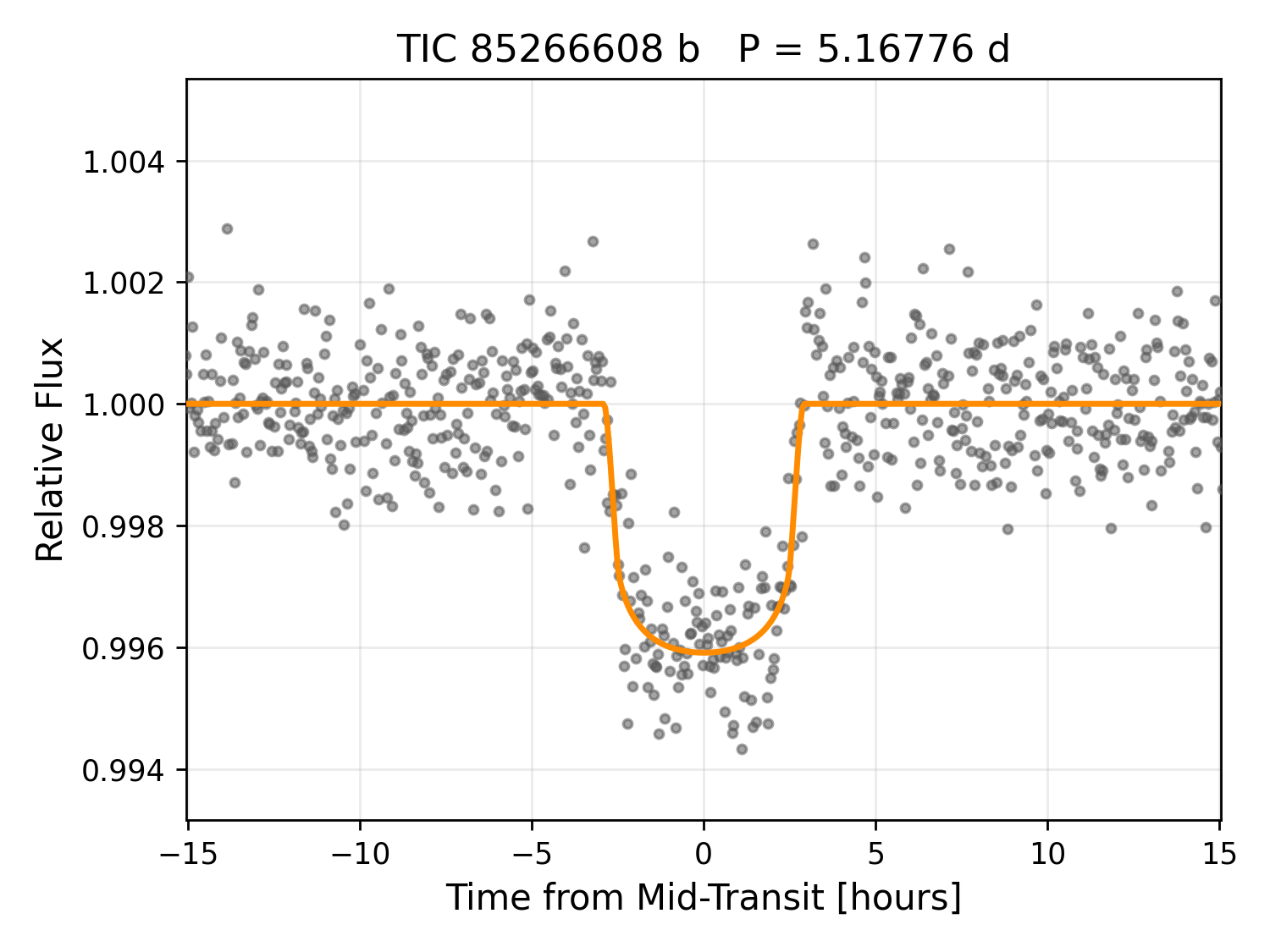}
        } &
        \subfloat[TIC 88902249\label{fov:45}]{
            \includegraphics[width=0.18\linewidth]{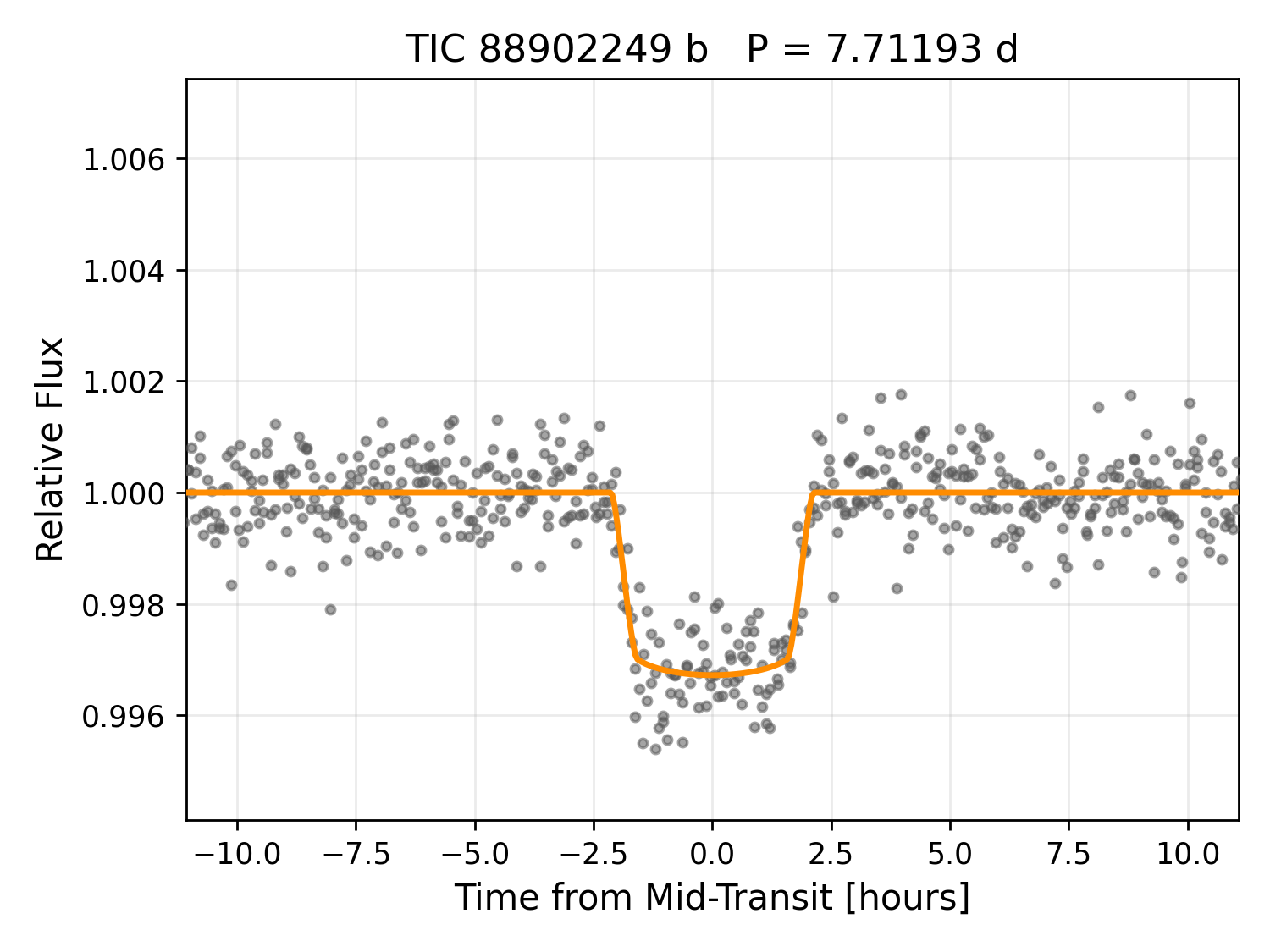}
        }
        \\
\end{tabular}
    \vspace{2pt}
    \caption{Selected Transits}
    \label{fig:appendix_transit}
\end{figure}

\begin{figure}[h]
    \ContinuedFloat
    \centering
    \setlength{\tabcolsep}{1.0pt} 
    \renewcommand{\arraystretch}{1.5}
    \begin{tabular}{ccccc} 
        \subfloat[TIC 149470723\label{fov:76}]{
            \includegraphics[width=0.18\linewidth]{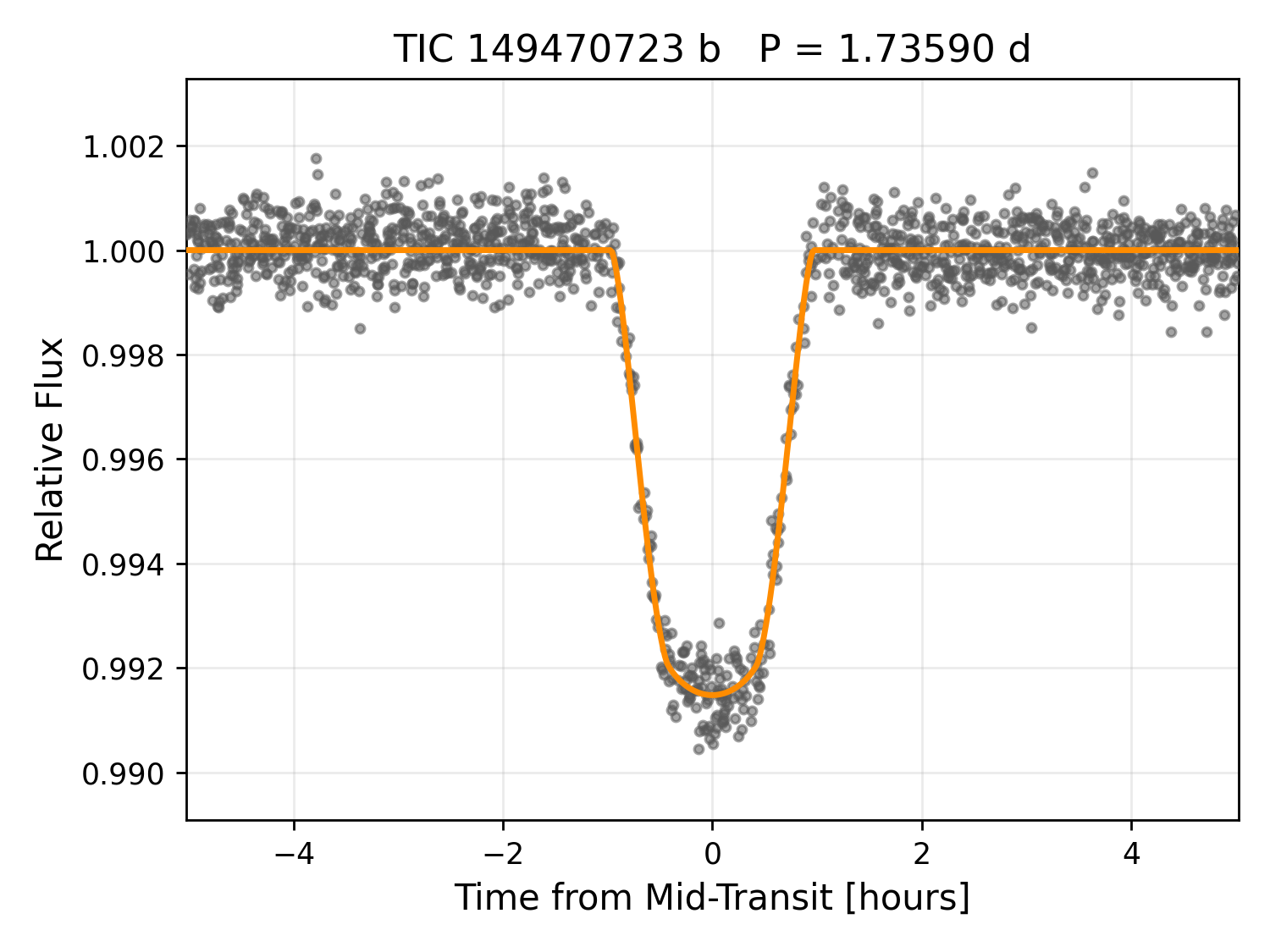}
        } &
        \subfloat[TIC 152476657\label{fov:77}]{
            \includegraphics[width=0.18\linewidth]{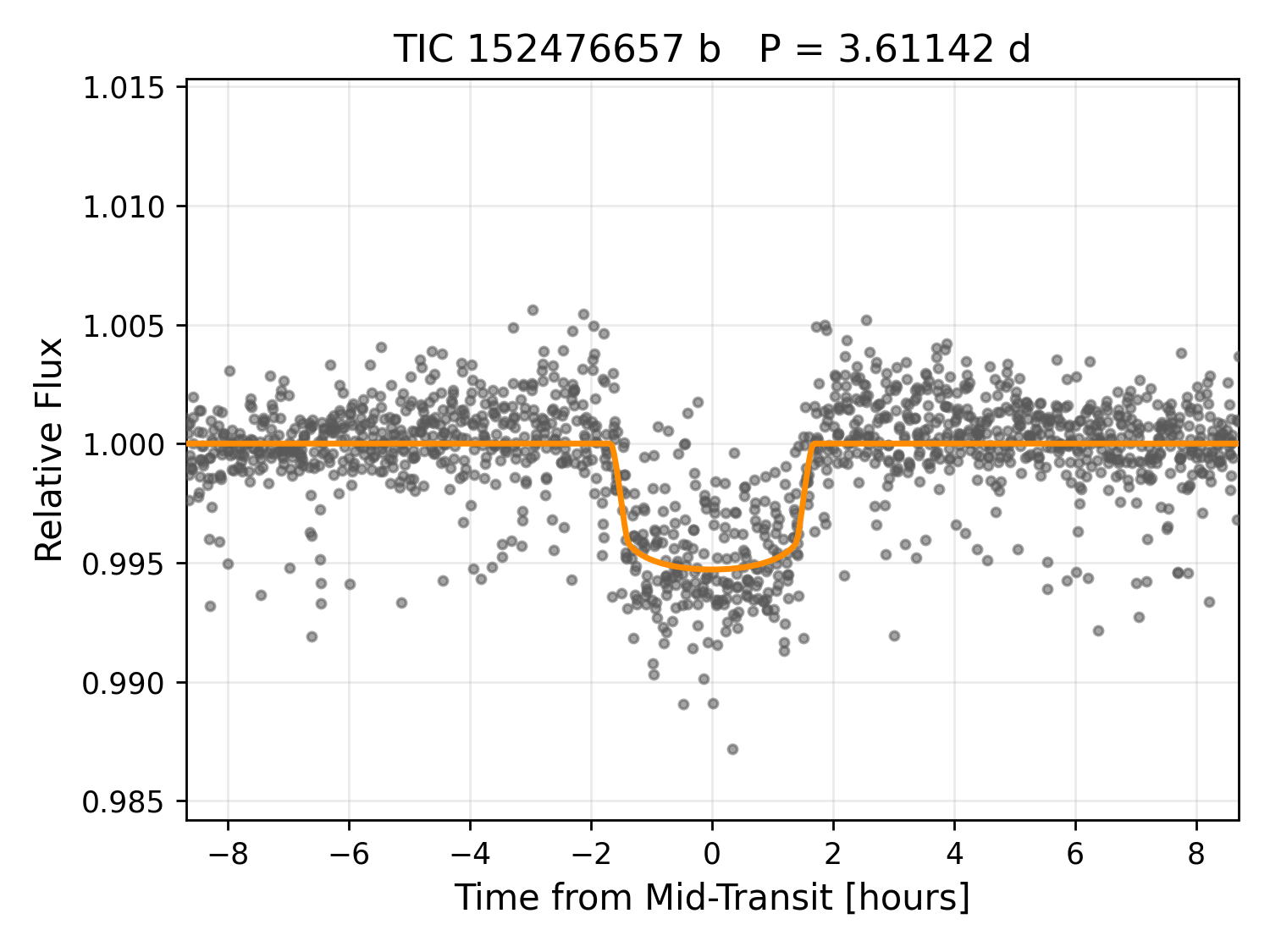}
        } &
        \subfloat[TIC 157999664\label{fov:78}]{
            \includegraphics[width=0.18\linewidth]{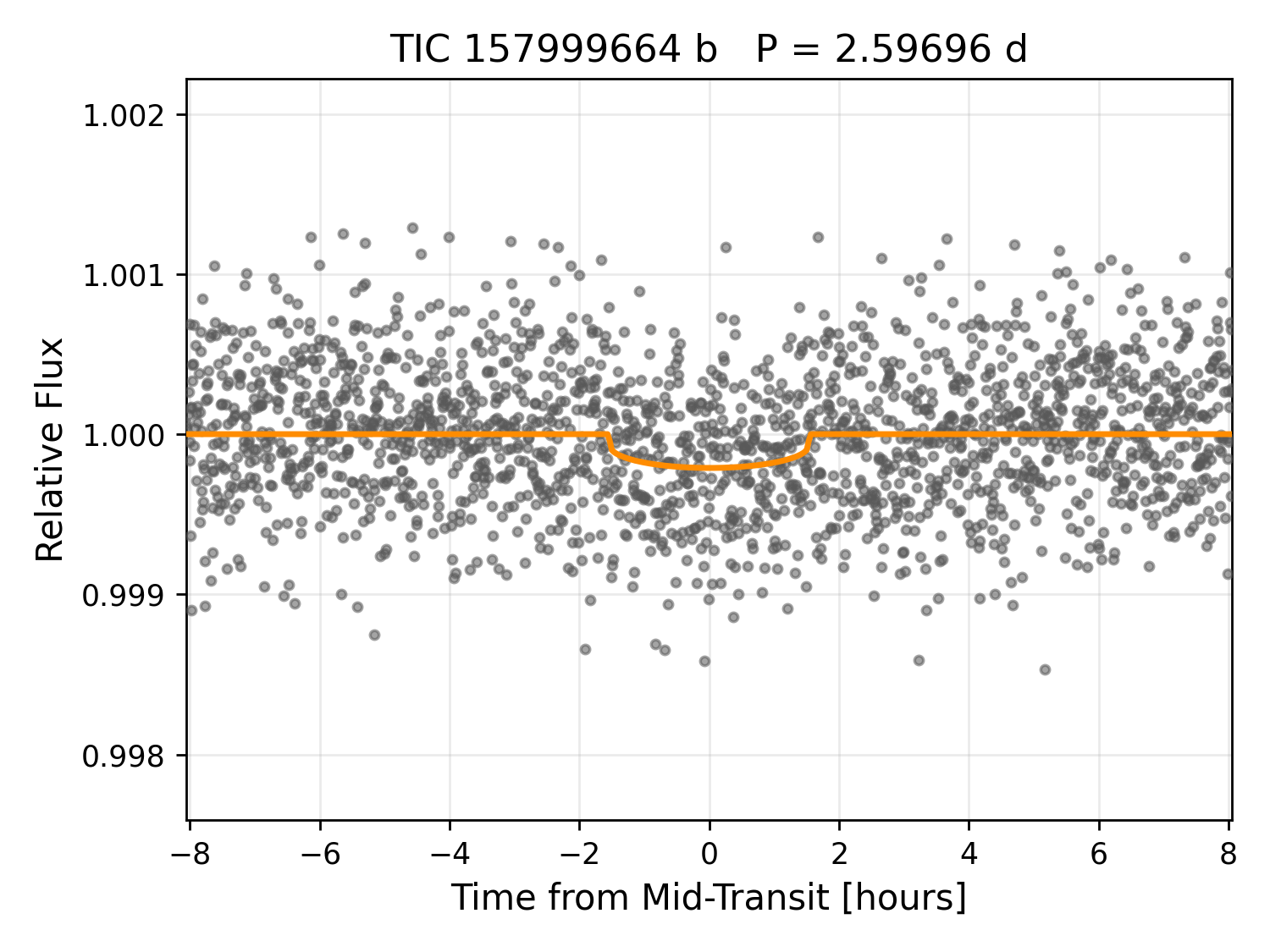}
        } &
        \subfloat[TIC 159742538\label{fov:79}]{
            \includegraphics[width=0.18\linewidth]{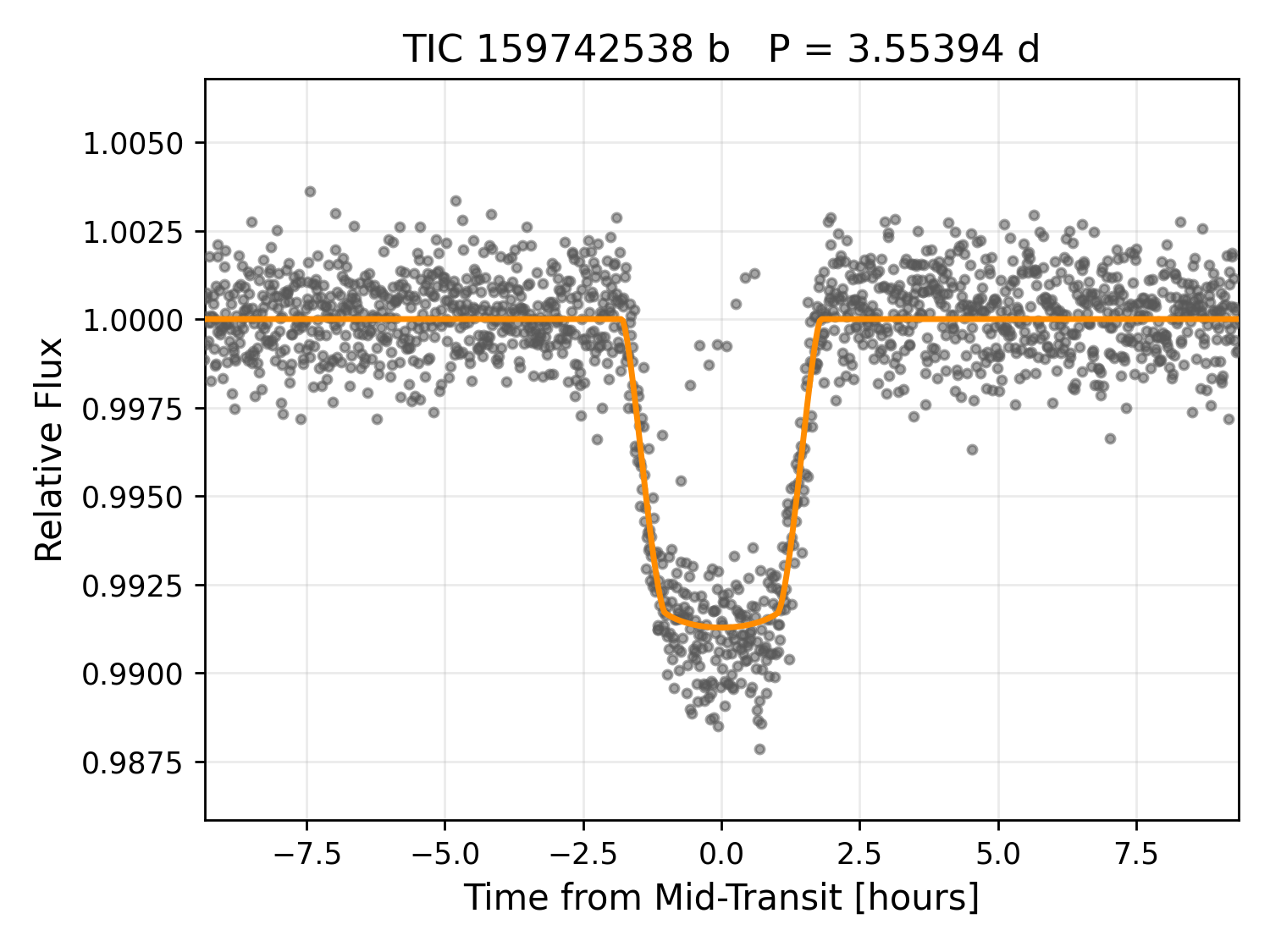}
        } &
        \subfloat[TIC 160432093\label{fov:80}]{
            \includegraphics[width=0.18\linewidth]{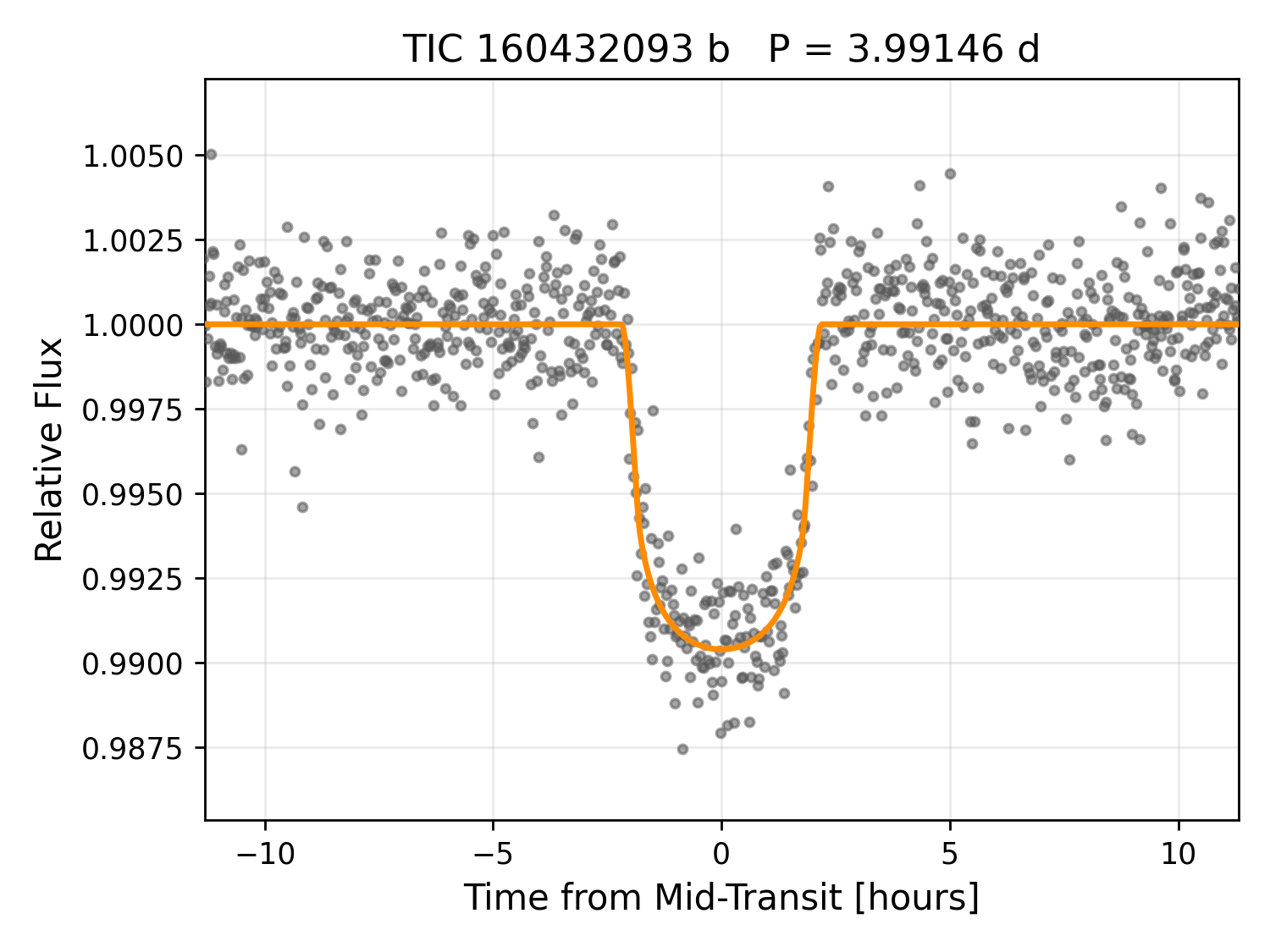}
        }
        \\

        \subfloat[TIC 161155463\label{fov:81}]{
            \includegraphics[width=0.18\linewidth]{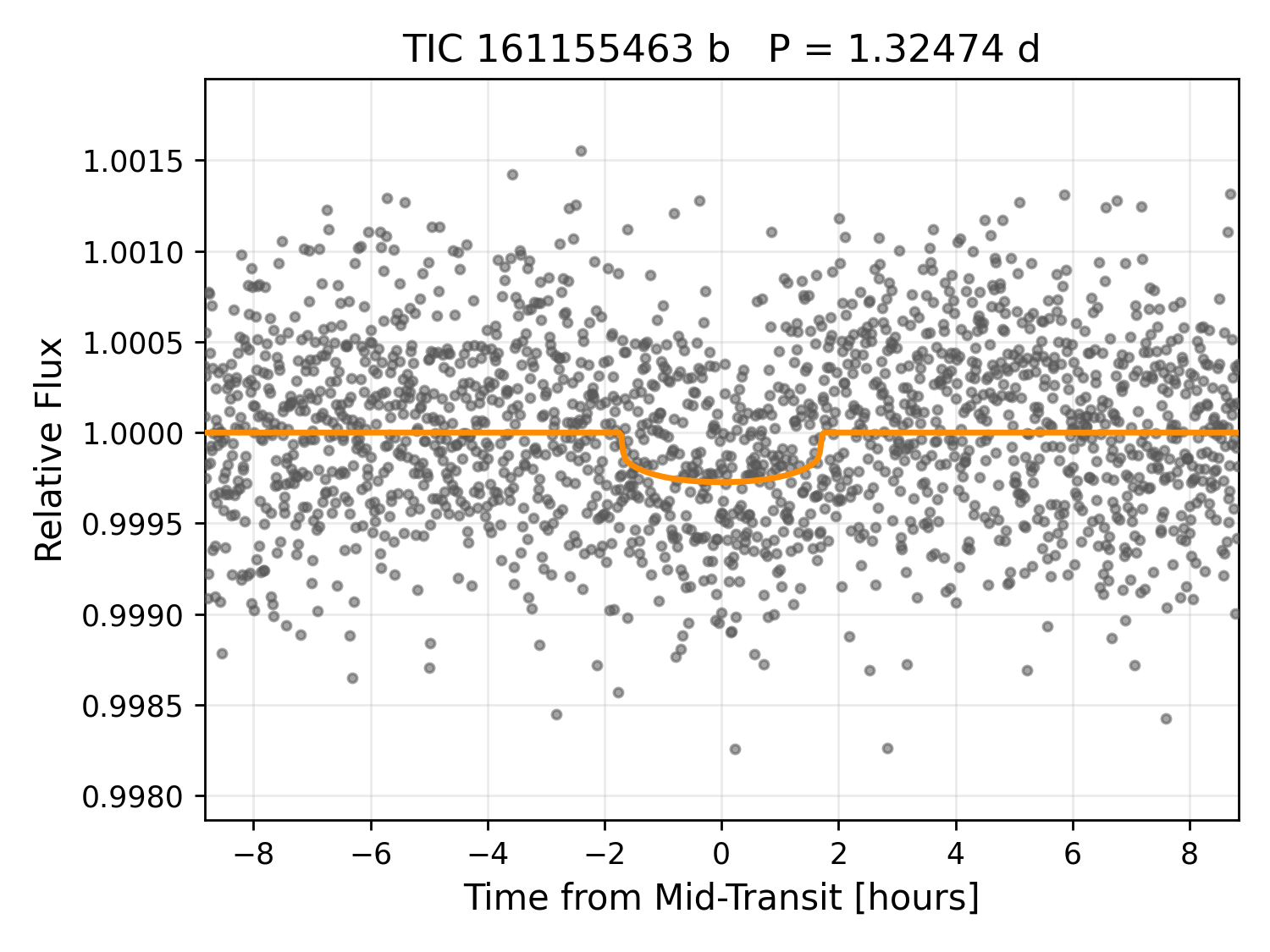}
        } &
        \subfloat[TIC 161479090\label{fov:82}]{
            \includegraphics[width=0.18\linewidth]{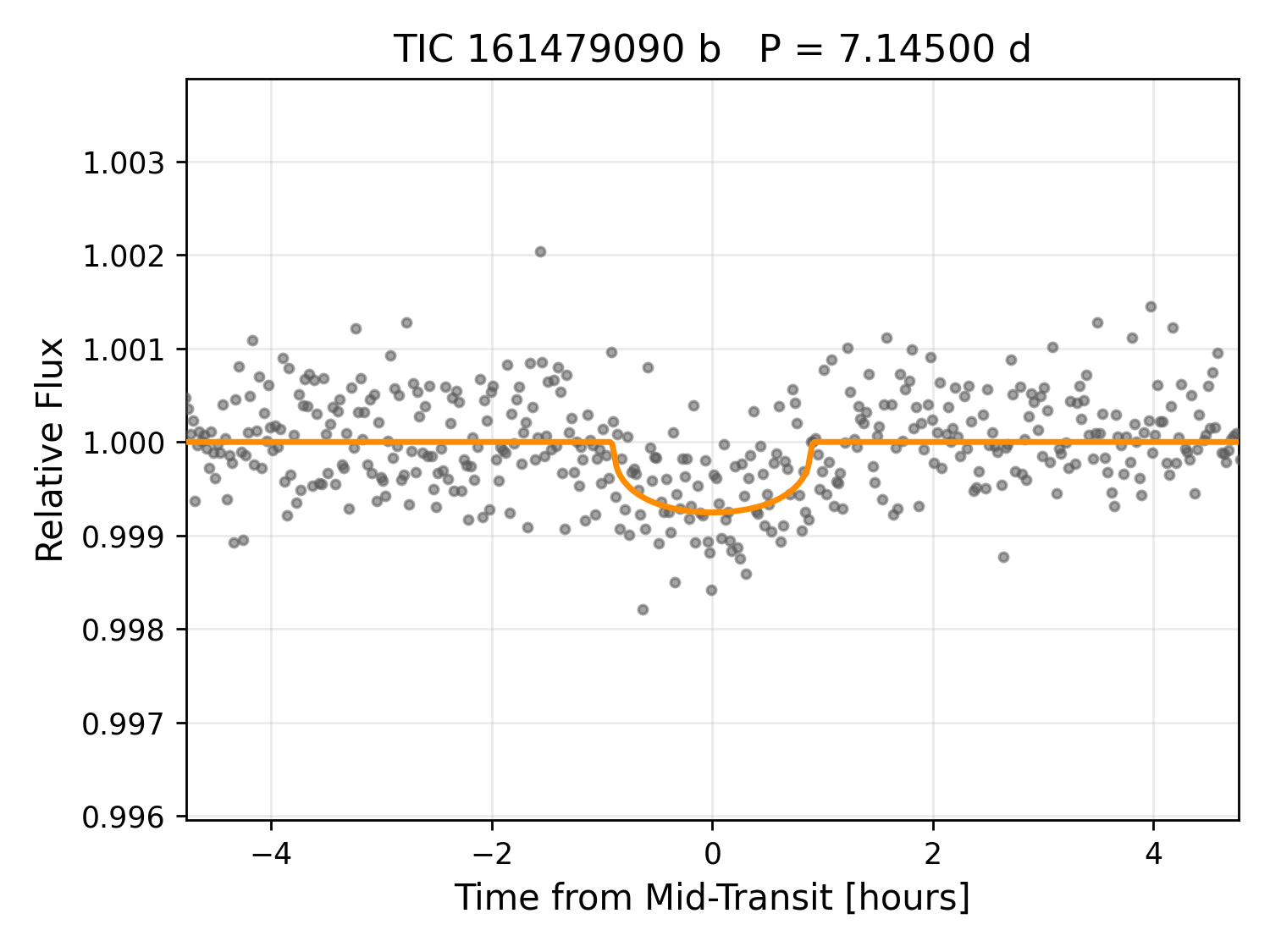}
        } &
        \subfloat[TIC 165297570\label{fov:83}]{
            \includegraphics[width=0.18\linewidth]{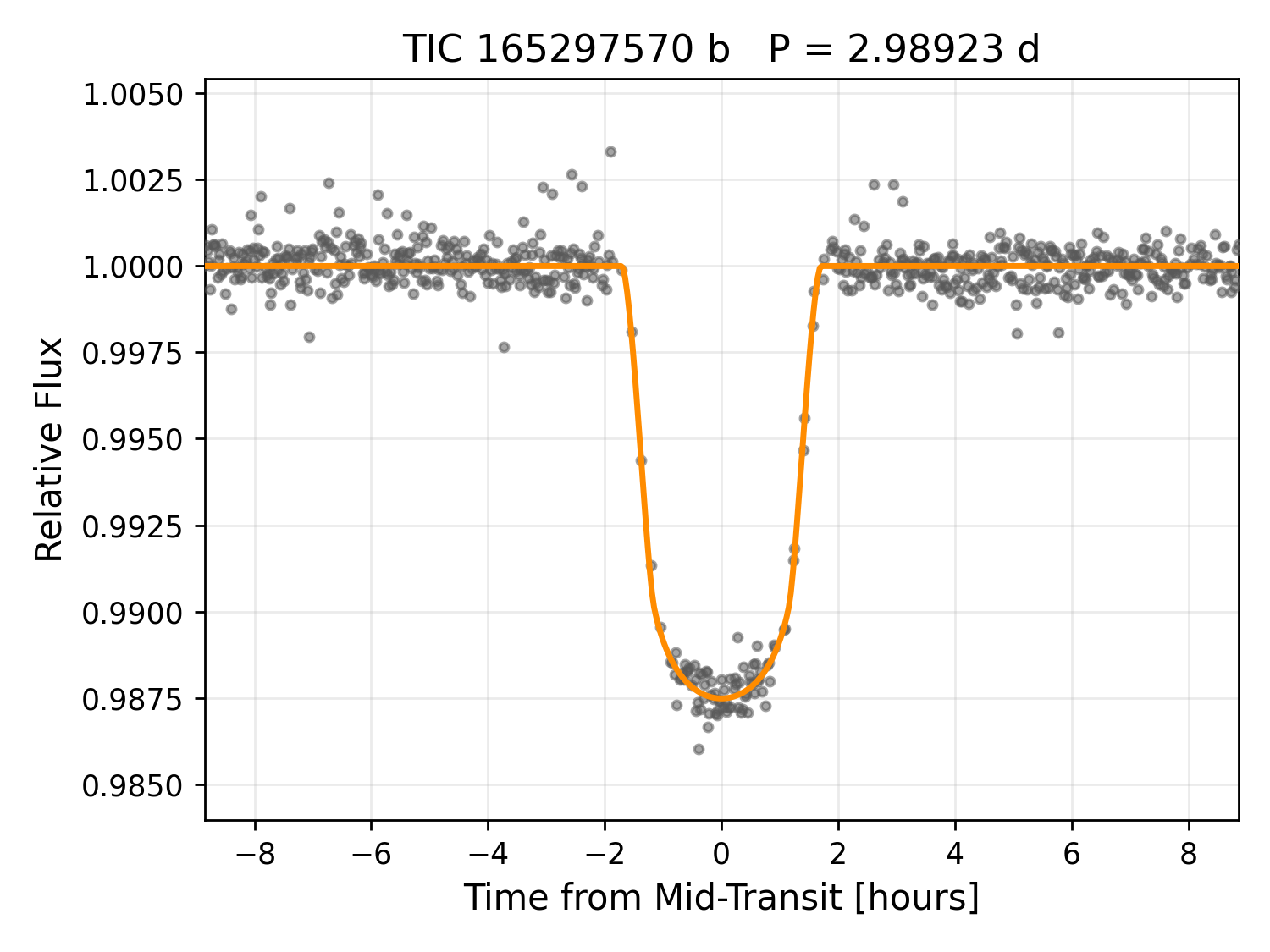}
        } &
        \subfloat[TIC 166527623\label{fov:84}]{
            \includegraphics[width=0.18\linewidth]{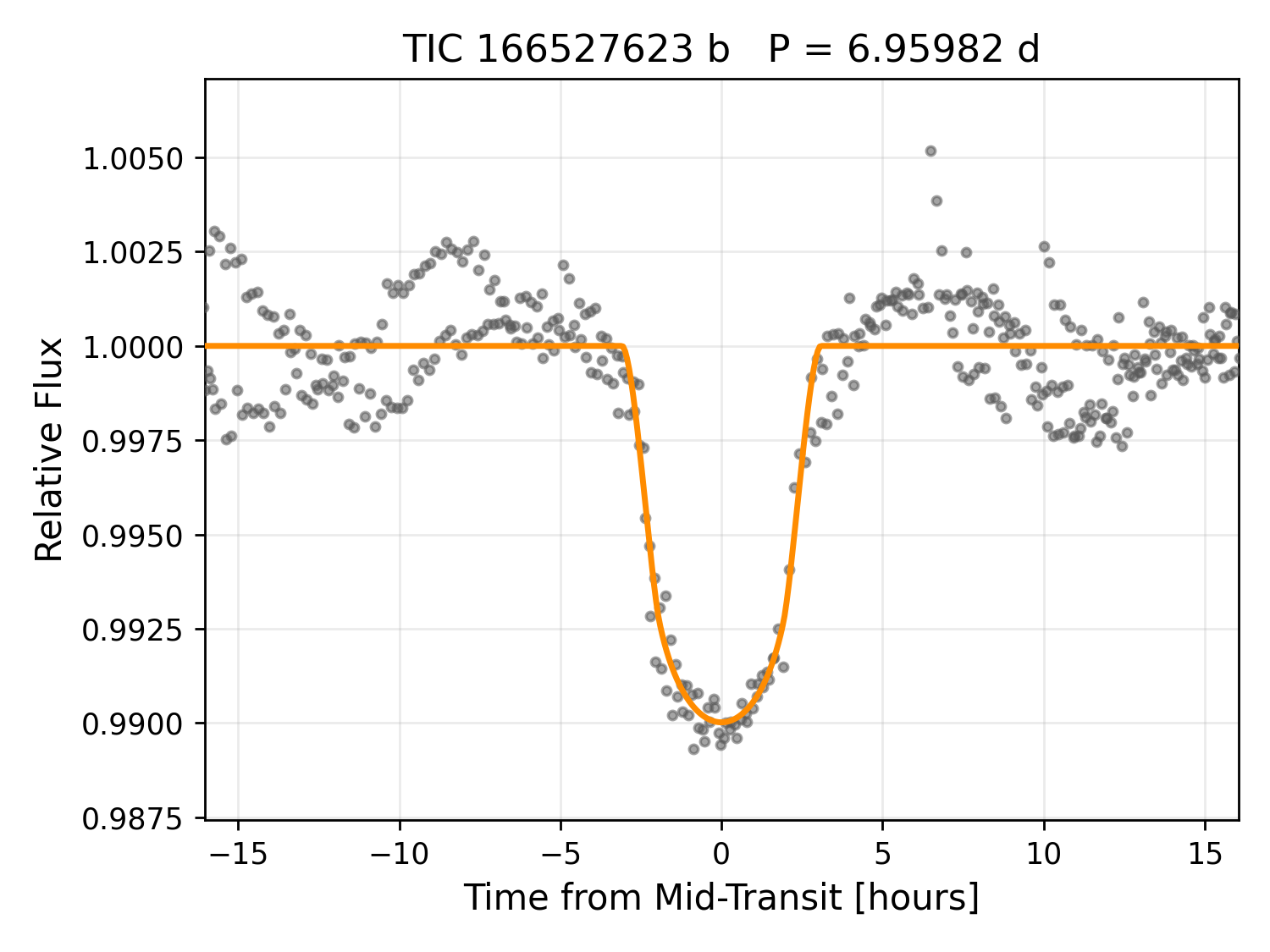}
        } &
        \subfloat[TIC 166836920\label{fov:85}]{
            \includegraphics[width=0.18\linewidth]{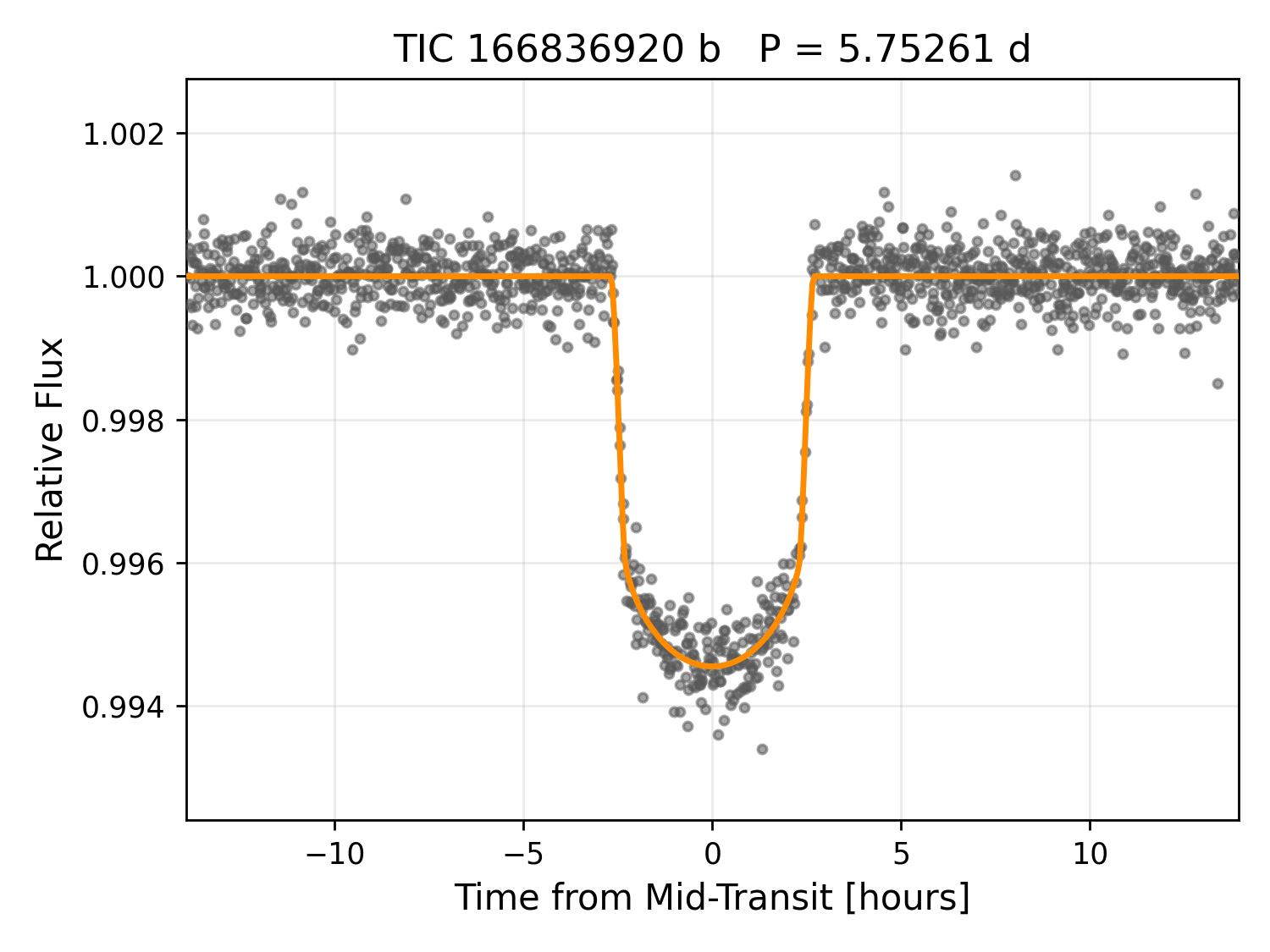}
        }
        \\
\end{tabular}
    \vspace{2pt}
    \caption{Selected Transits}
    \label{fig:appendix_transit}
\end{figure}

\begin{figure}[h]
    \ContinuedFloat
    \centering
    \setlength{\tabcolsep}{1.0pt} 
    \renewcommand{\arraystretch}{1.5}
    \begin{tabular}{ccccc} 
        \subfloat[TIC 240844466\label{fov:115}]{
            \includegraphics[width=0.18\linewidth]{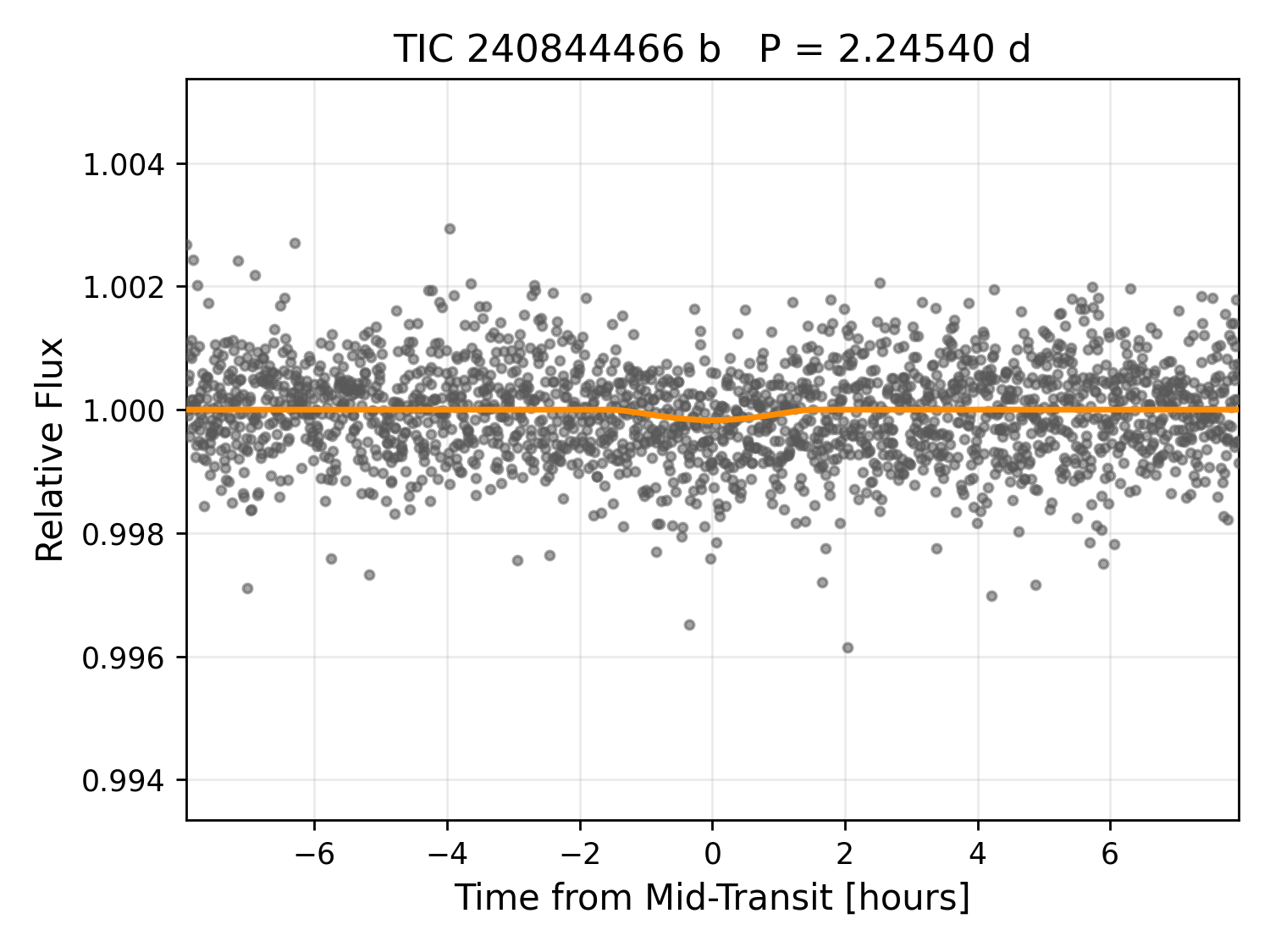}
        } &
        \subfloat[TIC 244089109\label{fov:116}]{
            \includegraphics[width=0.18\linewidth]{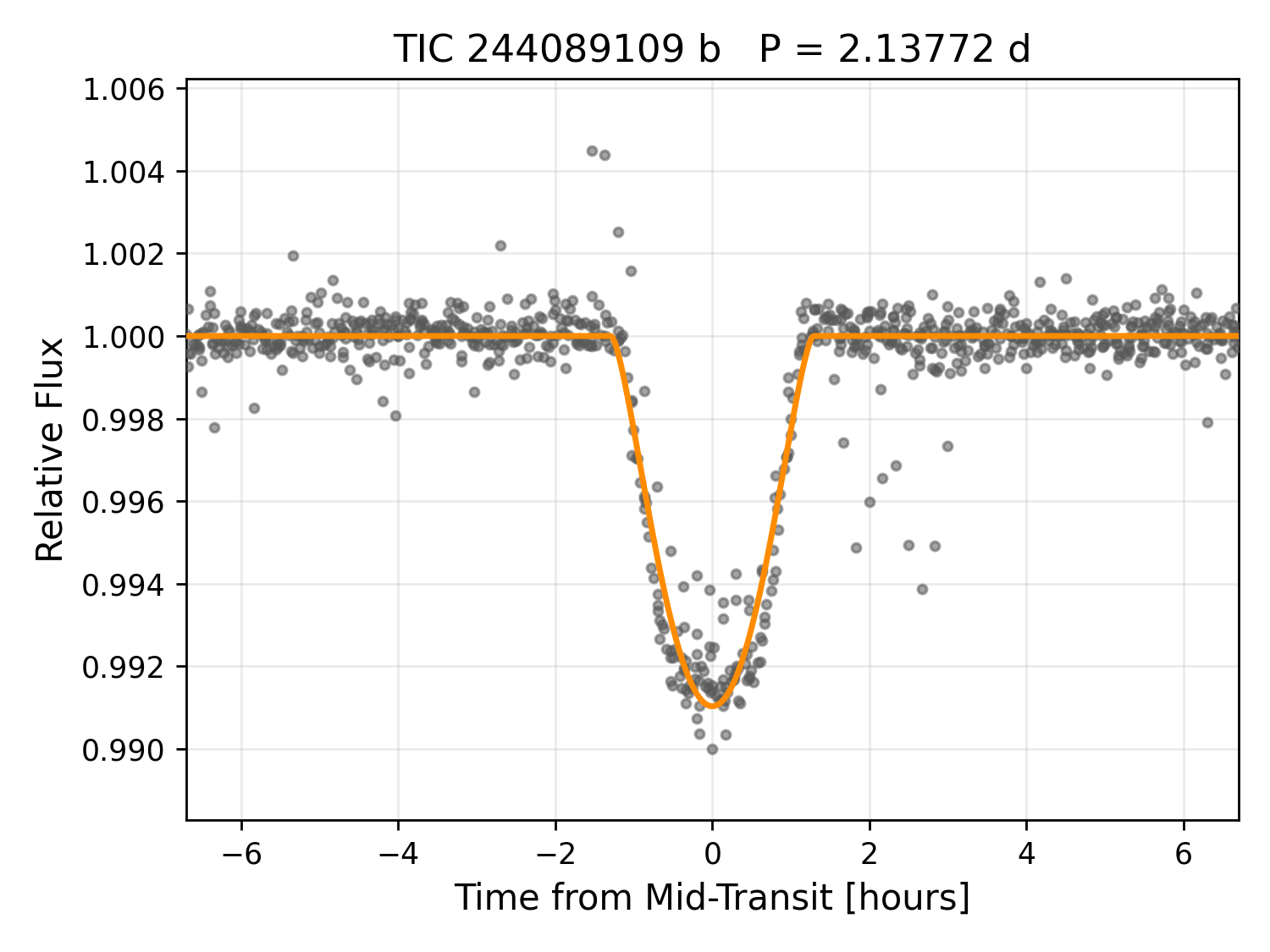}
        } &
        \subfloat[TIC 250519787\label{fov:117}]{
            \includegraphics[width=0.18\linewidth]{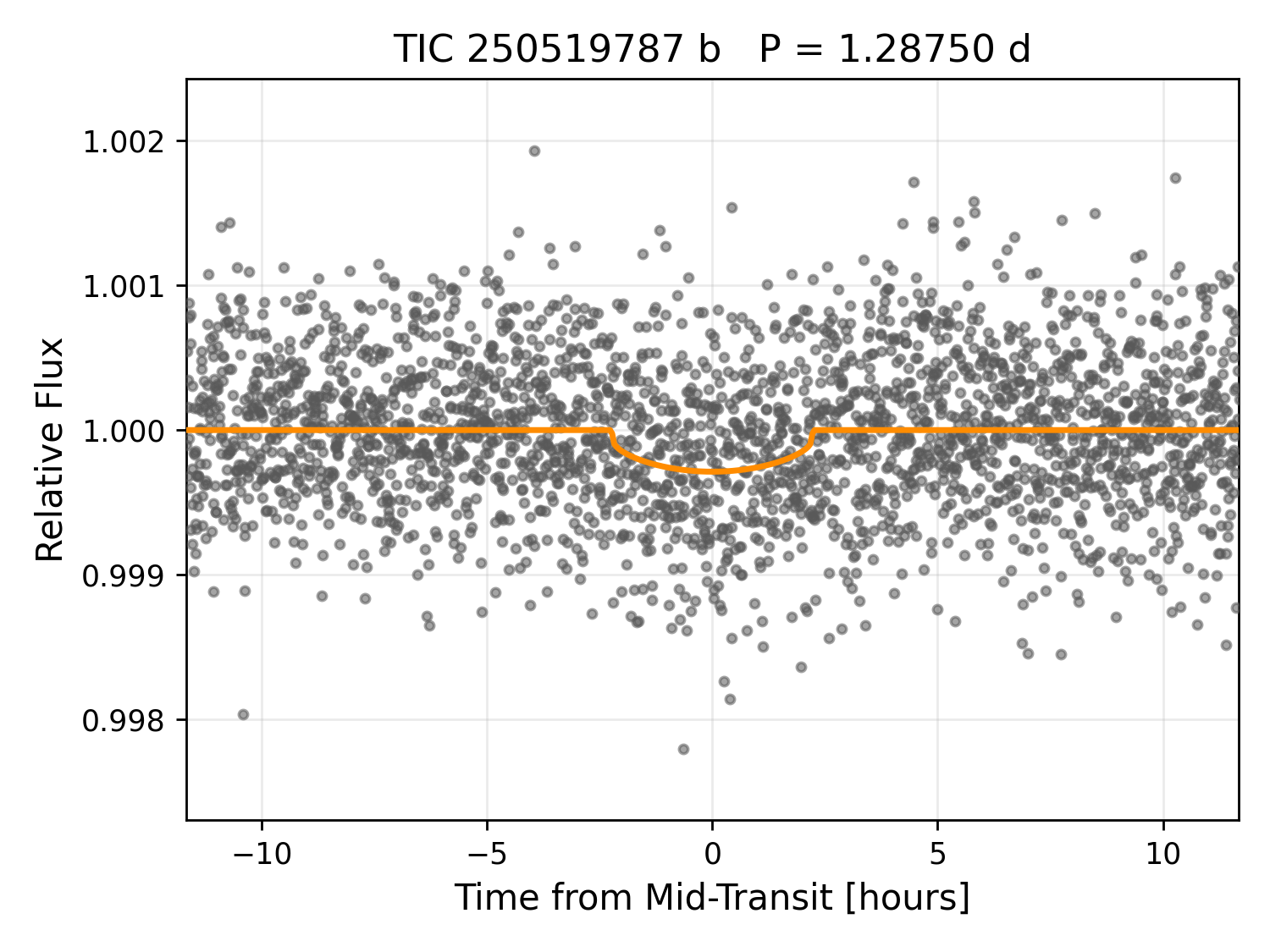}
        } &
        \subfloat[TIC 252616865\label{fov:118}]{
            \includegraphics[width=0.18\linewidth]{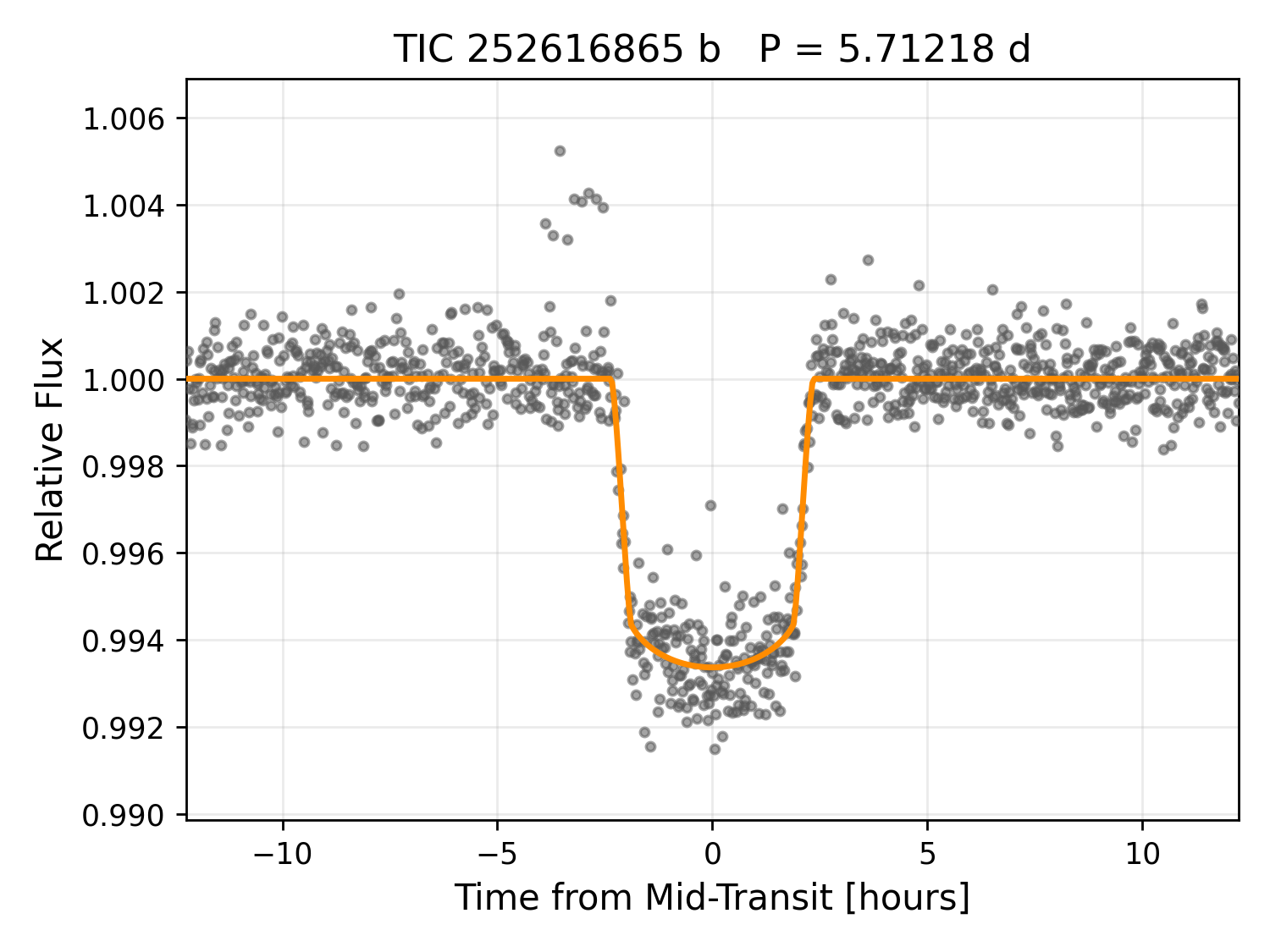}
        } &
        \subfloat[TIC 260757945\label{fov:119}]{
            \includegraphics[width=0.18\linewidth]{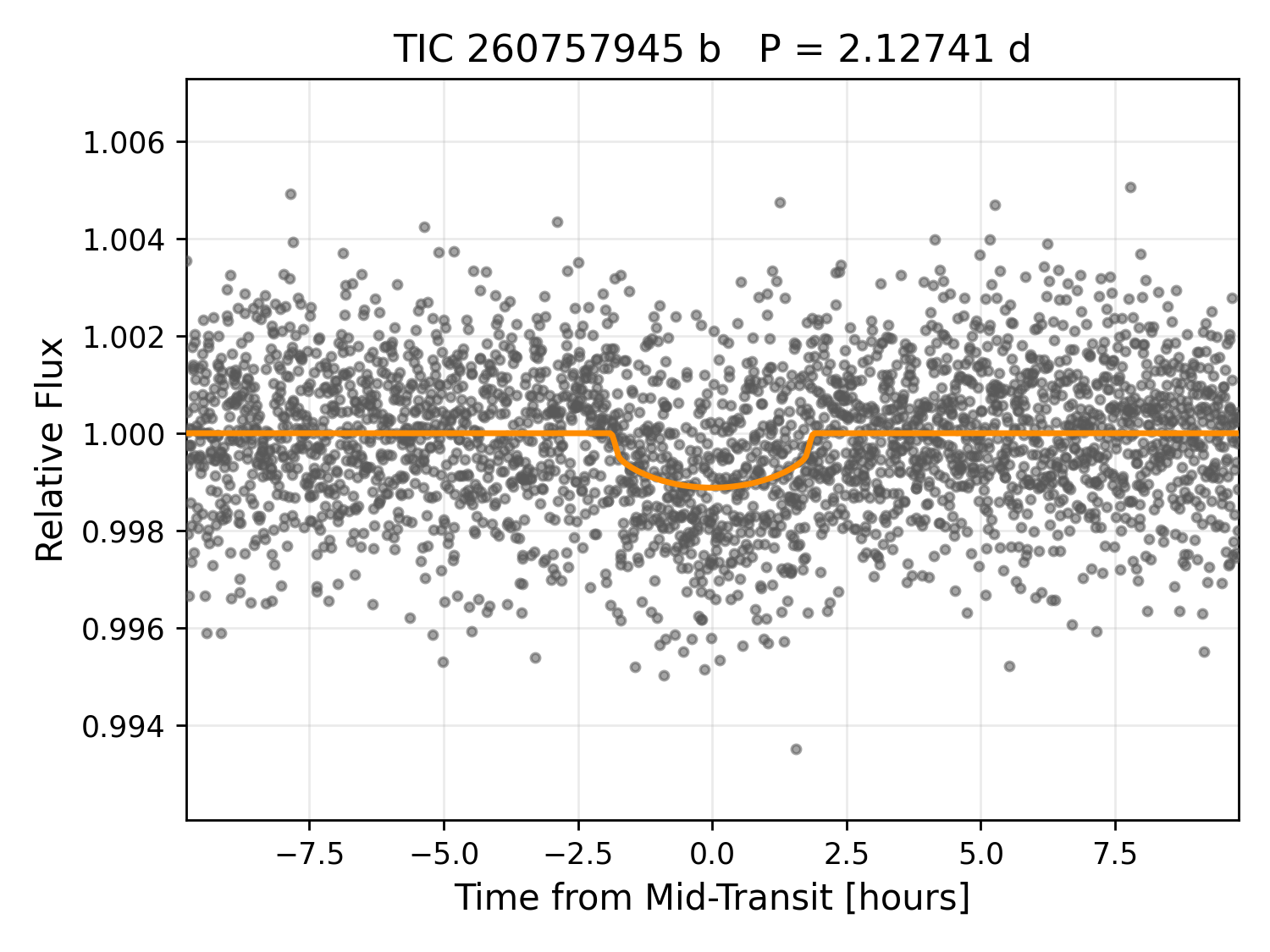}
        }
        \\

        \subfloat[TIC 262746281\label{fov:120}]{
            \includegraphics[width=0.18\linewidth]{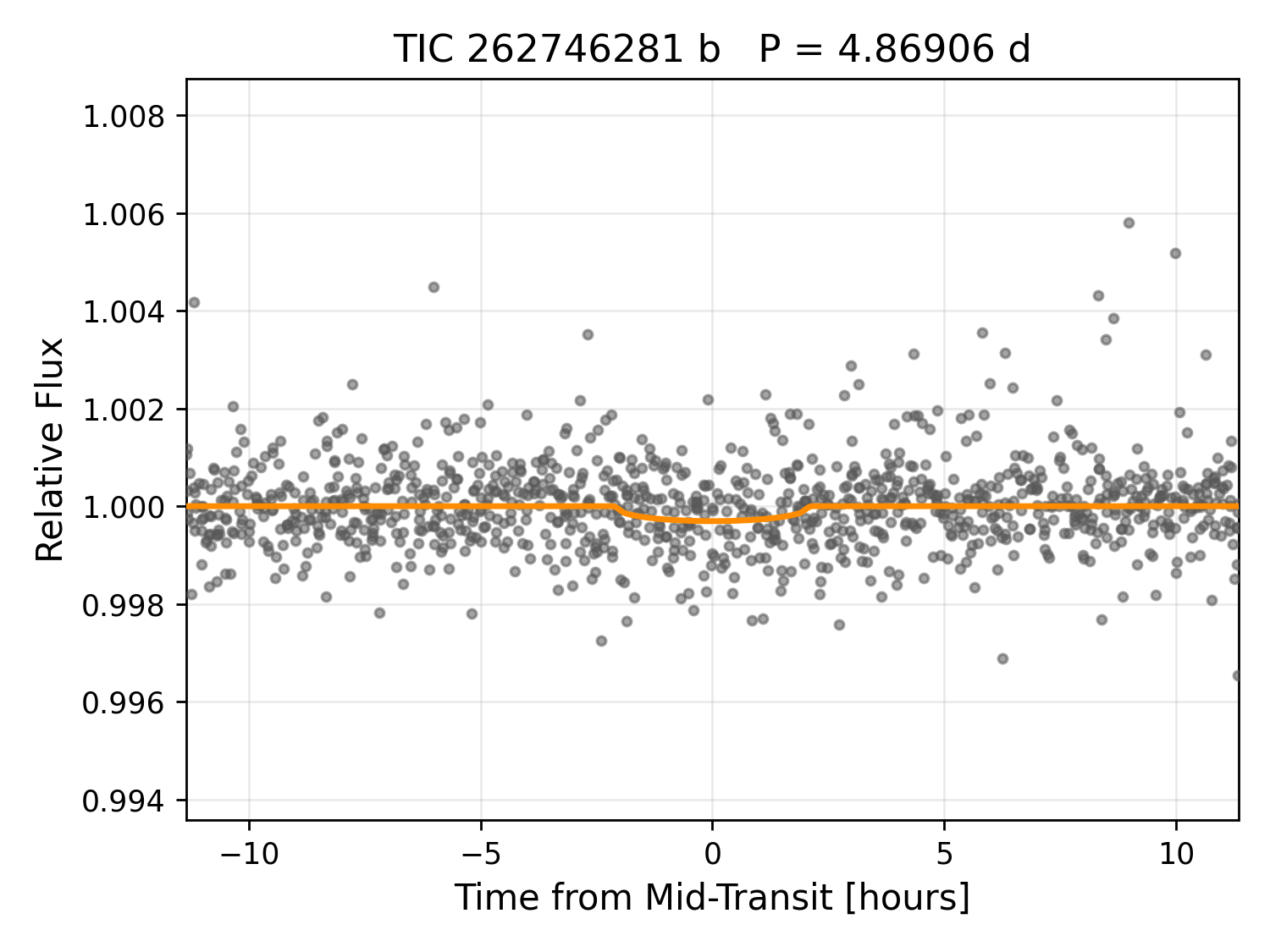}
        } &
        \subfloat[TIC 264831542\label{fov:121}]{
            \includegraphics[width=0.18\linewidth]{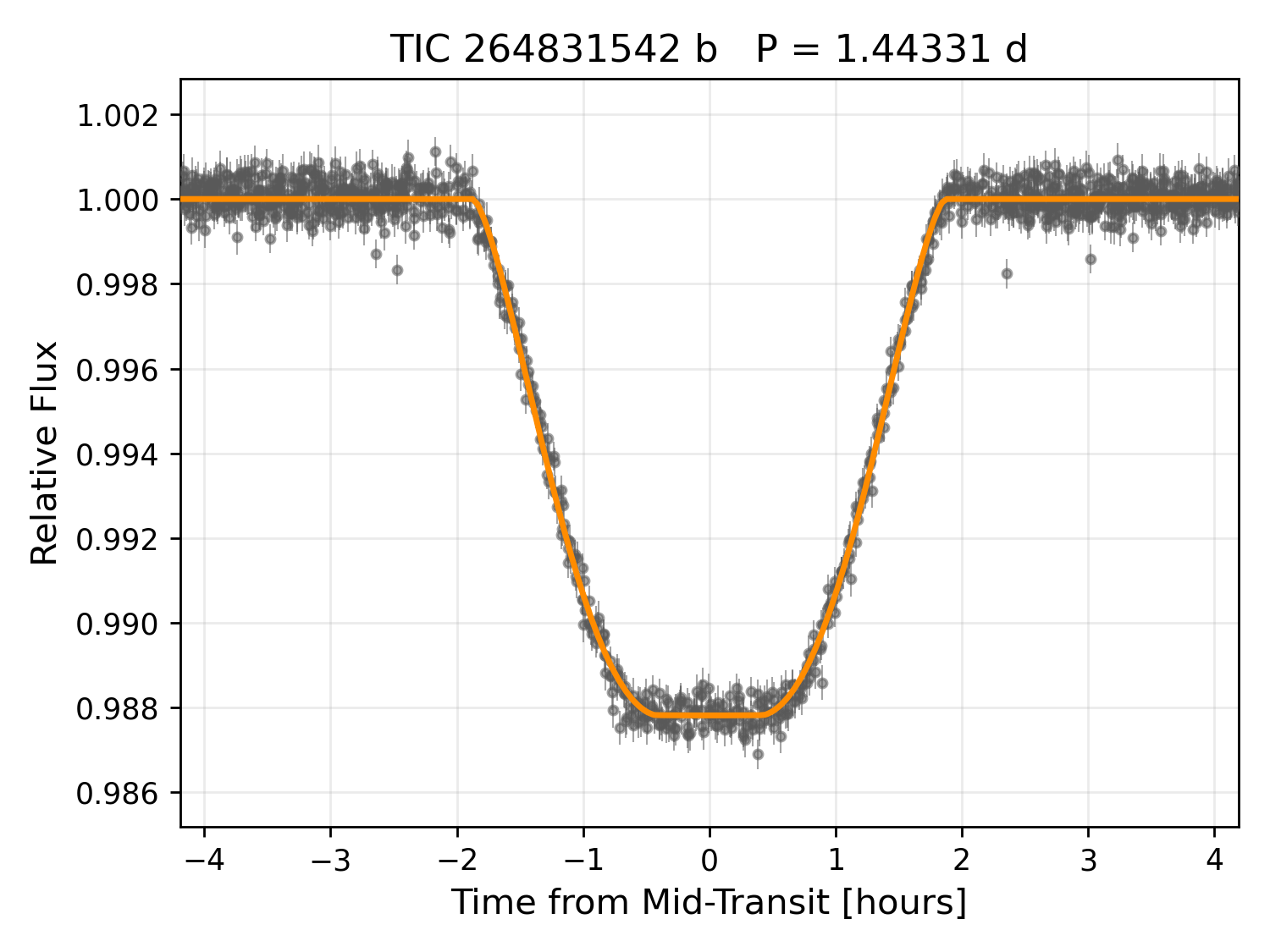}
        } &
        \subfloat[TIC 266193331\label{fov:122}]{
            \includegraphics[width=0.18\linewidth]{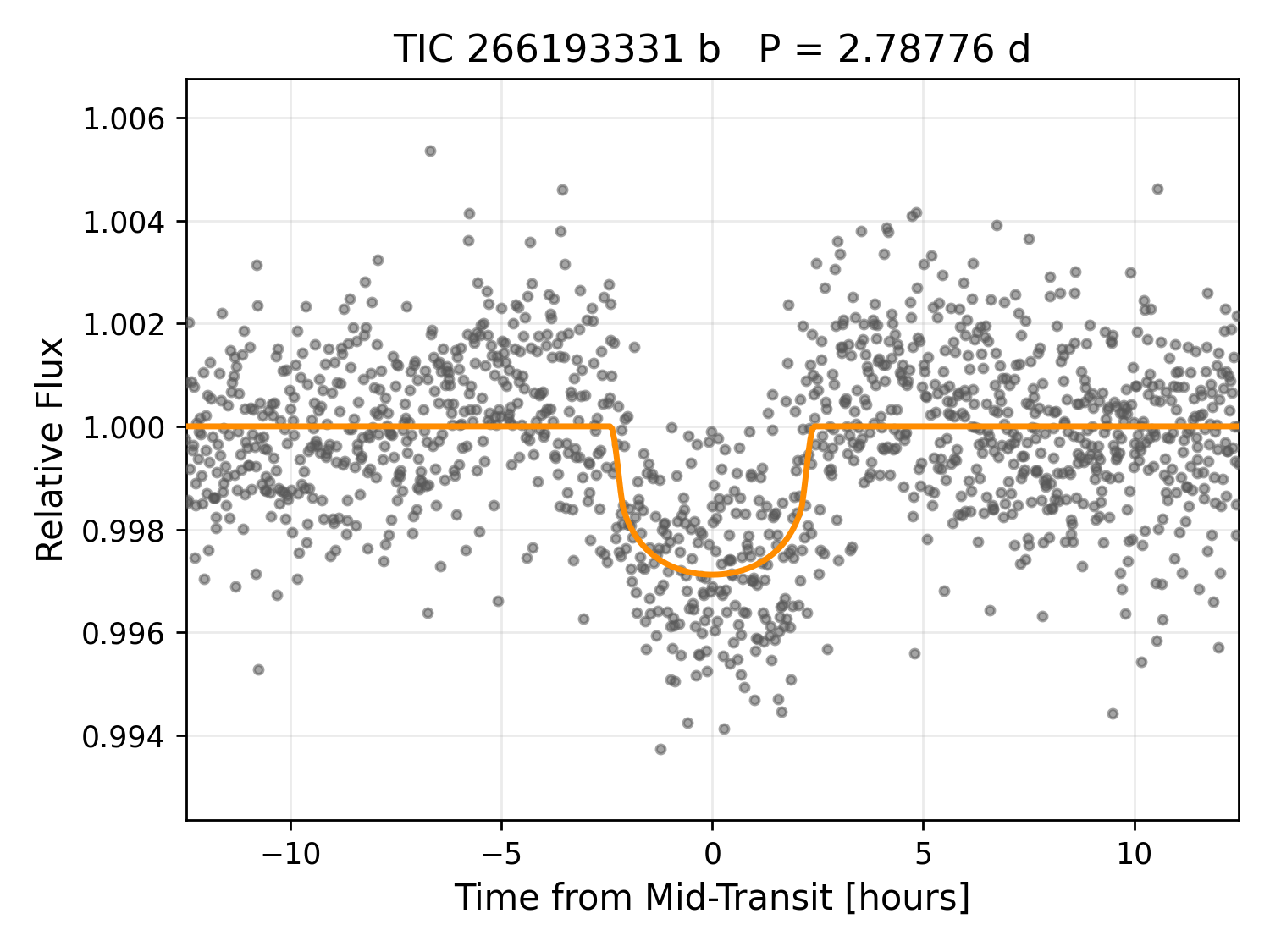}
        } &
        \subfloat[TIC 268403451\label{fov:123}]{
            \includegraphics[width=0.18\linewidth]{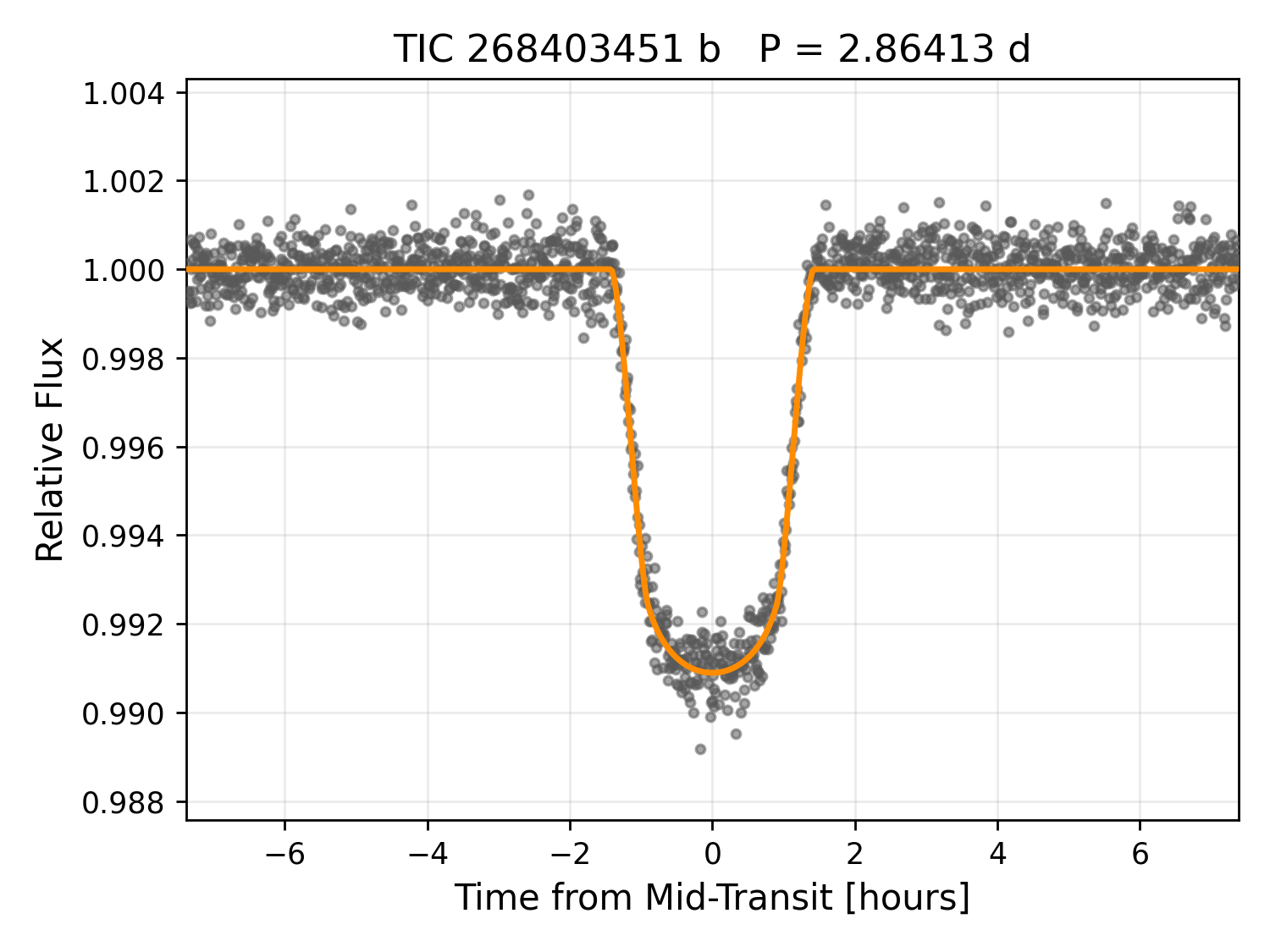}
        } &
        \subfloat[TIC 268511656\label{fov:124}]{
            \includegraphics[width=0.18\linewidth]{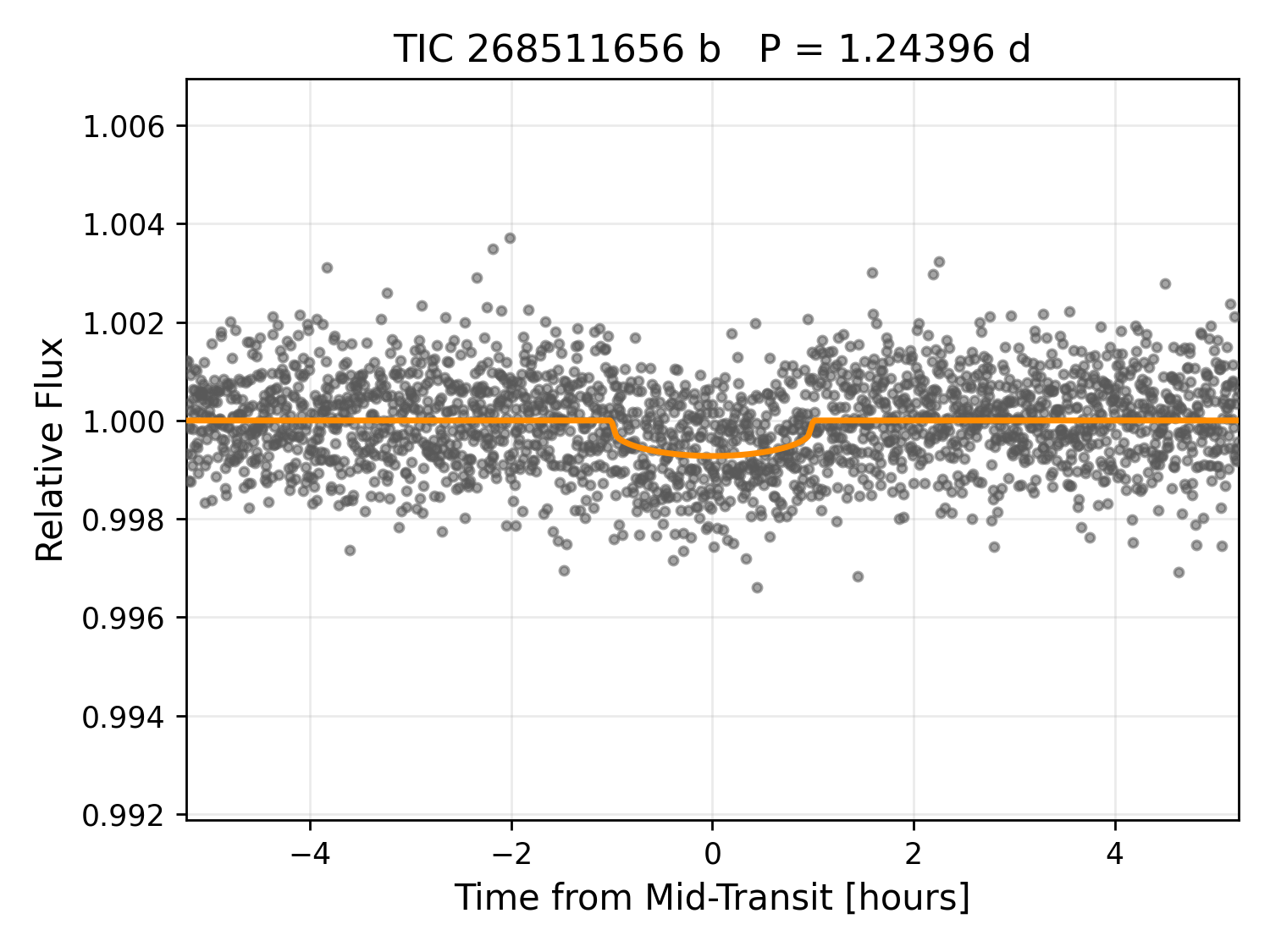}
        }
        \\
\end{tabular}
    \vspace{2pt}
    \caption{Selected Transits}
    \label{fig:appendix_transit}
\end{figure}

\begin{figure}[h]
    \ContinuedFloat
    \centering
    \setlength{\tabcolsep}{1.0pt} 
    \renewcommand{\arraystretch}{1.5}
    \begin{tabular}{ccccc} 
        \subfloat[TIC 347013211\label{fov:155}]{
            \includegraphics[width=0.18\linewidth]{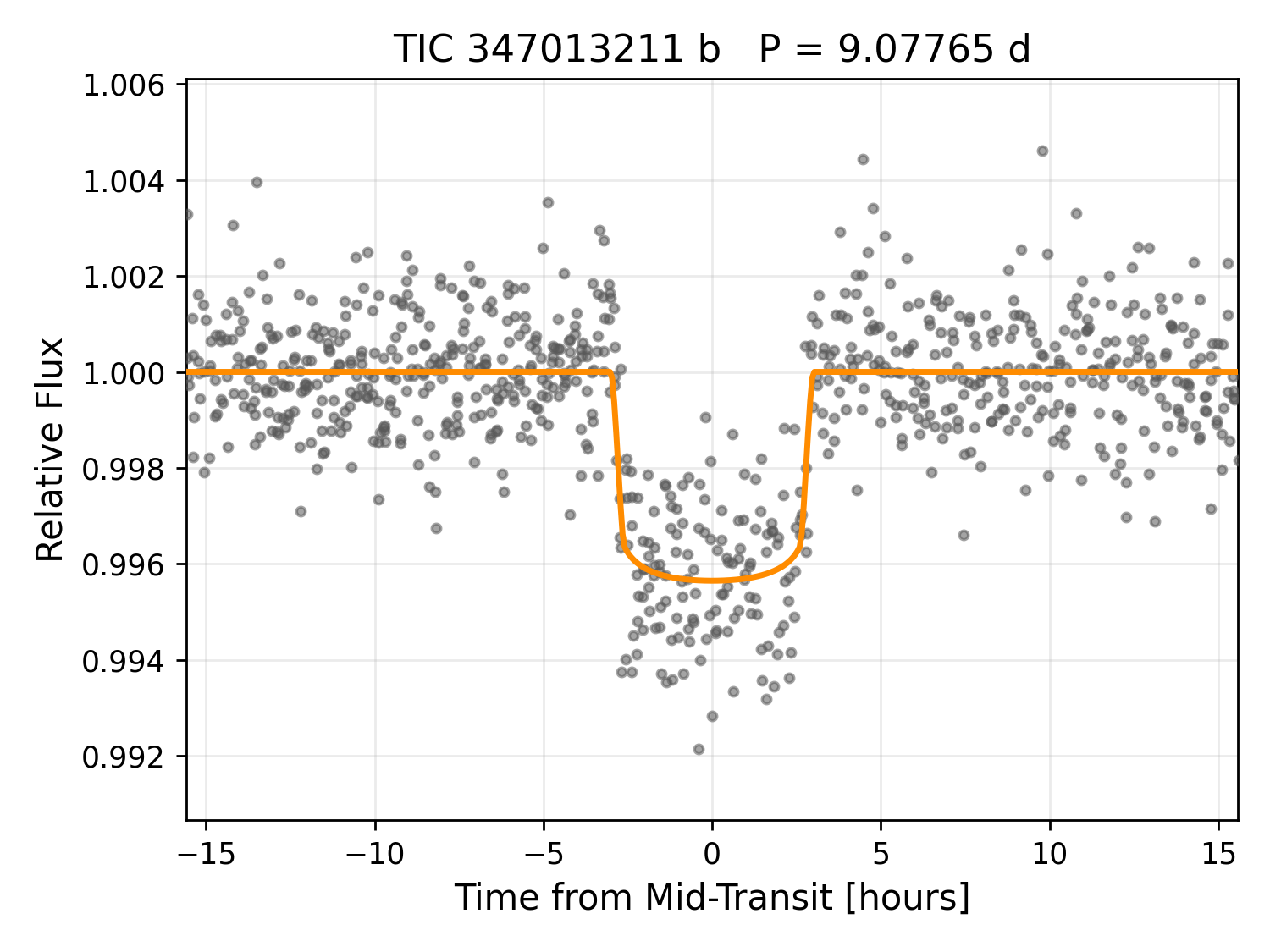}
        } &
        \subfloat[TIC 349827430\label{fov:156}]{
            \includegraphics[width=0.18\linewidth]{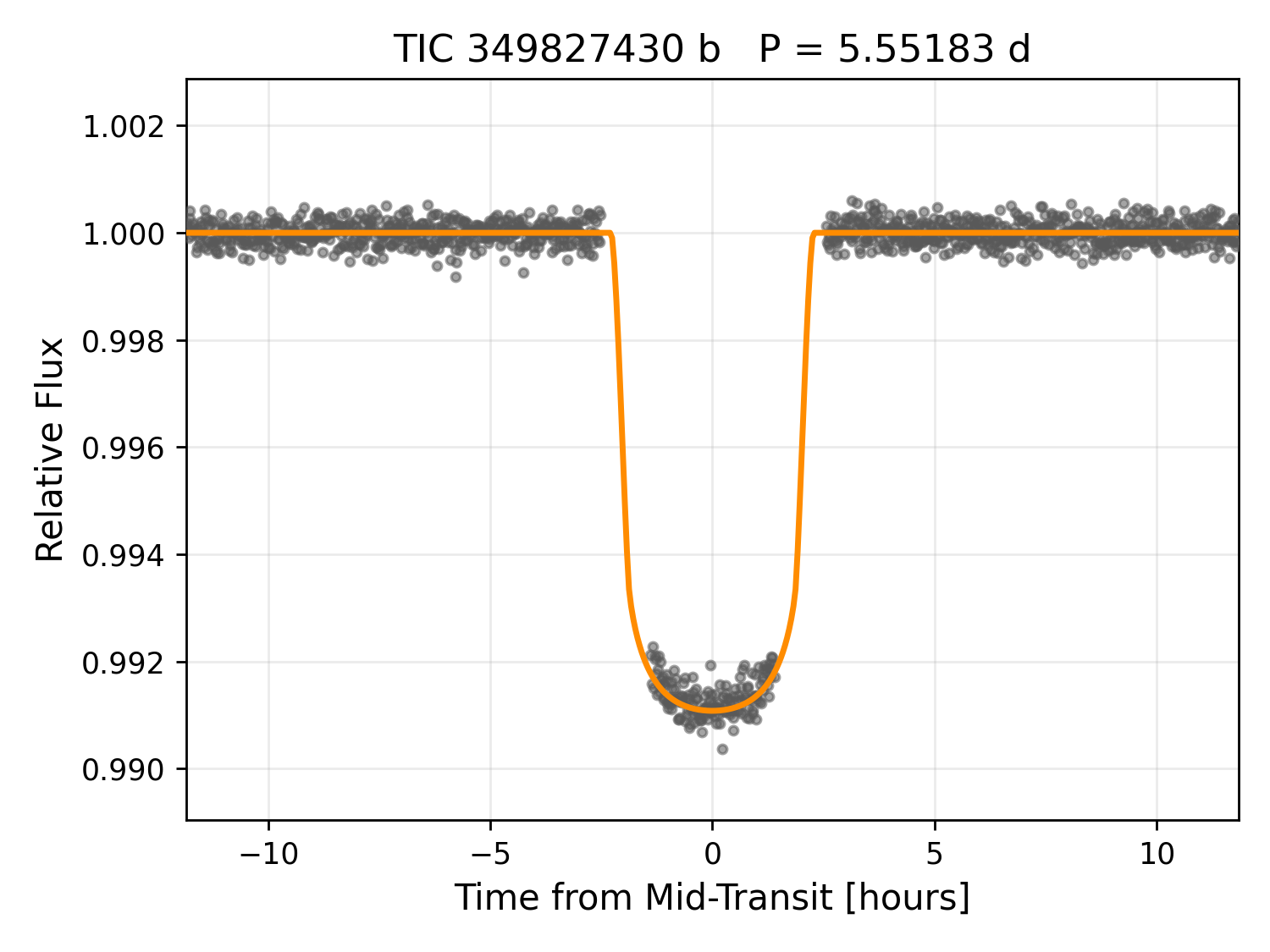}
        } &
        \subfloat[TIC 350020859\label{fov:157}]{
            \includegraphics[width=0.18\linewidth]{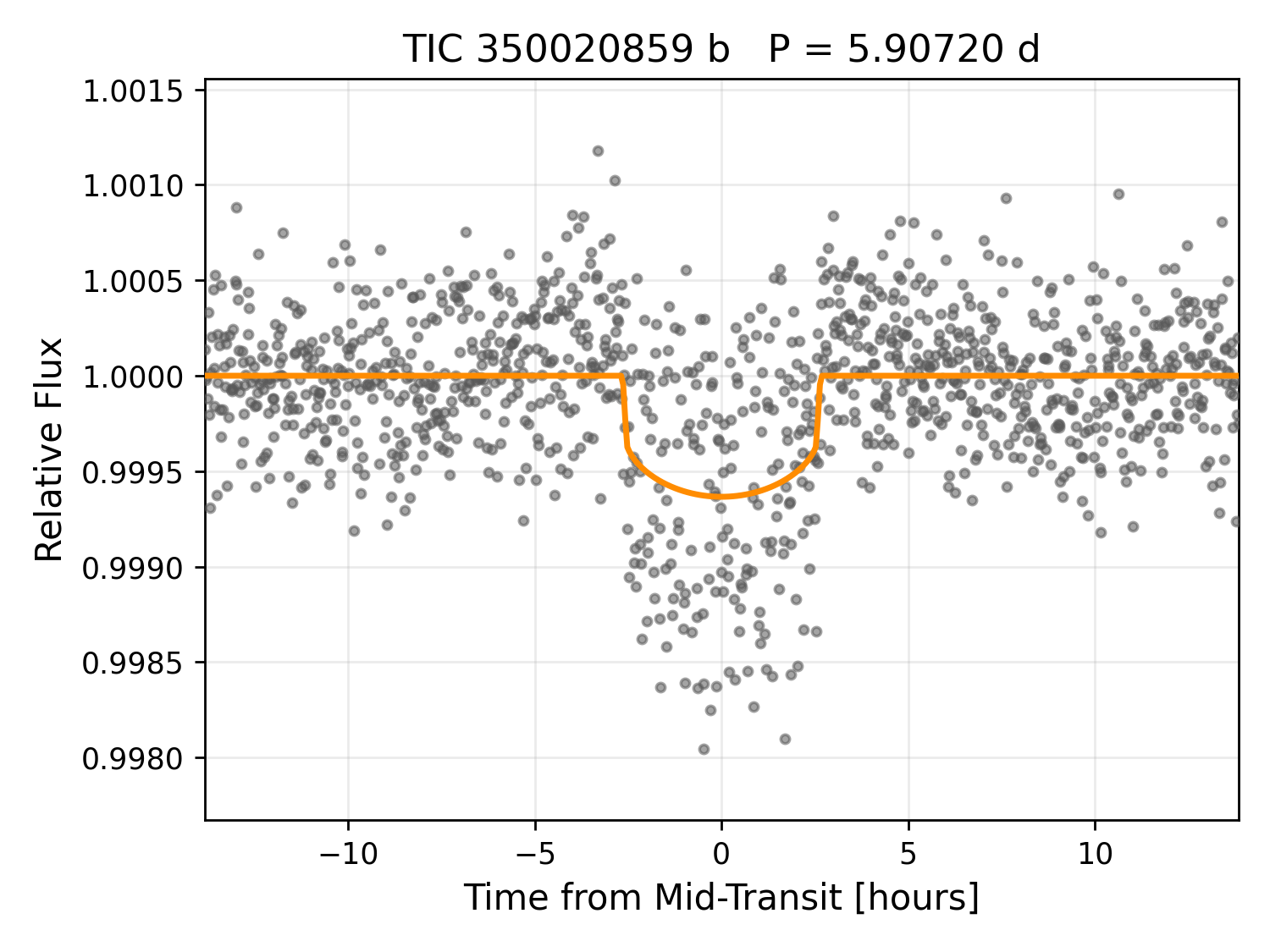}
        } &
        \subfloat[TIC 352626125\label{fov:158}]{
            \includegraphics[width=0.18\linewidth]{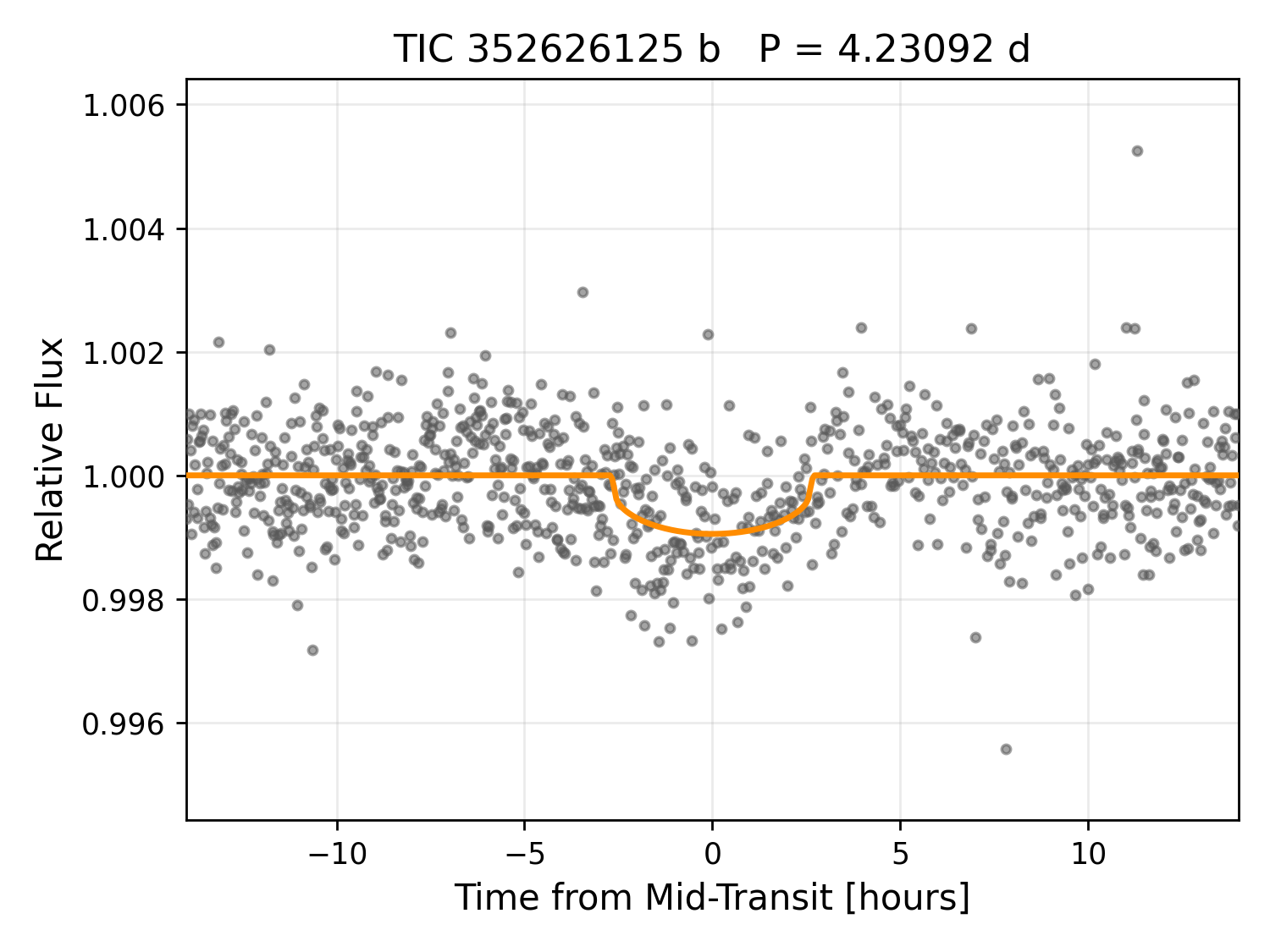}
        } &
        \subfloat[TIC 354031707\label{fov:159}]{
            \includegraphics[width=0.18\linewidth]{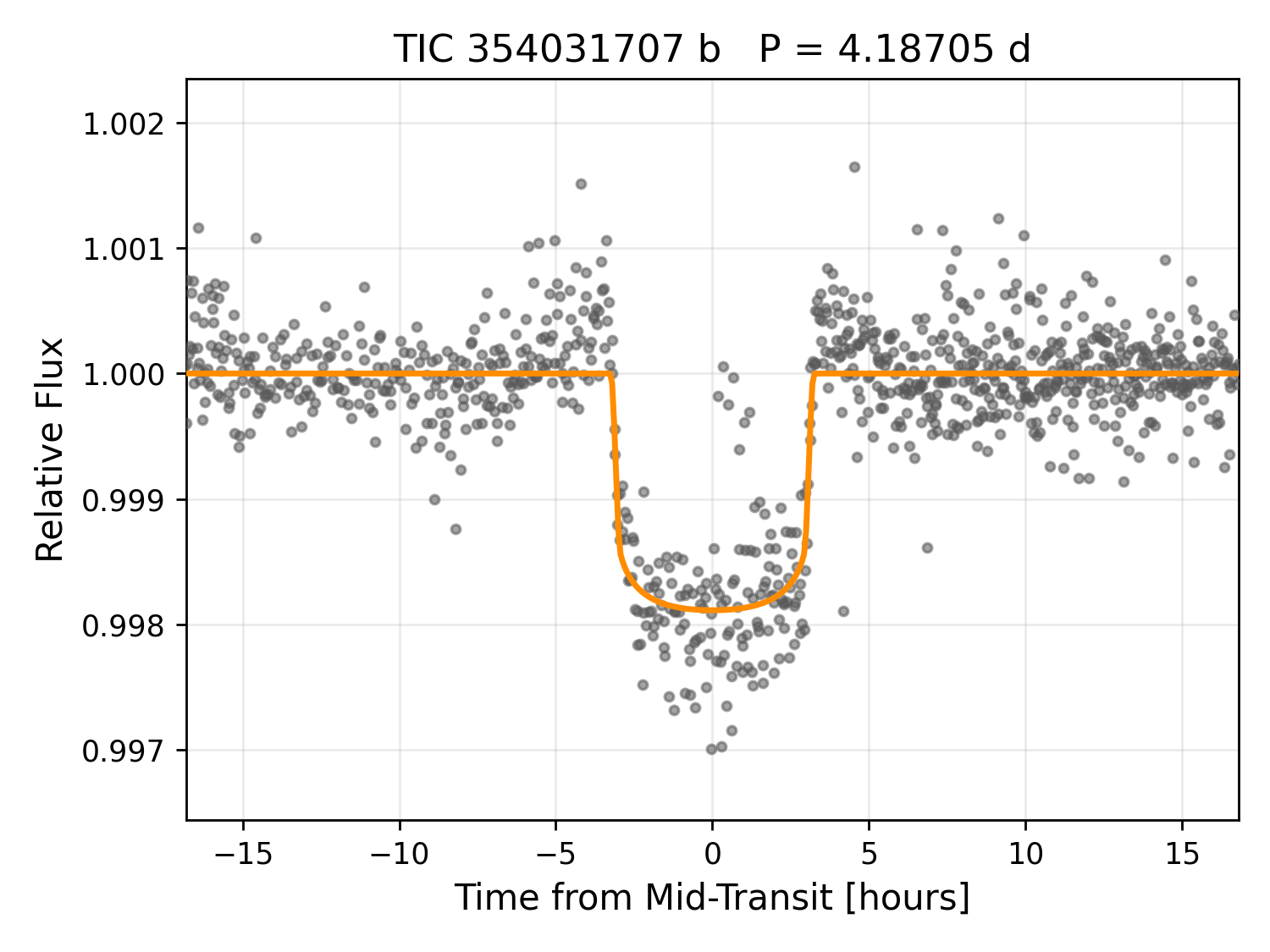}
        }
        \\

        \subfloat[TIC 354469661\label{fov:160}]{
            \includegraphics[width=0.18\linewidth]{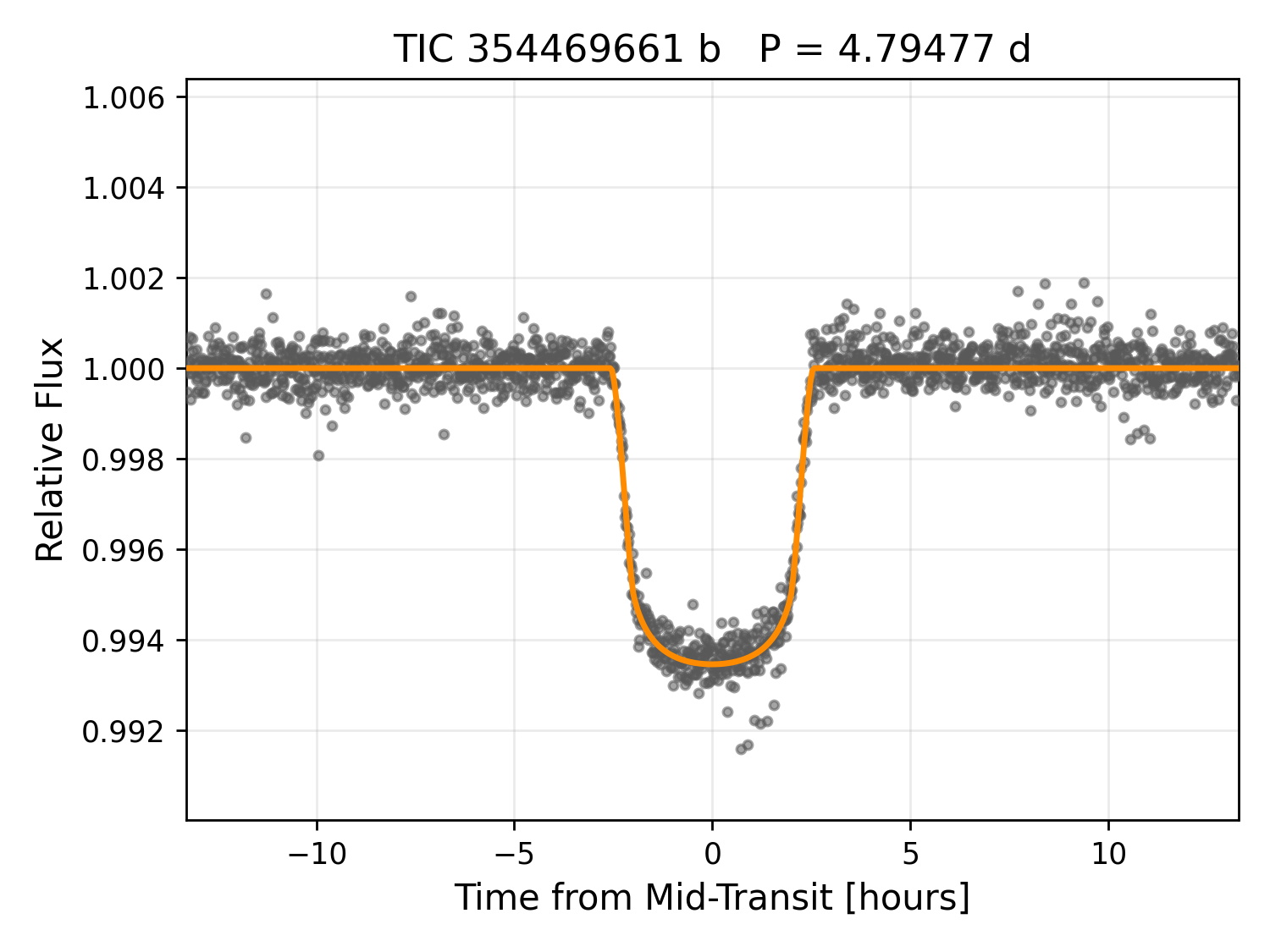}
        } &
        \subfloat[TIC 354660816\label{fov:161}]{
            \includegraphics[width=0.18\linewidth]{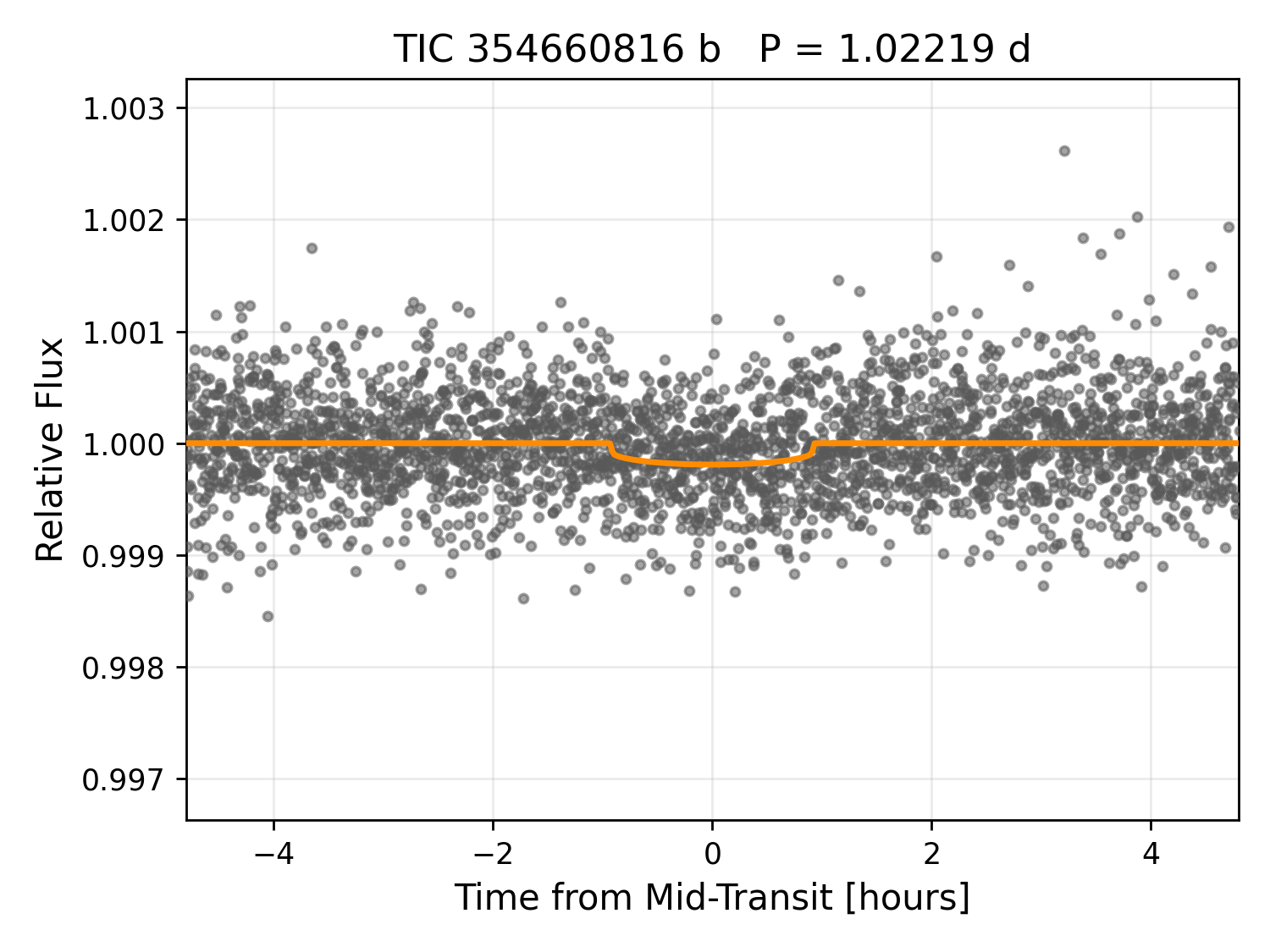}
        } &
        \subfloat[TIC 355374037\label{fov:162}]{
            \includegraphics[width=0.18\linewidth]{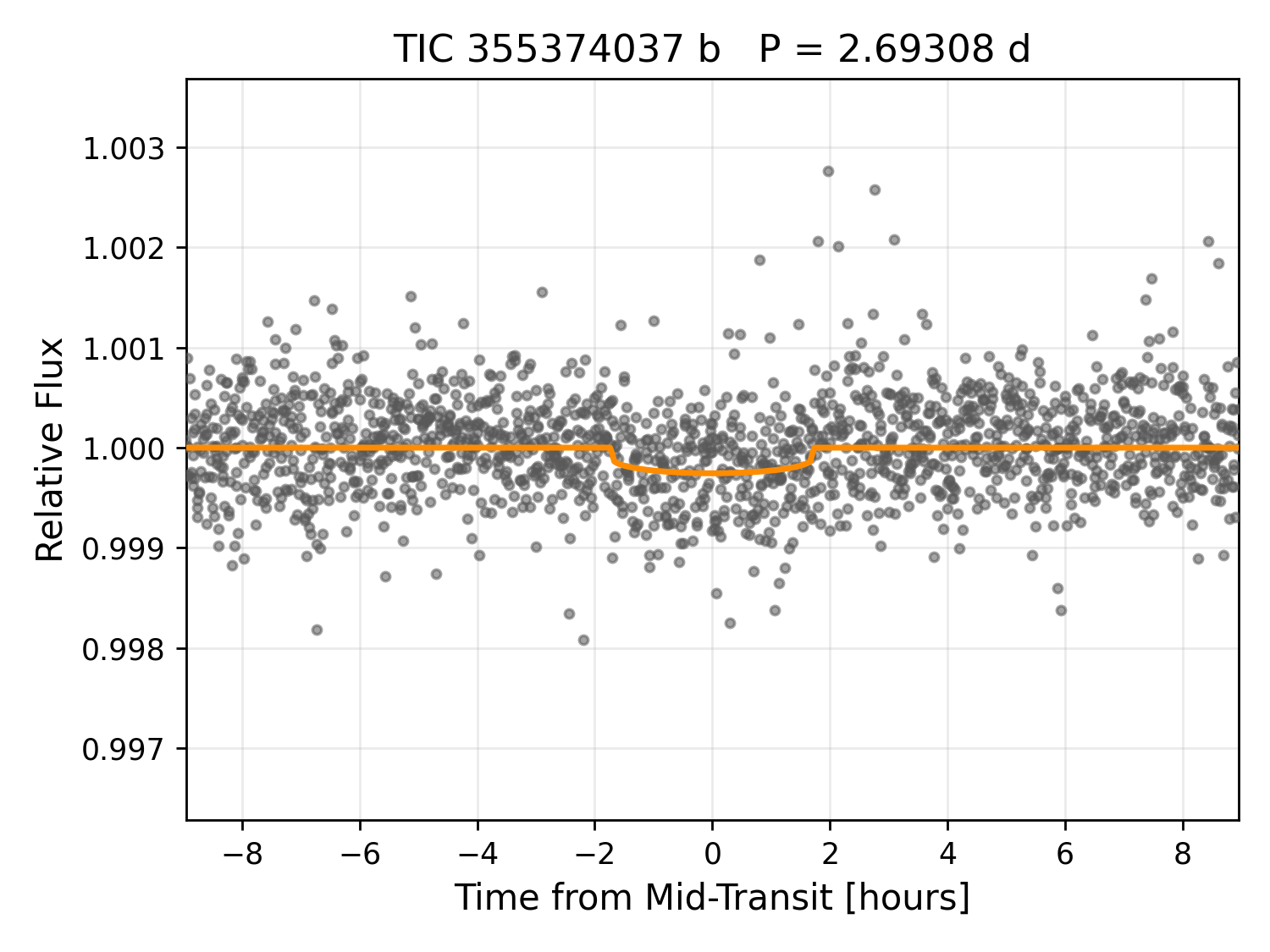}
        } &
        \subfloat[TIC 355408846\label{fov:163}]{
            \includegraphics[width=0.18\linewidth]{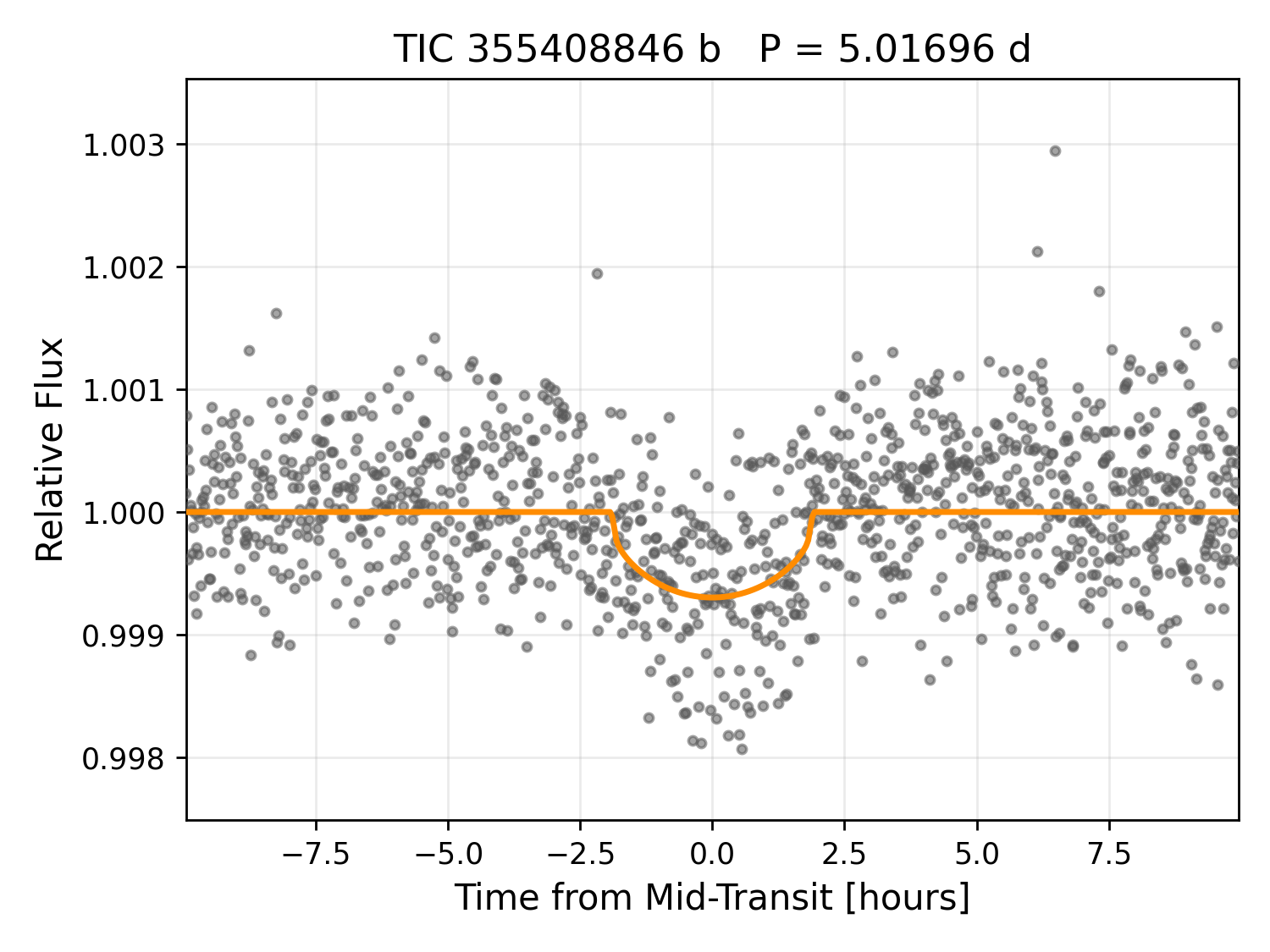}
        } &
        \subfloat[TIC 355841438\label{fov:164}]{
            \includegraphics[width=0.18\linewidth]{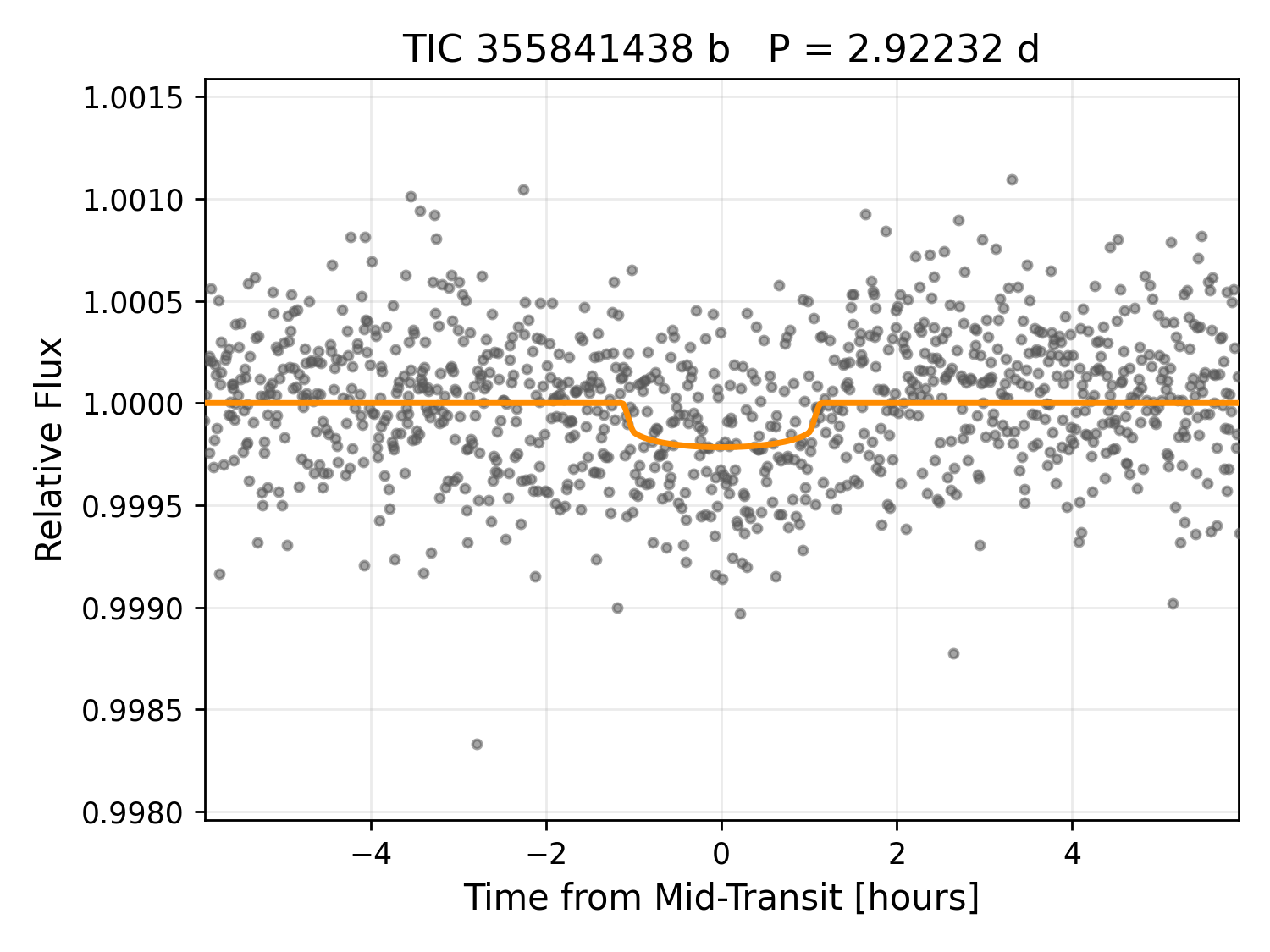}
        }
        \\
\end{tabular}
    \vspace{2pt}
    \caption{Selected Transits}
    \label{fig:appendix_transit}
\end{figure}

\begin{figure}[h]
    \ContinuedFloat
    \centering
    \setlength{\tabcolsep}{1.0pt} 
    \renewcommand{\arraystretch}{1.5}
    \begin{tabular}{ccccc} 
        \subfloat[TIC 409533090\label{fov:194}]{
            \includegraphics[width=0.18\linewidth]{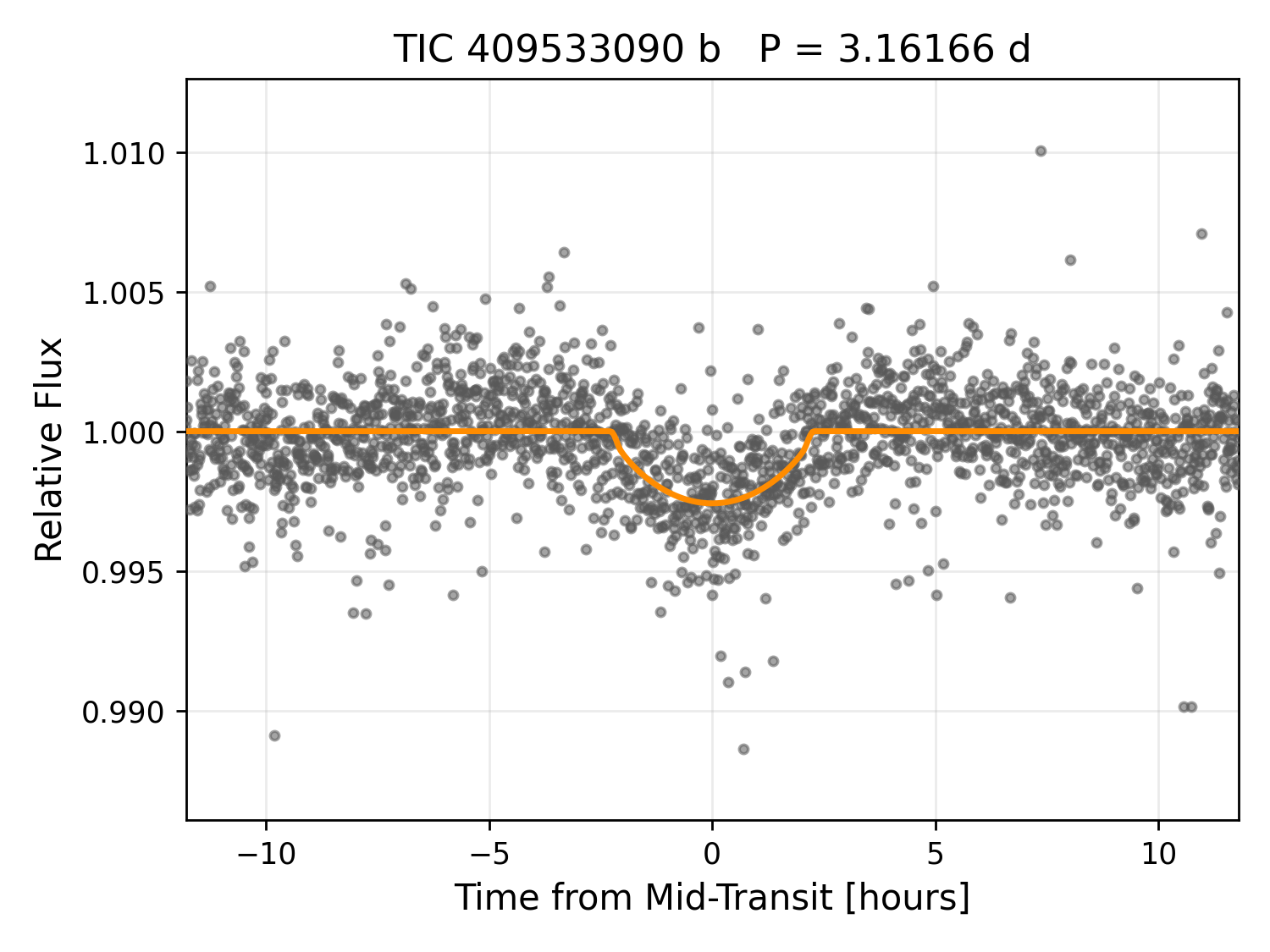}
        } &
        \subfloat[TIC 409594381\label{fov:195}]{
            \includegraphics[width=0.18\linewidth]{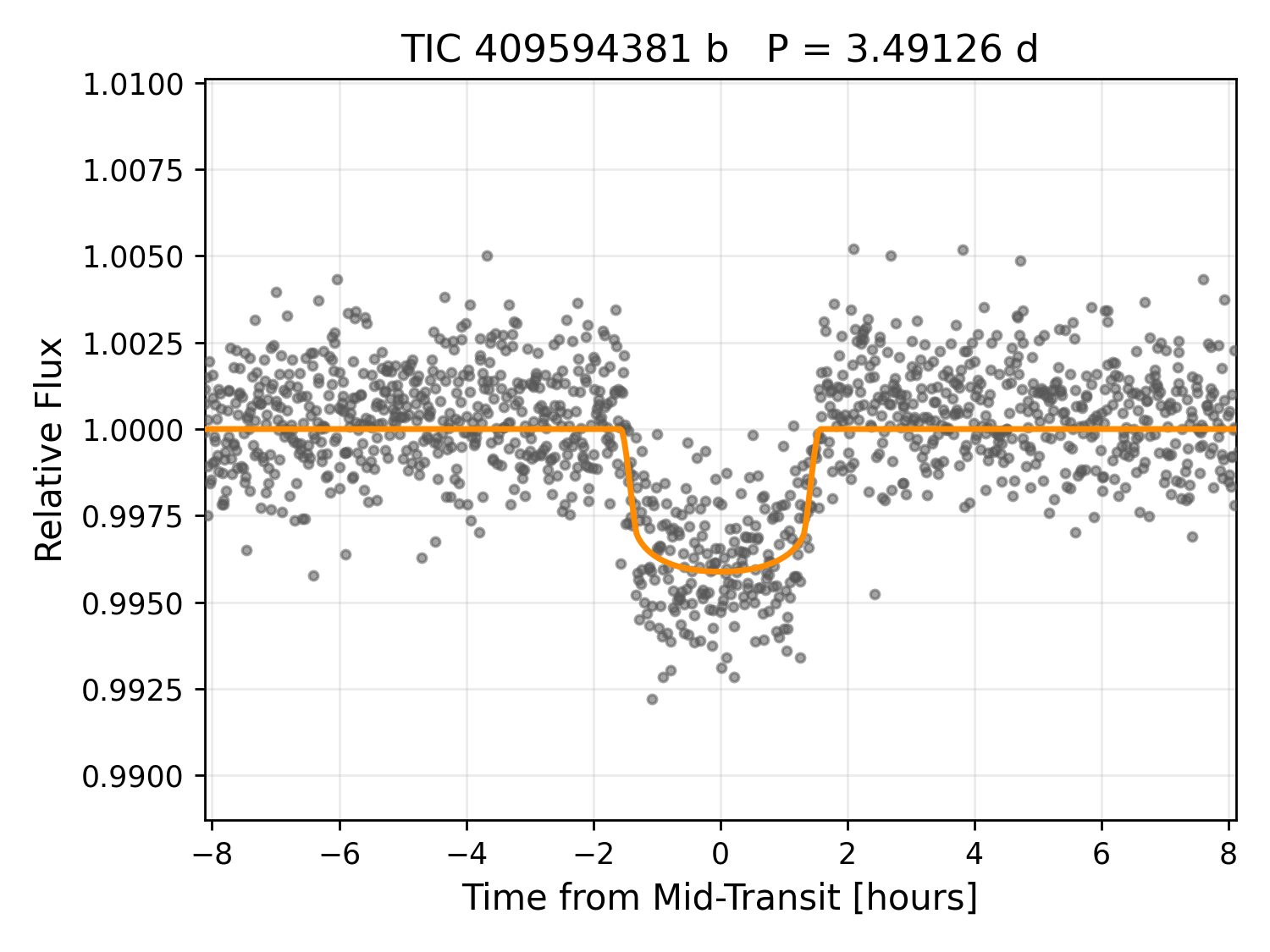}
        } &
        \subfloat[TIC 410317773\label{fov:196}]{
            \includegraphics[width=0.18\linewidth]{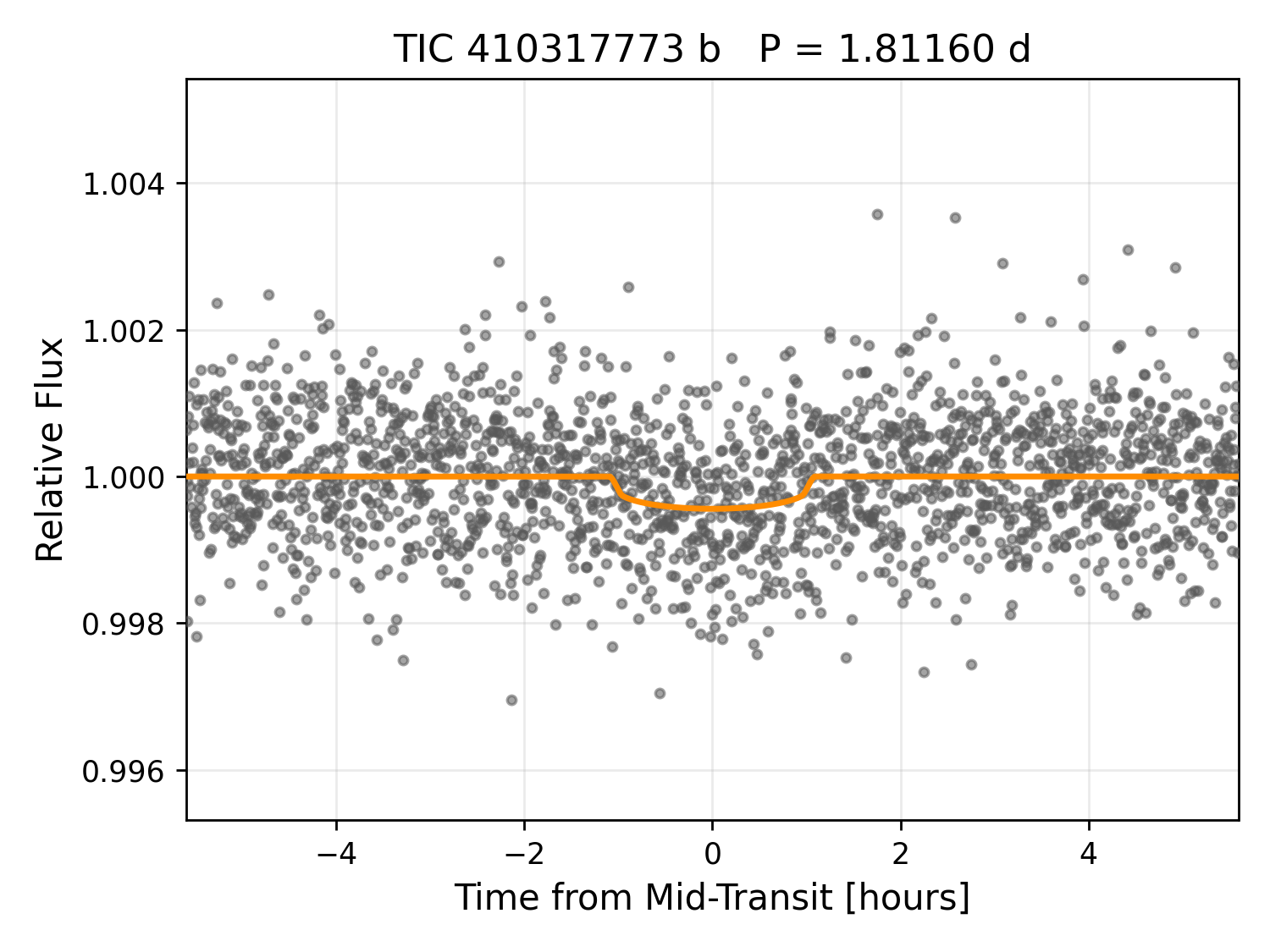}
        } &
        \subfloat[TIC 411608801\label{fov:197}]{
            \includegraphics[width=0.18\linewidth]{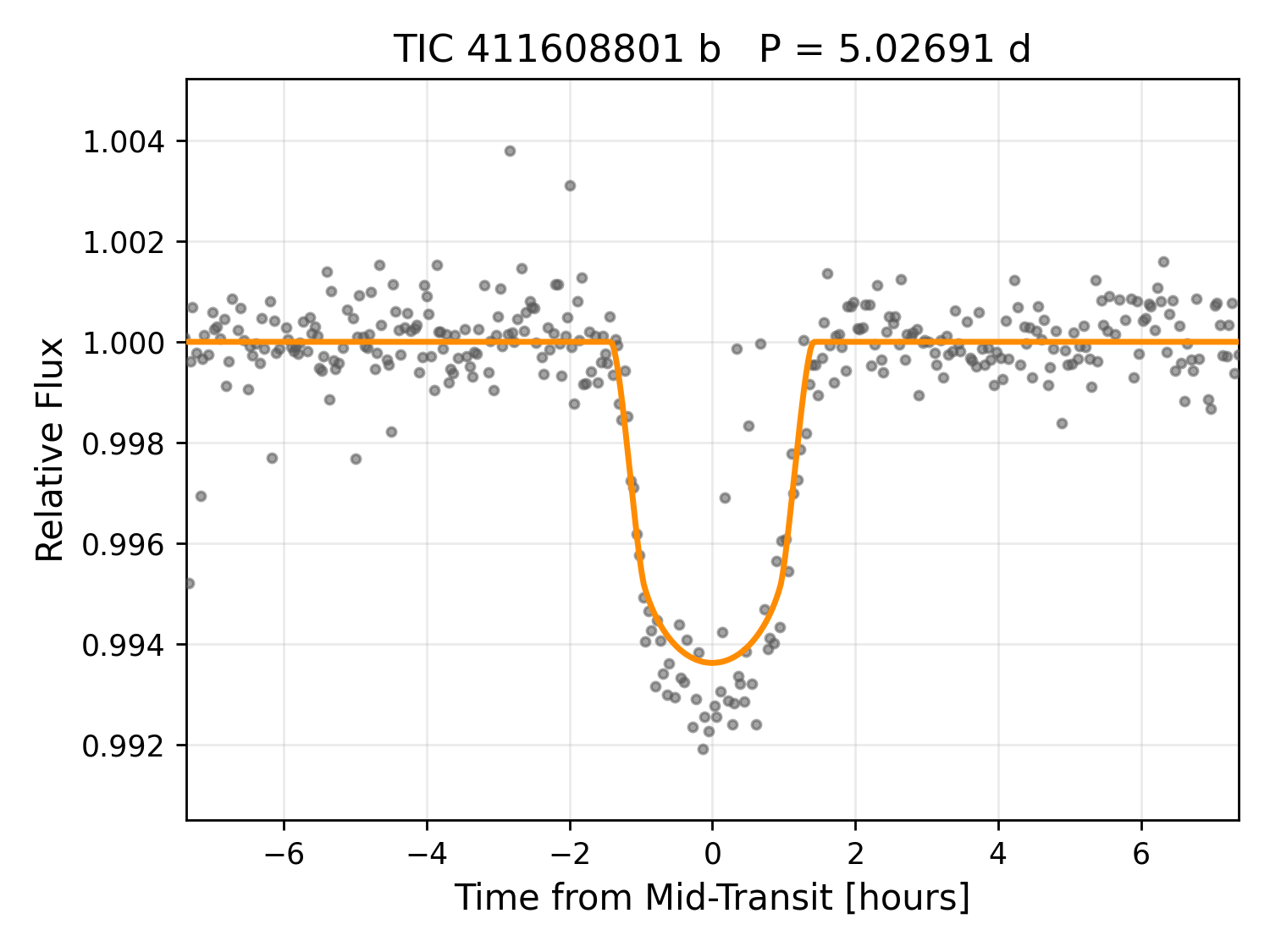}
        } &
        \subfloat[TIC 417646390\label{fov:198}]{
            \includegraphics[width=0.18\linewidth]{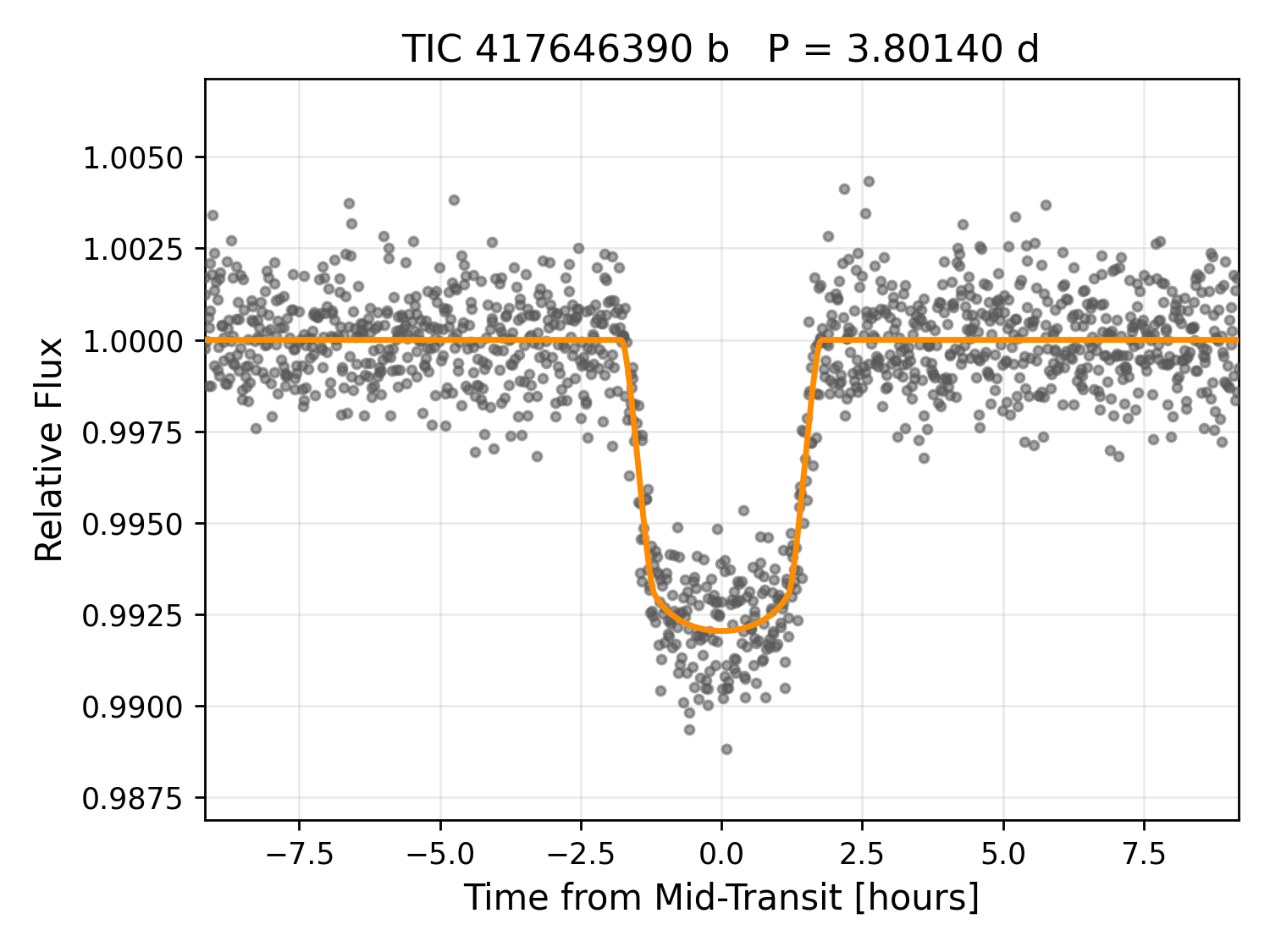}
        }
        \\

        \subfloat[TIC 419878733\label{fov:199}]{
            \includegraphics[width=0.18\linewidth]{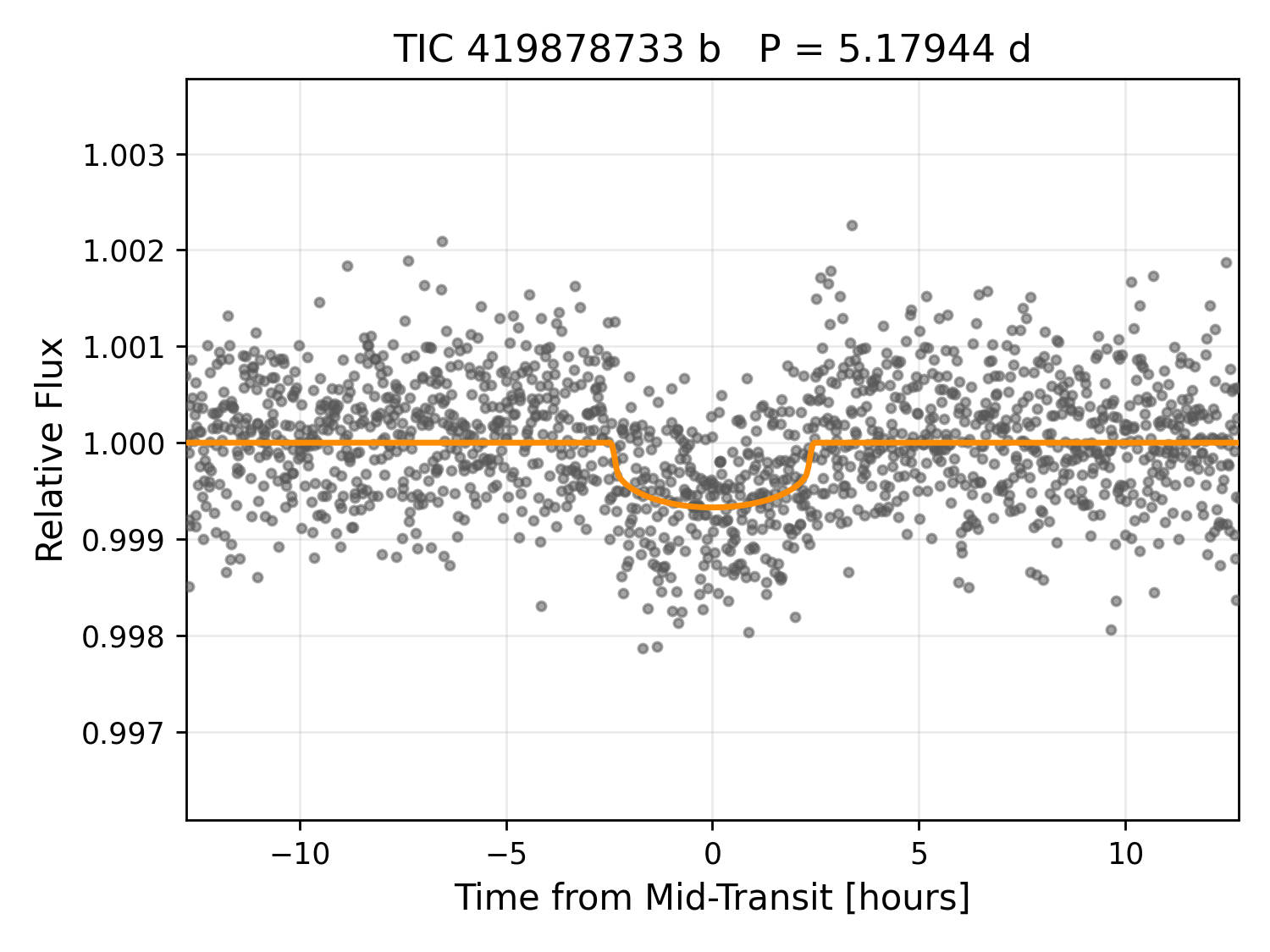}
        } &
        \subfloat[TIC 422655579\label{fov:200}]{
            \includegraphics[width=0.18\linewidth]{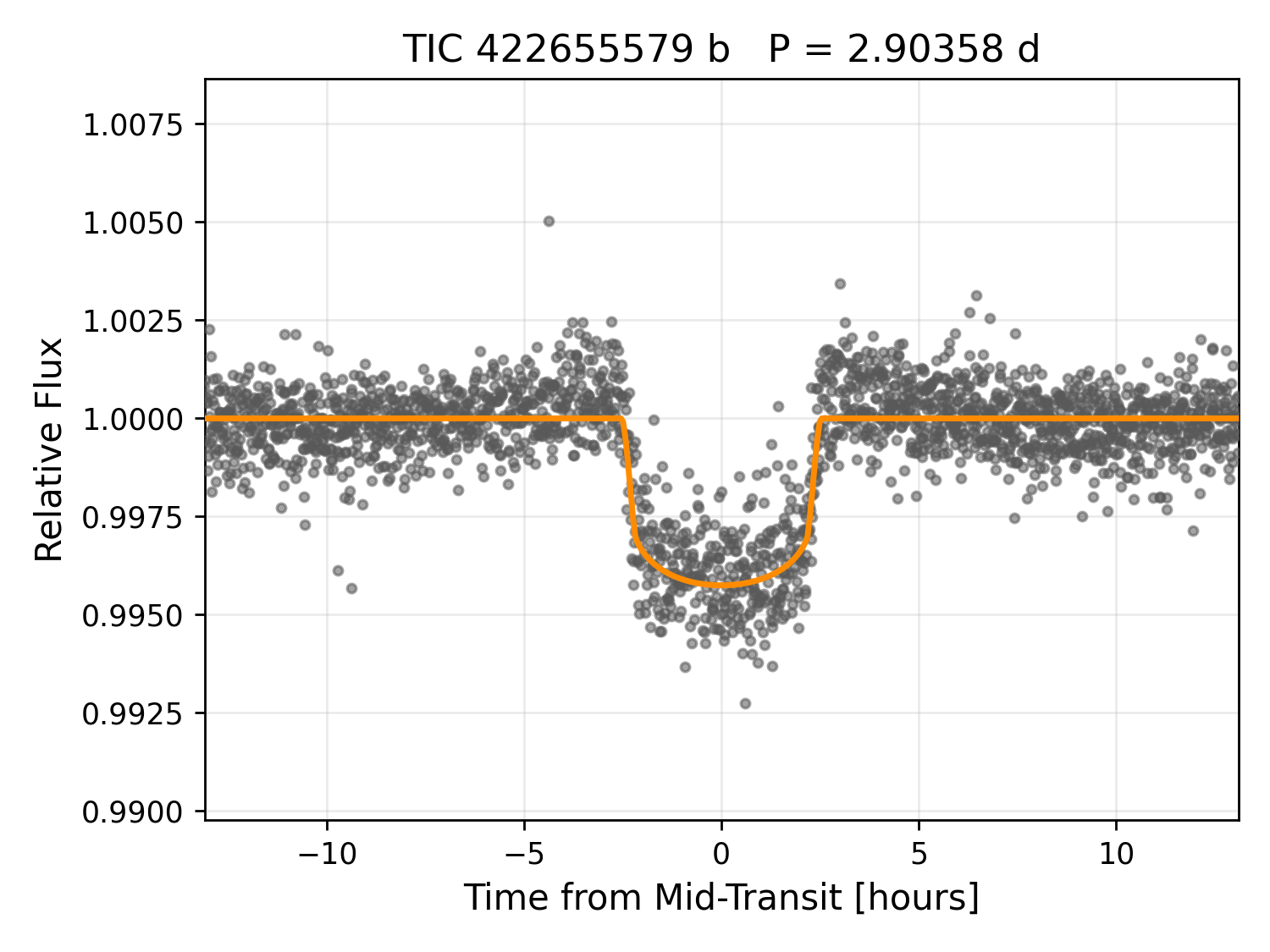}
        } &
        \subfloat[TIC 425206121\label{fov:201}]{
            \includegraphics[width=0.18\linewidth]{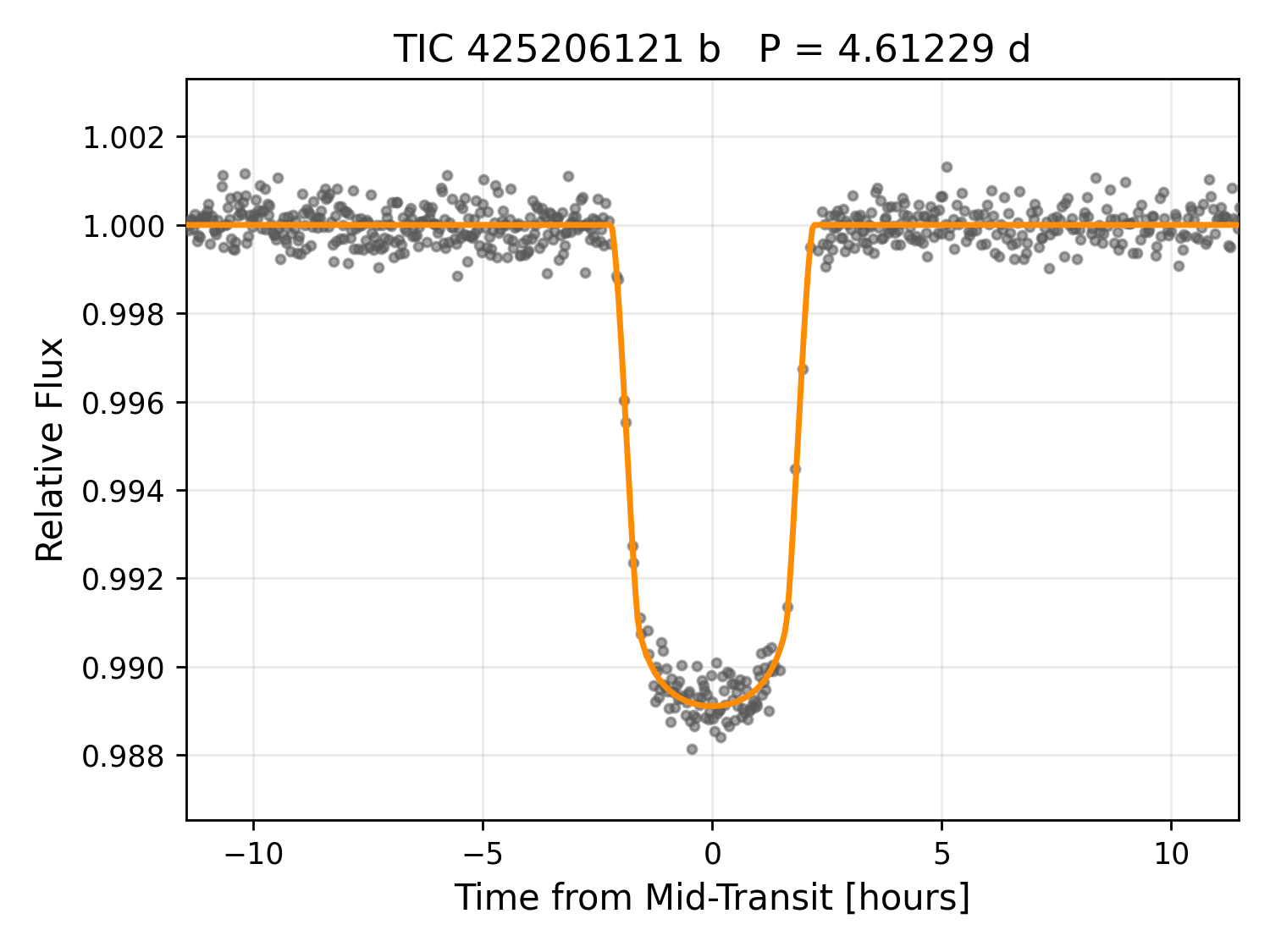}
        } &
        \subfloat[TIC 427761355\label{fov:202}]{
            \includegraphics[width=0.18\linewidth]{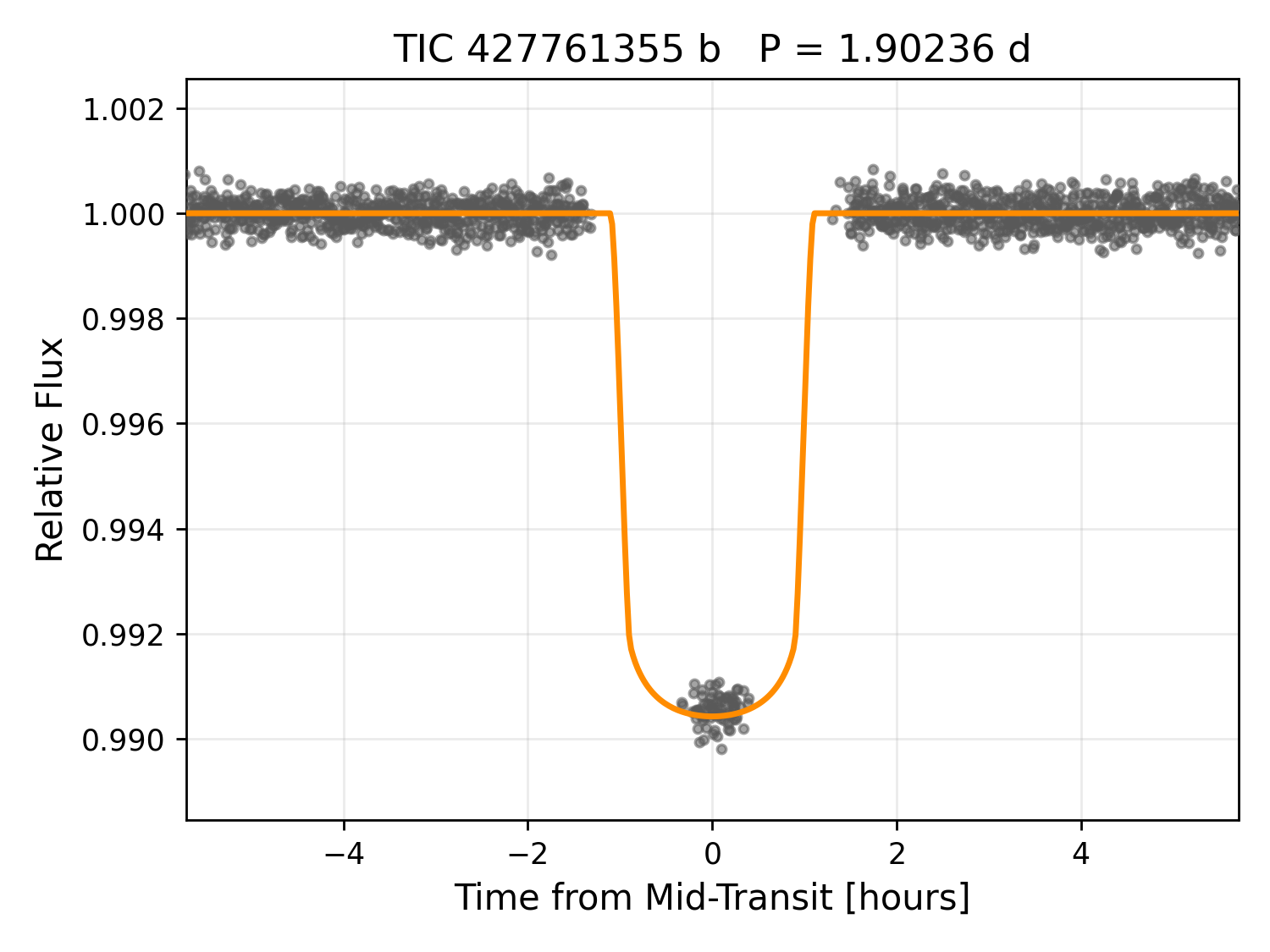}
        } &
        \subfloat[TIC 427994565\label{fov:203}]{
            \includegraphics[width=0.18\linewidth]{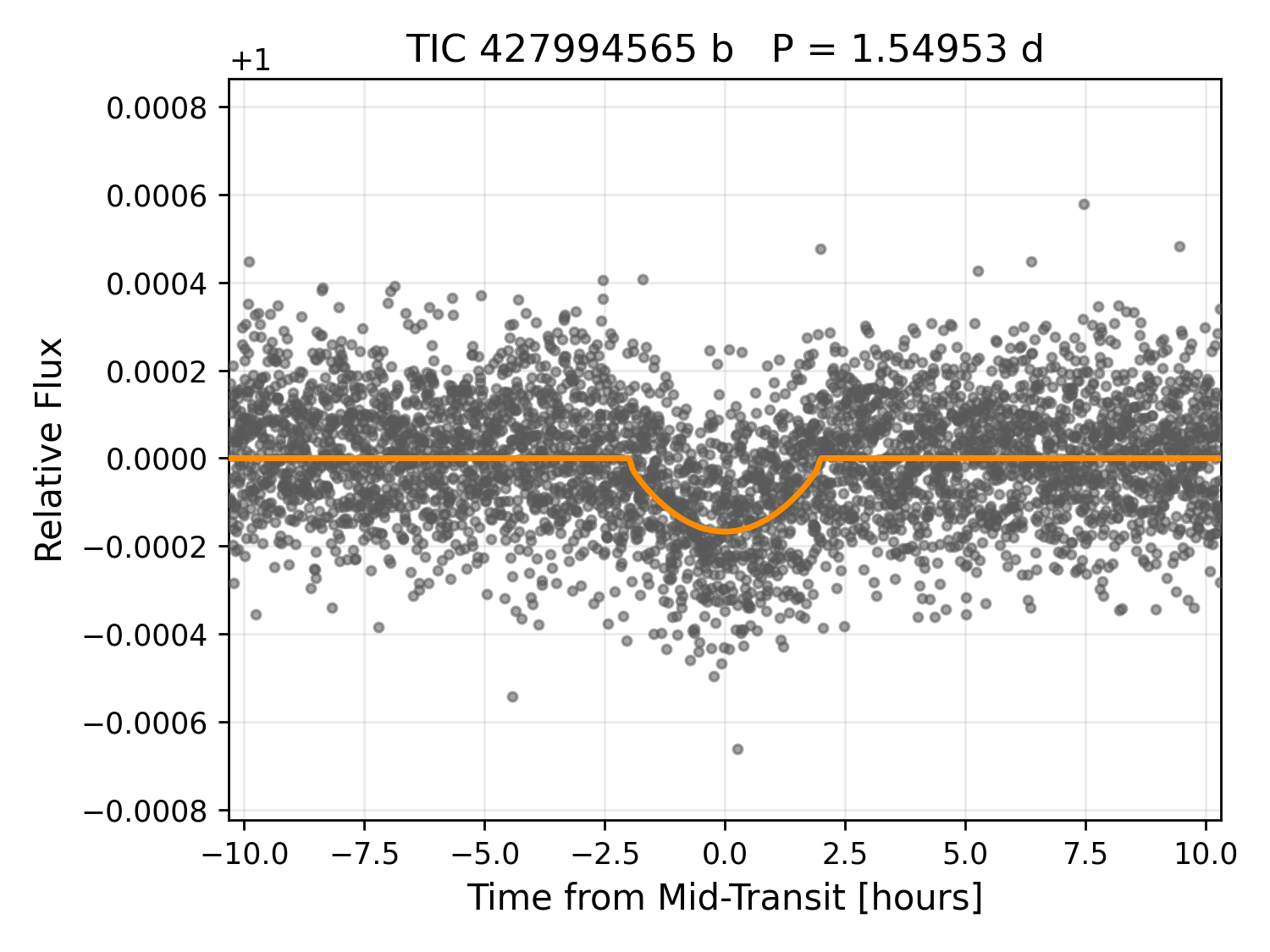}
        }
        \\
\end{tabular}
    \vspace{2pt}
    \caption{Selected Transits}
    \label{fig:appendix_transit}
\end{figure}

\begin{landscape}
\section{Planet and orbital parameters of TOI from Fig.~6}

\small
\renewcommand{\arraystretch}{1.50}

\begin{longtable}{cccccccc}
\caption{Planet and orbital parameters.}
\label{tab:toi_params} \\
\toprule
TIC Id &
\textit{TESS} Sectors &
\begin{tabular}[c]{@{}c@{}}Orbital Period\\ (days)\end{tabular} &
\begin{tabular}[c]{@{}c@{}}Planet Radius\\ ($R_\oplus$)\end{tabular} &
Impact Parameter $b$ &
\begin{tabular}[c]{@{}c@{}}Inclination Angle $i$\\ (deg)\end{tabular} &
\begin{tabular}[c]{@{}c@{}}Semi-major Axis\\ (AU)\end{tabular} &
\begin{tabular}[c]{@{}c@{}}False Positive\\ Probability\end{tabular} \\
\midrule
\midrule
\endfirsthead
\caption[]{Planet and orbital parameters (continued).} \\
\toprule
TIC Id &
\textit{TESS} Sectors &
\begin{tabular}[c]{@{}c@{}}Orbital Period\\ (days)\end{tabular} &
\begin{tabular}[c]{@{}c@{}}Planet Radius\\ ($R_\oplus$)\end{tabular} &
Impact Parameter $b$ &
\begin{tabular}[c]{@{}c@{}}Inclination Angle $i$\\ (deg)\end{tabular} &
\begin{tabular}[c]{@{}c@{}}Semi-major Axis\\ (AU)\end{tabular} &
\begin{tabular}[c]{@{}c@{}}False Positive\\ Probability\end{tabular} \\
\midrule
\endhead
\midrule
\multicolumn{8}{r}{Continued on next page}
\endfoot
\bottomrule
\multicolumn{8}{c}{%
  \parbox{0.9\linewidth}{\footnotesize\centering
  False positive probability (FPP) calculated using \textit{TRICERATOPS}. For confirmed planets, it is assumed to be 0.}%
}
\endlastfoot
TIC 100566492 & [27] & $4.95500^{+0.00060}_{-0.00058}$ & $18.54^{+0.26}_{-0.29}$ & $0.21^{+0.18}_{-0.15}$ & $88.19^{+1.26}_{-1.74}$ & $0.06^{+0.00}_{-0.00}$ & $0$ \\ \hline
TIC 101037590 & [58] & $8.70757^{+0.00090}_{-0.00082}$ & $13.17^{+0.90}_{-0.55}$ & $0.91^{+0.01}_{-0.01}$ & $85.83^{+0.24}_{-0.30}$ & $0.09^{+0.00}_{-0.01}$ & $0$ \\ \hline
TIC 101721385 & [52, 53] & $4.62770^{+0.00007}_{-0.00007}$ & $14.21^{+0.74}_{-0.54}$ & $0.80^{+0.06}_{-0.04}$ & $85.98^{+0.38}_{-0.79}$ & $0.08^{+0.00}_{-0.01}$ & $0$ \\ \hline
TIC 10394697 & [40, 52] & $1.84406^{+0.00003}_{-0.00003}$ & $3.97^{+3.97}_{-0.65}$ & $0.87^{+0.59}_{-0.18}$ & $80.26^{+4.13}_{-22.22}$ & $0.04^{+0.02}_{-0.02}$ & $0.2 \pm 0.0013$ \\ \hline
TIC 104024556 & [37] & $2.02202^{+0.00008}_{-0.00007}$ & $20.55^{+0.34}_{-0.39}$ & $0.72^{+0.03}_{-0.04}$ & $80.38^{+0.99}_{-0.82}$ & $0.04^{+0.00}_{-0.00}$ & $0$ \\ \hline
TIC 106940971 & [39] & $5.41235^{+0.00099}_{-0.00109}$ & $10.08^{+0.31}_{-0.30}$ & $0.30^{+0.23}_{-0.20}$ & $88.19^{+1.26}_{-1.77}$ & $0.06^{+0.00}_{-0.01}$ & $0.011 \pm 0.002$ \\ \hline
TIC 108940329 & [34] & $7.26472^{+0.00058}_{-0.00060}$ & $14.38^{+0.48}_{-0.51}$ & $0.48^{+0.17}_{-0.26}$ & $88.40^{+0.95}_{-0.84}$ & $0.13^{+0.01}_{-0.01}$ & $0.038 \pm 0.023$ \\ \hline
TIC 112755167 & [27] & $3.67367^{+0.00045}_{-0.00046}$ & $19.02^{+0.49}_{-0.90}$ & $0.64^{+0.07}_{-0.17}$ & $83.72^{+2.17}_{-1.23}$ & $0.05^{+0.01}_{-0.00}$ & $0.27 \pm 0.014$ \\
\hline
\end{longtable}

\end{landscape}

\end{appendix}

\end{document}